\documentclass[nohyper,11pt,letterpaper]{JHEP3}
\usepackage{amsmath,amssymb,bm,graphicx}

\newcommand{\be}{\begin{equation}}
\newcommand{\ee}{\end{equation}}

\newcommand{\ud}{\text{d}}
\newcommand{\totd}[2]{\frac{\ud {#1}}{\ud {#2}}}
\newcommand{\MM}{\mathcal{M}}

\newcommand{\df}[1]{{\bm {#1}}}
\newcommand{\ed}{\text{\bf{d}}}
\newcommand{\eu}[1]{\df{e}^{#1}}
\newcommand{\teu}[1]{\df{\tilde e}^{#1}}
\newcommand{\heu}[1]{\df{\hat e}^{#1}}

\newcommand{\eud}[2]{{e^{#1}}_{#2}}
\newcommand{\heud}[2]{{{\hat e }^{#1}}_{\;\;#2}}
\newcommand{\wud}[2]{{\df{\omega}^{#1}}_{#2}}
\newcommand{\wudd}[3]{{\omega^{#1}}_{{#2}{#3}}}
\newcommand{\wddd}[3]{\omega_{{#1}{#2}{#3}} }
\newcommand{\woddd}[3]{{\mathring{\omega}}_{{#1}{#2}{#3}}}

\newcommand{\wodd}[2]{{\df{\mathring{\omega}}}_{{#1}{#2}}}
\newcommand{\Dod}[1]{ \mathring{\nabla}_{#1} }
\newcommand{\DDo}{\mathring{\bigtriangleup}}
\newcommand{\Rodddd}[4]{ \mathring{R}_{ {#1}{#2}{#3}{#4} } }
\newcommand{\Rodd}[2]{ \mathring{R}_{ {#1}{#2} } }
\newcommand{\Rodu}[2]{ {{\mathring{R}}_{#1}}^{{\;\;#2}} }
\newcommand{\Ro}{ \mathring{R} }
\newcommand{\thud}[2]{{\df{\theta}^{#1}}_{#2}}
\newcommand{\dete}{{\rm det}(e)}
\newcommand{\dethe}{{\rm det}(\hat e)}
\newcommand{\deteM}{{\rm det}\left(e_{\MM}\right)}

\newcommand{\xiuu}[2]{ \xi^{{#1} {#2}} }
\newcommand{\xiud}[2]{ {\xi^{#1}}_{#2} }
\newcommand{\xidd}[2]{ \xi_{{#1} {#2}} }
\newcommand{\dotxidd}[2]{ {\dot\xi}_{{#1} {#2}} }
\newcommand{\xiu}[1]{ \df{\xi}^{#1} }
\newcommand{\xid}[1]{ \df{\xi}_{#1} }

\newcommand{\mavg}[1]{ {\bm\langle} {#1} {\bm\rangle} }
\newcommand{\mavgn}[2]{ {\bm\langle} {#2} {\bm\rangle}_{#1} }

\newcommand{\Lap}{ \bigtriangleup }

\title{Oxidised cosmic acceleration}

\author{Daniel H. Wesley \\
Centre for Theoretical Cosmology \\
DAMTP, Cambridge University\\
Wilberforce Road, Cambridge CB3 0WA\\
United Kingdom \\
E-mail: \email{\tt D.H.Wesley@damtp.cam.ac.uk}\\
}

\date{\today}

\abstract{
We give detailed proofs of several new no-go theorems for constructing flat four-dimensional accelerating universes from warped dimensional reduction.  
These new theorems improve upon previous ones by weakening the energy conditions, by including time-dependent compactifications, and by treating accelerated expansion that is not precisely de Sitter.  
We show that de Sitter expansion violates the higher-dimensional null energy condition (NEC) if the compactification manifold $\MM$ is one-dimensional, if its intrinsic Ricci scalar $\Ro$ vanishes everywhere, or if $\Ro$ and the warp function satisfy a simple limit condition.  
If expansion is not de Sitter, we establish threshold equation-of-state parameters $w$ below which accelerated expansion must be transient.
Below the threshold $w$ there are bounds on the number of e-foldings of expansion.
If $\MM$ is one-dimensional or $\Ro$ everywhere vanishing, exceeding the bound implies the NEC is violated. 
If $\Ro$ does not vanish everywhere on $\MM$, exceeding the bound implies the strong energy condition (SEC) is violated.
Observationally, the $w$ thresholds indicate that experiments with finite resolution in $w$ can cleanly discriminate between different models which satisfy or violate the relevant energy conditions.
}

\preprint{DAMTP-2008-11}

\begin{document}

\newpage
\section{Introduction}

Epochs of cosmic acceleration play essential roles in modern cosmological models.  Observations of type Ia supernovae (SNIa) \cite{Riess:1998cb,Perlmutter:1998np,Riess:2004nr}, the cosmic microwave background (CMB) \cite{Spergel:2006hy}, and other measurements indicate that the universe is currently undergoing a period of accelerated expansion \cite{Weinberg:1988cp,Sahni:1999gb,Padmanabhan:2002ji,Peebles:2002gy,Copeland:2006wr}.  The inflationary paradigm uses a period of nearly-de Sitter acceleration early in cosmic history to explain the flatness of the present universe, and to predict a nearly scale-invariant spectrum of primordial density perturbations.  A complete cosmological model based on more fundamental physics must at least accommodate, and should at best explain, these epochs of cosmic acceleration.  

In this work we show that accommodating accelerating universes in models with extra spatial dimensions requires the higher-dimensional theory to violate either the strong or null energy conditions (SEC or NEC, respectively).  The results proven here are releated to some well-known ``no-go" theorems, which show that \emph{static} compactifications cannot yield pure de Sitter solutions, or accelerating Friedmann-Robertson-Walker (FRW) universes, unless the SEC is violated \cite{Gibbons:1985,Maldacena:2000mw}.   Here we describe two ways in which these no-gos can be improved, as briefly summarised in \cite{Wesley:2008de}.  First, in some cases the energy condition can be weakened from the SEC to the NEC, which makes the theorems much more powerful.  Second, these theorems apply when the compactification manifold is time-dependent, and expansion is not precisely de Sitter but is described by an effective equation of state parameter $w > -1$.  Results concerning expansion which is nearly -- but not exactly -- de Sitter are essential for comparison to observation, since real measurements of $w$ can never determine it with infinite precision.

The no-go theorems we present here depend on the intrinsic curvature of the compactification manifold $\MM$.  We divide the possibilities for $\MM$ into two categories:
\begin{itemize}

\item \emph{Curvature-free}:  Compact manifolds with vanishing intrinsic Ricci scalar (Specifically the intrinsic Ricci scalar for $g^{(k)}_{\alpha\beta}$ in  (\ref{e:KKmetric})).  This category includes all models with a single extra dimension, such as braneworlds \cite{Randall:1999ee,Randall:1999vf}.  It includes flat tori, such as those constructed as quotients $R^k/\Lambda$, with $\Lambda$ a lattice, and tori with nonnegative Ricci scalar \cite{Schoen:1979a,Schoen:1979b,Gromov:1980}.  Ricci-flat special holonomy manifolds with exactly $SU(n)$, $Sp(n)$, $G_2$ and $Spin(7)$ holonomy are also included \cite{Besse:1987,Joyce:2000}.  Therefore it includes the Calabi-Yau spaces and $G_2$ seven-folds that are important for realistic dimensional reductions of string and M theory \cite{Candelas:1985en,GSW2,Pol2}.

\item \emph{Curved}: Compact manifolds with non-vanishing intrinsic Ricci scalar.  If $\MM$ is Ricci-flat, then its only continuous isometries are Abelian: therefore a number of models that realise four-dimensional non-Abelian gauge symmetries through Kaluza-Klein reduction are in this category.  The curved category contains a subcategory of manifolds whose Ricci curvature and warp function satisfy the ``bounded average condition" we define in Section \ref{ss:WarpeddeSitter}.  Especially strong results hold for de Sitter expansion in this subcategory.

\end{itemize}
We describe additional technical assumption regarding $\MM$ below.
With this classification of compactification manifolds $\MM$, we prove new no-go theorems which claim:
\begin{itemize}

\item For curvature-free $\MM$, de Sitter expansion implies that the NEC is violated by the higher dimensional theory. (See Appendix \ref{ss:RFdeSitter}).

\item For curved $\MM$ which satisfy the bounded average condition, de Sitter expansion implies the higher-dimensional theory violates the NEC. (See Section \ref{ss:WarpeddeSitter}).

\item There is a threshold $w$, and below the threshold a bound on the number of e-foldings of expansion.  Exceeding the bound violates the NEC when $\MM$ is curvature-free, or the SEC when $\MM$ is curved.    (See Sections \ref{ss:RFQuintessence} and \ref{ss:CurvedQuintessence}).

\end{itemize}
The e-folding bounds are given by (\ref{e:RFefoldBoundNoCuts}), (\ref{e:RFefoldBounds410noCut}) and (\ref{e:CurvedEfoldBoundsNoCut}) and illustrated in Figures \ref{f:NECeFolds} and \ref{f:TuneCurves}. Some of their properties are summarised in Table \ref{t:Summary}.  In our conclusions (Section \ref{s:Conclusions}) we discuss some examples from the literature which illustrate the theorems.

\begin{table}
\begin{tabular*}{\textwidth}{@{\extracolsep{\fill}} || c || c | c || c | c || c ||}
\hline
 Dim. & \multicolumn{2}{c||}{Strong energy condition (SEC)} & \multicolumn{2}{c||}{Null energy condition (NEC)} & $w$ from $\Lambda$ \\
     $k$    &  $w$ for $N \le 1$ & $w$ for transient & $w$ for $N \le 1$ &  $w$ for transient    &  $w_k$\\
\hline\hline
 1 &   n/a    &   n/a    & $-0.848$ & $-0.778$ & $-0.778$ \\
 2 & $-0.569$ & $-0.394$ & $-0.792$ & $-0.667$ & $-0.667$ \\
 3 & $-0.596$ & $-0.408$ & $-0.764$ & $-0.600$ & $-0.600$ \\
 4 & $-0.613$ & $-0.417$ & $-0.747$ & $-0.562$ & $-0.556$ \\
 5 & $-0.624$ & $-0.424$ & $-0.885$ & $-0.858$ & $-0.524$ \\
 6 & $-0.632$ & $-0.429$ & $-0.822$ & $-0.751$ & $-0.500$ \\
 7 & $-0.637$ & $-0.433$ & $-0.780$ & $-0.667$ & $-0.481$ \\
 8 & $-0.642$ & $-0.437$ & $-0.750$ & $-0.600$ & $-0.467$ \\
\hline
\end{tabular*}
\label{t:Summary}
\caption{Summary of some no-go results for $k$ extra dimensions with $1 \le k \le 8$ (formulae for all other $k$ are given in the text). For each $k$ there are two pairs of columns.  The first is concerned with the strong energy condition, which is relevant when nothing is assumed about the curvature of the compactification manifold.  The second focuses on the null energy condition, relevant for curvature-free compactification manifolds.  (There is also a null energy condition no-go theorem for curved manifolds and exact de Sitter expansion.) The first column in each pair gives the four-dimensional $w$ below which less than one e-folding is possible without violating the corresponding energy condition.  The second column in each pair gives the four-dimensional $w$ below which the number of allowed e-foldings $N$ must be finite.  In the final column the four-dimensional $w_k$ obtained by compactifying a higher-dimensional cosmological constant is given.  For $w \ge w_k$ it is always possible for the higher-dimensional theory to satisfy the NEC. A perfectly ``optimal" no-go theorem would exclude eternal acceleration for all $w < w_k$, so comparing the last and penultimate columns is an indication of how optimal the theorems are.  The SEC columns for $k=1$ are marked ``n/a" since  all one-dimensional manifolds are curvature-free.}
\end{table}

To prove the no-go theorems, we invert the Kaluza-Klein dimensional reduction procedure.  In the classic Kaluza-Klein reduction, shape and size parameters (moduli) of the compactification space $\MM$ appear as scalar fields in four dimensions.  If a suitable potential is present, then one of these fields could drive accelerated expansion.  From the time evolution of this field, we can work out the time-evolution of $\MM$.  This is sufficient information to compute the Einstein tensor $G_{MN}$, and the higher-dimensional Einstein equations give the corresponding stress-energy tensor $T_{MN}$.  This implicitly requires that the desired lower-dimensional dynamics satisfies the higher-dimensional equations of motion, so we are enforcing a ``consistent" Kaluza-Klein reduction in the sense of \cite{Cvetic:2000dm,Cvetic:2003jy,Gibbons:2003gp,Cvetic:2004km}.
To prove a no-go theorem we must account for much more general possibilities: there are many ways that $\MM$ could evolve which give the same four-dimensional cosmology, and the metric moduli may not even drive expansion at all.  Nonetheless the basic idea is the reverse of the usual Kaluza-Klein philosophy: instead of starting with a specified matter content in a higher-dimensional model and reducing to four dimensions, we go the other way.  Since the opposite of reduction is oxidation, (borrowing terminology from chemistry by way of $D=3$ coset models \cite{Breitenlohner:1987dg,Cremmer:1999du}) by studying cosmic acceleration in the context of a higher-dimensional theory, we are studying oxidised cosmic acceleration.

The technique of proof for $w > -1$ rests on the concept of an ``optimal" higher-dimensional solution.  Our tool is a one-parameter family of averages on $\MM$. For each no-go, we construct scalar quantities with the property that an energy condition is violated if their averages over $\MM$ are negative.  Introducing a parameter $v$ related to the breathing mode deformation of $\MM$, we show that satisfying the energy conditions leads to inequalities that place a lower bound on $\ud v / \ud t$ and upper bounds on $v^2$.  The lower bound on $\ud v / \ud t$ depends on the choice of average, and a number of ``arbitrary" functions such  the warp factor, four-dimensional equation of state, and specific metric deformations of $\MM$.  For a judicious choice of average, the lower bound for $\ud v / \ud t$ itself has a lower bound as a function of the various arbitrary functions. The corresponding minimum has a vanishing warp factor and deformations of $\MM$ frozen, save for the breathing mode.  Saturating this bound gives a differential equation for the breathing mode, with initial conditions set by the bound on $v^2$. The solution to this equation is the ``optimal" solution: any other choice of warp factor or metric deformation of $\MM$ yields fewer e-foldings of expansion.  In this way we bound the results of arbitrary warp factors or time-dependence of $\MM$ by studying a much simpler unwarped compactification with only breathing-mode dynamics.

No-go theorems that probe NEC violation are very useful, for it is the weakest of the energy conditions \cite{HawkingEllis}.  It asserts that
\be\label{e:NECdefinition}
T_{MN} n^M n^N \ge 0
\ee
for any null vector $n^M$. The NEC is not violated by any known matter field, or by unitary two-derivative quantum field theories.\footnote{The NEC can be violated pointwise in quantum field theory (QFT) by the Casimir effect.  Nonetheless there is evidence that an averaged NEC is respected in QFT, and this has been proven in some circumstances.  In this work we only study averaged NEC violation, and from this point of view the NEC should be respected by ``reasonable" QFTs.}   Often NEC violation signals a pathology in the underlying theory, and it seems likely that most ``well-behaved" theories should satisfy the NEC.  There are some rigorous formulations of this belief, where NEC violation is shown to lead to superluminal propagation, instabilities, and violations of unitarity or causality \cite{Tipler:1976bi,Tipler:1977eb,Friedman:1993ty,Cline:2003gs,Hsu:2004vr,Dubovsky:2005xd,Buniy:2006xf}.  Certainly the NEC forbids a number of solutions to Einstein's equations with strange properties: traversable wormholes \cite{Morris:1988cz,Visser:2003yf}, superluminal ``warp drives" \cite{Alcubierre:1994tu,Krasnikov:1995ad,Everett:1997hb,Pfenning:1997wh,Olum:1998mu,Low:1998uy}, time machines \cite{Morris:1988tu,Hawking:1991nk}, universes with big rip singularities \cite{Caldwell:1999ew,Caldwell:2003vq}, and pathologies with gravitational thermodynamics \cite{Rubakov:2004eb,Dubovsky:2004sg,Dubovsky:2006vk,ArkaniHamed:2007ky,Eling:2007qd}
are possible with NEC-violating ``exotic" matter.\footnote{Actually a number of these solutions only require matter which violates the weak energy condition (WEC).  Since NEC violation implies WEC violation, then if NEC-violating matter exists these solutions are not forbidden.}  (If one considers non-Einstein gravity, these conclusions may differ: for example, the NEC can be violated by a scalar field in Brans-Dicke gravity in Jordan frame \cite{Boisseau:2000pr} without allowing wormholes \cite{Bronnikov:2006pt}). Whether exotic NEC-violating possibilities should be allowed is perhaps a matter of taste.  It would be very interesting if accommodating cosmic acceleration implies the exotic matter required by these solutions must exist.

Some previous no-go theorems \cite{Gibbons:1985,Maldacena:2000mw} showed that de Sitter expansion or accelerating cosmologies obtained from static compactifications must violate the SEC, which asserts that 
\be\label{e:SECdefinition}
R_{MN} t^M t^N =
\left( T_{MN} - \frac{1}{D-2} T g_{MN} \right) t^M t^N \ge 0
\ee
for any non-spacelike vector $t^M$ in $D$ spacetime dimensions.
 The SEC is a much stronger energy condition than the NEC.  There are perfectly consistent systems that violate the SEC: perhaps the simplest example is a scalar field with mass term.   For this reason it seems that violating the SEC does not necessarily imply a pathology.  Nonetheless, the SEC is a useful energy condition for it is satisfied by a large class of higher-dimensional models.  The classical M theory action and a variety of supergravities all satisfy the SEC.  The no-go theorems presented here indicate that if such models are to accommodate cosmic acceleration, then the classical theory is not enough.  Acceleration must be due to quantum effects, or to other objects in the theory (such as D branes, which violate the SEC) in an essential way.  
While static compactifications which satisfy the SEC are already known to exclude accelerating universes (which have $R_{00} < 0$), we show here that even time-dependent compactifications must violate the SEC to obtain nearly-de Sitter accelerated expansion.

The new no-go theorems have important consequences for experiments which seek to measure the effective $w$ with precision, such as SNIa searches, weak lensing surveys, CMB measurements, and large-scale structure observations  \cite{Huterer:2001yu,Huterer:2000mj,Maor:2001ku,Tegmark:2003ud,Upadhye:2004hh,Ishak:2005we}.  From a purely four-dimensional viewpoint, there is no way to distinguish between a cosmological constant $\Lambda$ and other models purely by measurements of $w$.  For example, by using a ``slow-rolling" scalar field and flattening its potential, one can engineer a model in which $w$ approaches $w=-1$ arbitrarily closely, and so cannot be distinguished from $\Lambda$ by any experiment with finite resolution in $w$.  This statement rests on the assumption that one can make the potential as flat as one pleases.  The no-go theorems indicate that there are thresholds in $w$, and if we wish to push $w$ below these thresholds, we must violate an energy condition in the higher-dimensional theory.  As the thresholds are crossed, nothing significant happens in the four-dimensional theory, but the no-go theorems show that something significant must happen in the higher-dimensional theory from which it derives.  The existence of these thresholds indicates that finite resolution in $w$ is enough to give us important information about the higher-dimensional theory.  If a dark energy model derives from a fundamental physics model which satisfies the relevant energy condition, then a finite resolution in $w$ suffices to rule it out.

Whenever we make statements about the four-dimensional scale factor, we always refer to the scale factor in the Einstein conformal frame.  This frame is uniquely defined as the one in which the four-dimensional action has the usual Einstein-Hilbert form 
\be
S_{\rm 4D} = \frac{1}{2\ell_4^2}\int R(g) \sqrt{-g} \, \ud^4 x + \text{other terms}
\ee
or equivalently as the frame in which the four-dimensional Planck length $\ell_4$ is constant.  Actions in which Ricci terms appear as $g(\phi) R$, with $\phi$ other fields in the theory, or $f(R)$ models can all be cast into Einstein frame form through a suitable conformal redefinition of the four-dimensional metric.\footnote{In some contexts a different definition of the four-dimensional metric is made. For example, when branes are present the four-dimensional metric is often defined as the induced metric on the brane \cite{Shiromizu:1999wj,DeWolfe:1999cp,Brax:2004xh}, which is not the same as the Einstein frame metric used here.}  

We assume that the higher-dimensional action is of Einstein-Hilbert form.  This includes the actions for supergravities and the (low-energy approximations to) superstring theories, for just as in the four-dimensional case, if the higher dimensional Ricci scalar appears in a term of the form $f(R)$ or $g(\phi)R$ then we can bring it into Einstein-frame form through a suitable conformal redefinition  of the higher-dimensional metric.  When we discuss violations of the energy conditions in the higher-dimensional theory, we always mean the energy conditions as applied in the higher-dimensional Einstein frame.  We can have arbitrary additional matter fields in the theory, but we assume that derivatives of the metric only appear in the Ricci term, which ensures that it captures all of the contributions to the ``kinetic energy" associated with deformations of the compactification space $\MM$.  In string and supergravity models the action often contains higher powers of the curvature, which could conceivably be important for keeping the higher-dimensional theory consistent in the face of apparent NEC violation.  This is a logical possibility but perhaps an unlikely one, for each additional power of the curvature comes along with inverse powers of the higher-dimensional Planck length $\ell_{4+k}$, and so these terms will be suppressed by positive powers of $H \ell_{4+k}$.  Nonetheless it would be interesting to learn that this crude argument is wrong, for this would indicate these terms must play an essential role in a consistent model of cosmic acceleration.

We make mild assumptions about the compactification manifold $\MM$.  The manifold $\MM$ is allowed to have boundaries, but only those that arise from orbifolding.  We thus assume that $\MM$ is compact and closed, or that $\MM = \MM' / G$ where $\MM'$ is closed and compact and $G$ is a group which acts on $\MM'$.  In the latter case, we take all calculations to be carried out on the covering space $\MM'$.  The warp factor must be sufficiently well-behaved so that integration by parts is possible and that the four-dimensional Planck mass is finite.  Distributional stress-energy sources such as branes are covered here, as are certain types of singularities in the curvature or warp factor.  Our arguments here rely on averaging quantities over $\MM$, so we require the curvature and warp term to have finite integrals with the weighted measures introduced in Section \ref{ss:MetricParam}.  When the four dimensional universe is not exactly de Sitter, we make an additional assumption about the time-evolution of $\MM$.  The assumption amounts to excluding volume-preserving transformations of a certain type.  When the moduli space approximation applies, we show  the restriction we make is merely a gauge choice.  If no such restriction is made, the four-dimensional effective theory has apparent ghost modes and so the four-dimensional interpretation is breaking down.  For the scalar sector to have a positive-definite kinetic term some restriction on the allowed fluctuations of $\MM$ is necessary.  It is entirely possible that there are nonetheless consistent time-dependent reductions without making quite this restriction. For the no-go theorems proven here to be inapplicable to a specific time-dependent scenario, one must show that the corresponding restriction is not equivalent to ours under choices of gauge or coordinate transformations.

We organise this paper as follows.  
In Section \ref{s:Simple} we illustrate, using a simple breathing-mode example, how NEC violation is related to cosmic acceleration.  We also define the averaging processes and notation that we employ throughout the rest of this paper.
In Section \ref{s:RicciFlat} we study the case where $\MM$ is curvature-free.  We derive threshold values of $w$ and the NEC contraints on the number of e-foldings for $w$ below the threshold.  In Section \ref{s:Curved} we study the case where $\MM$ is curved, 
and give transience constraints similar to the curvature-free case, but with the SEC instead of the NEC.  We also show that when $\MM$ satisfies the bounded average condition, then exact de Sitter expansion violates the NEC.  We present our conclusions in Section \ref{s:Conclusions}.

\section{Background}\label{s:Simple}

In this section we motivate the new no-go theorems, by showing how NEC violation and cosmic acceleration are related in a simple breathing-mode compactification in Section \ref{ss:BreathingMode}.  We then describe the averaging procedures that we use, and the scalar mode restriction that we impose, in Section \ref{ss:MetricParam}.

\subsection{A simple example -- the breathing mode case}\label{ss:BreathingMode}

In the general cases studied in later sections, the averaging techniques make the physics somewhat less than transparent.  
However, a substantial part of the physics involved in  the Kaluza-Klein oxidation can be understood through models in which only the breathing mode (dilation) of $\MM$ is dynamical, and where the corresponding scalar field drives accelerated expansion in four dimensions.  The discussion here compliments other work discussing breathing-mode dynamics and energy condition violation \cite{Carroll:2001ih,Teo:2004hq,Wesley:2006jn}.

We first show that the null energy condition must be violated if the extra dimensions are flat and static.  We consider $k$ extra dimensions, with a factorizable $(4+k)$-dimensional metric $\ud s_{4+k}^2$ given by 
\be
\ud s^2_{4+k} = \ud s^2_4 + \ud s^2_k
\ee
For definiteness we take $\ud s_k^2$ to be the flat metric $\ud s_k^2 = \delta_{\alpha\beta} \ud y^\alpha \ud y^\beta$ on a $k$-torus, and the  four-dimensional space to be a flat FRW universe with proper time coordinate $t$ and metric
\be
\ud s^2_4 = -\ud t^2 + a(t)^2 ( \ud x_1^2 + \ud x_2^2 + \ud x_3^2 )
\ee
We assume that the four-dimensional cosmology has a simple power-law scale factor $a(t) \sim t^p$, though similar conclusions can be obtained if $a(t)$ varies with time in an arbitrary fashion \cite{Wesley:2006jn}.  The four-dimensional universe is accelerating, with $\ddot a/a > 0$, if either $p > 1$ or if $p<0$.  The $(4+k)$-dimensional Einstein equations are
\be
t^2 G_{00} = 3p^2, \quad t^2 G_{\mu\nu} =  p(2-3p) \delta_{\mu\nu}, \quad
t^2 G_{\alpha\beta} = 3 p (1-2p) \delta_{\alpha\beta} 
\ee
The pressure $G_{\alpha\beta}$ along the $k$ extra dimensions is negative whenever $p<0$ or $p > 1/2$, and in the pure de Sitter limit $p \to \infty$ the pressure is exactly twice the negative of the energy density. In fact, this stress-energy  violates the NEC  whenever the four-dimensional universe is accelerating.  To see this, we use the NEC definition (\ref{e:NECdefinition}), and consider the 
null vector $n^M = (1, 0,0,0, \hat u)$ with $\hat u$ a unit vector.  The Einstein equations imply
\be\label{e:FrozenExtraDTMN}
t^2 T^{D}_{MN} n^M n^N = 3p(1-p)
\ee
which is negative, indicating NEC violation, whenever the four-dimensional universe is accelerating.

The presence of this negative pressure can be understood heuristically.  If we we have three spatial dimensions evolving as power laws in time, it is natural that the extra dimensions evolve as power laws as well. We can explore this possibility using the Kasner metric
\be
\ud s_{Kasner}^2 = -\ud t^2 + \sum_{j=1}^{3+k} t^{2p_j} \ud x^2_j
\ee
by taking $p_1=p_2=p_3=p$, and noting that the volume of the extra dimensional space goes like $t^q$ with $q=\sum_{j=4}^{3+k} p_j$.
This metric is a solution of the vacuum Einstein equations if the Kasner conditions
\be
\sum_{j=1}^{3+k} p_j = 3p + q = 1 \qquad \sum_{j=1}^{3+k} p^2_j =  3p^2 +\sum_{j=4}^{3+k} p_j^2 = 1
\ee
are satisfied.  The first condition implies that if $p>1/3$ then $q<0$.  This indicates that in the absence of other forces, when the three noncompact dimensions are expanding rapidly the other directions have a tendency to contract.  To counteract this, a negative pressure component is required -- just as a cosmological constant with negative pressure counteracts contraction in a FRW universe.  When the three noncompact dimensions are undergoing accelerated expansion, then (\ref{e:FrozenExtraDTMN}) shows that the required negative pressure is so great that the NEC is violated.

\begin{figure}
  \begin{center}
  \includegraphics[width=1.0\textwidth]{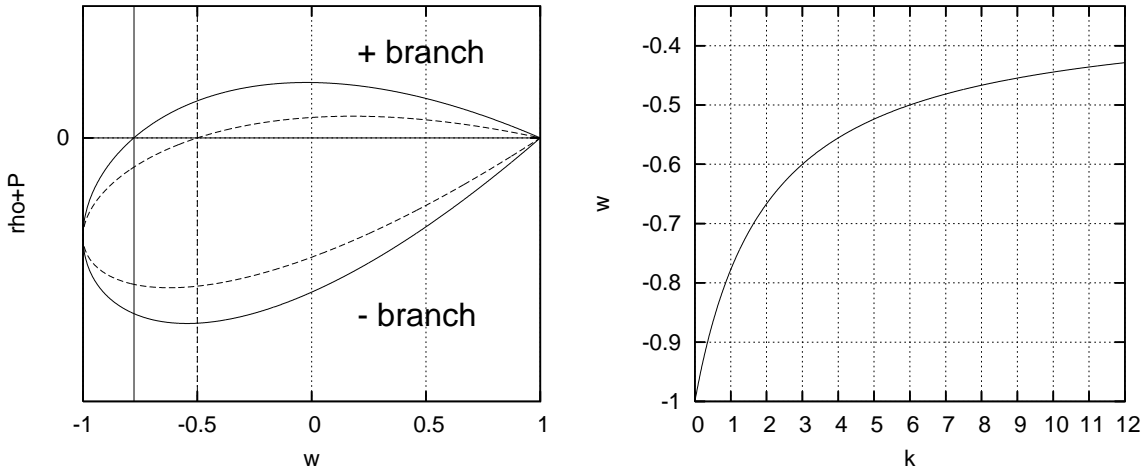}
  \end{center}
   \caption{Left panel: Higher-dimensional $\rho+P=T_{00} + T_{\alpha\beta}$ for breathing-mode solutions, with the $y$-axis in arbitrary units, and vertical lines are critical $w_k$ (\ref{e:CriticalWk}).  The cases $k=1$ (solid) and $k=6$ (dashed) are shown. The NEC is satisfied if $\rho+P > 0$,  which only happens for the  ``$+$" branch when $w_k < w < 1$, and for the ``$-$" branch when $w > 1$.  Right panel: critical $w_k$ for various dimensions, which is the $w$ that arises from compactification of a positive cosmological constant on a $k$-torus, and also the lower bound for $w$ that can be obtained by breathing-mode dynamics satisfying the NEC.}
     \label{f:breathing}
\end{figure}

One strategy to evade this problem is to allow the extra-dimensional space to evolve with time.  This possibility can be explored with the metric
\be\label{eq:simple_D_metric}
\ud s^2_{4+k} = A(\eta)^2 \left( -\ud \eta^2 + \ud x_1^2 + \ud x_2^2 + \ud x_3^2 \right) + \exp \left[ \frac{2c}{k} \psi(\eta) \right] \ud s_k^2
\ee
where $A(\eta)$ is the scale factor of the four-dimensional universe,  as measured in the $(4+k)$-dimensional Einstein frame.  It is convenient to use conformal time $\eta$, since the conformal transformation required in Kaluza-Klein reduction will not take the metric out of conformal-time form.  (This is not true if the metric (\ref{eq:simple_D_metric}) is expressed using the proper time $t$).  The volume of the extra dimensional space is parameterised by $\psi$, and $c$ is a constant given by
\be\label{e:DefOfc}
c = \sqrt{\frac{2k}{k+2}}
\ee
which is chosen so $\psi$ appears as a minimally coupled and canonically normalised scalar field in the four-dimensional Einstein frame: thus $\psi$ is the universal Kaluza-Klein breathing mode modulus. The four-dimensional Einstein-frame scale factor $a(\eta)$ is 
\be\label{eq:simple_KK_conform}
a(\eta) = e^{c\psi/2} A(\eta) 
\ee
If we assume that unspecified physics generates an effective potential for $\psi$ in four dimensions, then after Kaluza-Klein reduction the equations of motion of the Friedmann universe imply
\be
\rho+P = 3(1+w) H^2 = \left(\totd \psi \eta \right)^2
\ee
Assuming as before that the four-dimensional universe has constant $w=P/\rho$, this implies
\be
\psi(\eta) = \pm  \frac{2\sqrt{3(1+w)}}{1+3w} \ln \eta + \psi_0
\ee
where the two branches correspond to the extra dimensional space growing or shrinking with time.  Knowing $\psi$ and $a(\eta)$ one can solve for $A(\eta)$ using (\ref{eq:simple_KK_conform}), and therefore all of the  $(4+k)$-dimensional metric (\ref{eq:simple_D_metric}) is determined (up to a choice of branch) by the four-dimensional parameter $w$. Up to a positive function $F$ of $\eta$ and $w$ the stress energy tensor has components
\be
T_{00} = - 
T_{\mu\nu} = F (1-w), \quad
T_{\alpha\beta} = -F(1-w) \left[ 2 \mp \sqrt{\frac{3(1+w)}{2}\frac{(k+2)}{k}} \right] \delta_{\alpha\beta}
\ee
Therefore using the same null vector as in (\ref{e:FrozenExtraDTMN}) we find the NEC requires
\be
\pm \sqrt{\frac{3(1+w)}{2}\frac{(k+2)}{k}} \ge 1
\ee
which is only possible for one of the branches, and then only when the four-dimensional parameter $w$ satisfies $w \ge w_k$ with
\be\label{e:CriticalWk}
w_k = - \frac{k+6}{3(k+2)}
\ee
as illustrated in Figure \ref{f:breathing}.
One branch always violates the NEC in the higher-dimensional theory, and for $w < w_k$, both branches violate the NEC.
This limiting $w$ goes from its most negative value of $-7/9$ at $k=1$, and rises with $k$ to asymptote to $-1/3$.  We conclude that, with only breathing-mode dynamics and flat compactification manifolds, we cannot approach de Sitter expansion without violating the NEC.

The threshold values 
$w_k$ have a simple physical interpretation.  A  cosmological constant in the $(4+k)$-dimensional theory produces a potential
\be
V_\Lambda(\psi) = V_0 \exp{( c \psi )}
\ee
in the four-dimensional theory. Such exponential potentials support ``scaling solutions'' in which the scalar field system behaves as a perfect fluid with constant  $w$, given by
\be\label{e:wFromLambda}
w = \frac{c^2}{3}-1
\ee
Using (\ref{e:DefOfc}), the value of $w$ from (\ref{e:wFromLambda}) is precisely the same as the critical value $w_k$ (\ref{e:CriticalWk}).  The lowest value of $w$ achievable with NEC-satisfying breathing-mode dynamics on a Ricci-flat compactification manifold is precisely the one obtained by compactifying a $(4+k)$-dimensional cosmological constant.  Since a cosmological constant has the most negative pressure consistent with the NEC,  it is not surprising that it sets this threshold.

In this section we have shown that flat extra dimensions  with only breathing-mode dynamics allow accelerating  universes without violating the NEC, but since $w_k \ge -7/9$ these models cannot attain  anything close to de Sitter.  There are many different ingredients that one could add to this basic picture.  The internal manifold could distort in a more complicated fashion than the breathing mode, yielding different moduli field dynamics in the four-dimensional theory.  The extra dimensional space could be static and  some other field, or perhaps entirely different physics, could cause acceleration.  There is also the possibility of warp factors and non-factorisable spacetime metrics.  In this paper we deal with all of these possibilities and conclude that  none of these changes make a significant difference.   So long as the compactification manifold $\MM$ is curvature-free, then there is a critical $w$ below which the higher-dimensional theory must violate the NEC.  In some dimensions this critical $w$ is the same as the $w_k$ defined by compactification of the higher-dimensional cosmological constant, but in most dimensions the critical $w$ is lower.  This means that in the curvature-free case there is a ``gap" between the lowest $w$ that can be achieved while satisfying the NEC, and pure de Sitter expansion.

The only possibility which does not violate the NEC is curvature of the internal manifold $\MM$.  Up to now we have described internal manifolds $\MM$ that are Ricci flat.  If $\MM$ has curvature $\Rodd a b$ then additional terms appear in the $(4+k)$-dimensional Einstein equations.  These additional terms are\footnote{Here we are using the vielbein indices defined in the beginning of Section \ref{ss:MetricParam}.}
\be\label{e:SchemEinstein}
\delta {G_0}^0 = -\frac{1}{2} \Ro \qquad
\delta {G_m}^n = -\frac{1}{2} \Ro \,{\delta_m}^n\qquad
\delta {G_a}^b =  \Rodu a b - \frac{1}{2} \Ro \,{\delta_a}^b 
\ee
These curvature terms contribute to some of the NEC conditions.  The curvature $\Ro$ does not appear in the NEC condition with a null vector of the form $n^M = (1,\hat u,\vec 0)$, with $\hat u$ a three-dimensional unit vector.  But if a null vector such as $n^M = (1,\vec 0, \hat u)$ is chosen, with $\hat u$ a $k$-dimensional unit vector, then the NEC condition gains a contribution
\be
\delta T_{MN} n^M n^N = \Ro
\ee
If  $\Ro$ is adjustable, then by adjusting $\Ro$ we can ensure that the NEC is satisfied regardless of any other contributions.  However this solution seems to require a specific kind of fine-tuning.  Contributions to the Einstein equations from the evolution of the four-dimensional universe are of the order of $H^2$, with $H$ the four-dimensional Hubble parameter.  While $\Ro$ can be made very positive to keep the NEC satisfied, in order to satisfy the Einstein equations the stress-energy must be finely tuned so that it cancels off against the $\Rodd a b$ terms leaving a residue of order $H^2$.  This is the usual tuning problem with the cosmological constant.  While the cosmological constant can be accommodated with this tuning, it is not explained by it.  If we are prepared to tune the curvature of $\MM$ to $\sim 1$ meV, why not accept the tuning of the four-dimensional theory and dispense with a higher-dimensional explanation entirely?

It may be that solutions of this form are natural in the context of some  models.  Even so, we show that these accelerating solutions always fall afoul of the SEC if they approximate de Sitter expansion.  Curvature evades the NEC because  $\Rodd a b$ appears explicitly in the NEC inequalities.  Taking traces of the Einstein terms (\ref{e:SchemEinstein}), we can construct a one-parameter family of linear combinations of traces in which $\Ro$ does not appear.\footnote{Modulo an irrelevant overall rescaling by a positive coefficient.}  One member of this family is precisely the combination that defines the SEC.  The $00$-component of the SEC involves the combination
\be
{T_0}^0 - \frac{T}{2+k} = \frac{k+1}{k+2} {G_0}^0 - \frac{1}{k+2} \left( {G_m}^n {\delta_n}^m + {G_a}^b {\delta_b}^a \right)
\ee
which is independent of $\Rodd a b$.  We show below that it is possible to prove no-go theorems for SEC violation that are similar in spirit to those for the NEC.  The SEC is a stronger energy condition than the NEC, and can be violated in systems that are not pathological.  However it is satisfied by the classical eleven dimensional supergravity action that forms the low-energy limit of M theory, as well as by other supergravities.  While SEC violation does not indicate pathologies, it does indicate that any attempt to obtain an accelerating universe must go beyond the supergravity approximation and appeal to other essentially ``stringy" features, such as D or M branes, which can violate the SEC.

One of the surprising results proven here is that, even when $\Ro$ is tuned, it may still be impossible to satisfy the NEC when four-dimensional expansion is exactly de Sitter.   This is the case when there is a nonzero warp factor on $\MM$, so we defer discussion this result until Section \ref{ss:WarpeddeSitter}.  There are supergravity no-go theorems which forbid the presence of warp terms in supergravity compactifications when only $p$-form fluxes are present \cite{Maldacena:2000mw,de Wit:1986xg}.  Warping is only possible when SEC-violating extended objects with sufficently negative pressure are introduced.  It is precisely when we have nonzero warping that the SEC no-go extends to a NEC no-go for de Sitter cosmologies: so it seems that introducing extended objects that evade the de Sitter SEC no-go theorem can lead one afoul of the NEC no-go theorem.


\subsection{Metric parameterisation and averaging}\label{ss:MetricParam}

To extend the discussion of Section \ref{s:Simple} to a fully dynamical higher-dimensional metric, we need to parameterize the time-dependence of the metric and carry out the Kaluza-Klein reduction.  In this Section we introduce our parameterization of the higher-dimensional space an its time evolution. We also define a family of averages which will form an essential part of our argument in later sections.

The details of many related calculations are provided in the appendices.  In Appendix \ref{s:curvature}, we give the components of the curvature tensors in our chosen parameterization.  We use this in Appendix \ref{ss:EinsteinEquations} to derive the Einstein equations, after averaging, and to give the four-dimensional effective action after dimensional reduction.  For consistency of the Kaluza-Klein \emph{ansatz}, we must place a ``scalar mode restriction" on the evolution of the metric, which is equivalent to a gauge choice in the adiabatic limit.  This restriction is discussed in detail in Section \ref{ss:RestrictionDiscussion}.


Our relativity sign conventions are those of \cite{MTW}.  We use $X^M$ to denote coordinates on the full $(4+k)$-dimensional spacetime, with $M,N,\dots$ for coordinate indices and $A,B,\dots$ for tangent-space indices.  We use $t$ for (coordinate) time, $x$ for coordinates on the three ``large" dimensions, and $y$ for coordinates on $\MM$.  Greek indices $\mu,\nu,\dots$ and $\alpha,\beta,\dots$ are used for coordinate indices in the three large dimensions and on $\MM$, repectively.  While usually $\mu,\nu,\dots$ encompass $1,2,3$, occasionally they will be taken to go from $0,\dots,3$ where there is little risk of confusion.  Latin indices $m,n,\dots$ and $a,b,\dots$ are used for tangent-space indices on the three large dimensions and on $\MM$, respectively.

We take the $(4+k)$-dimensional metric $g_{MN}^{(4+k)}$ to have the form
\be\label{e:KKmetric}
g_{MN}^{(4+k)}\ud X^M \ud X^N = e^{2\Omega(t,y)} h^{(4)}_{\mu\nu}(t) \ud x^\mu \ud x^\nu  + g^{(k)}_{\alpha\beta}(t,y)\ud y^\alpha \ud y^\beta
\ee
Here $h^{(4)}_{\mu\nu}$ is the metric which  describes the four-dimensional universe after dimensional reduction.  It is conformally related to the Einstein frame metric as described below.  Since the target Einstein frame metric is a that of a flat FRW universe, $h^{(4)}_{\mu\nu}$ has the form
\be
h^{(4)}_{\mu\nu} \ud x^\mu \ud x^\nu = -N(t)^2 \ud t^2 + A(t)^2 \delta_{mn} \ud x^m \ud x^n
\ee
Here $g^{(k)}_{\alpha\beta}$ is the (intrinsic) metric of the extra-dimensional space $\MM$.  It can depend in an arbitrary way on the extra-dimensional coordinates $y^\alpha$ and on time, but the FRW symmetry means it cannot depend on $x^m$.  There is also a warp factor $\Omega(t,y^\alpha)$ which is consistent with the FRW symmetry.  This metric is certainly general enough to describe the vast majority of Kaluza-Klein compactifications.  Whether it is the \emph{most} general metric that dimensionally reduces to a flat FRW universe is a subtle question, which to some extent we address in Appendix \ref{ss:RestrictionDiscussion}.

For calculational convenience is is easier to study the $(4+k)$-dimensional spacetime using the Mauer-Cartan formalism \cite{MTW,Eguchi:1980jx}.  The metric is completely encoded in vielbeins $\eu A$ through 
\be
g_{MN}^{(4+k)}\,\ed X^M \otimes\ed X^N = \eta_{AB}\,
\eu A \otimes \eu B
\ee
with $\eta_{AB}$ the $(4+k)$-dimensional flat Minkowski metric.  The vielbeins are given by
\begin{subequations}\label{e:TheVielbeins}
\begin{align}
\eu 0 & = e^{\Omega(t,y)} N(t) \, \ed t \\
\eu m & = e^{\Omega(t,y)} A(t) \, \ed x^\mu \\
\eu a & = \eud a \alpha (t,y) \, \ed y^\alpha
\end{align}
\end{subequations}
and precisely reproduce the original metric (\ref{e:KKmetric}).  Throughout the rest of the paper tensor components are given with vielbein indices.  This is equivalent to referring the tensors to a non-coordinate basis defined by the components of the vielbeins.

Though we have no explicit parameterisation of the metric itself,  we can  parameterise its ``velocity" on the space of metrics.
The components of this velocity, projected along the tangent space of $\MM$, are denoted $\xi_{ab}$ and defined by 
\be
\totd {\eu a} {t} \Big|_{T\MM} =  {\xi^a}_b \eu b
\ee
In general each vielbein has time derivatives with components along both the time direction and along the $\MM$ direction, but we only label the components corresponding to $\xi_{ab}$.
It is useful to decompose this velocity $\xi_{ab}$ into a rotational part $\omega_{ab}$, a trace part $\xi$, and a shear component $\sigma_{ab}$ as
\be\label{e:VDecomp}
\xi_{ab}=  \omega_{ab}  + \frac{\delta_{ab}}{k} \xi  + \sigma_{ab} 
\ee
where $\omega_{ab}$ is antisymmetric, $\sigma_{ab}$ is symmetric and traceless, and $\xi$ is the trace.  The individual components are uniquely defined by
\be
\omega_{ab} = \xi_{[ab]}, \quad \xi = \delta^{ab}\xi_{ab}, \quad
\sigma_{ab} = \xi_{(ab)} - \frac{\delta_{ab}}{k}\xi
\ee
Not all of these quantities are physical, since there are coordinate transformations which preserve the form specified for the vielbeins.  In the metric representation this would be all the redundancy there is, but
in the Mauer-Cartan formalism there is also the freedom to rotate by position-dependent tangent space index transformations
\be
\eu a \to {\Phi^a}_b(t,y^\alpha) \eu b, \quad
\totd{}{t} \Phi_{ab} = A_{ab} = - A_{ba}
\ee
The matrix $A_{ab}$ has the same number of degrees of freedom and symmetries as $\omega_{ab}$ and can be chosen to set $\omega_{ab} = 0$.  Then (\ref{e:VDecomp}) is equivalent to the metric decomposition
\be
\frac{1}{2} \frac{\ud}{\ud t} g^{(k)}_{\alpha\beta} = 
\frac{1}{k} \xi g^{(k)}_{\alpha\beta} + \sigma_{\alpha\beta}
\ee
in which $\omega_{\alpha\beta}$ does not appear.
There is residual coordinate freedom affecting various components of $\xi$ and $\sigma_{ab}$, but we leave these components free.

After dimensional reduction we must define the Einstein frame metric in four dimensions, and in this work it is convenient to express the higher-dimensional Einstein equations in terms of four-dimensional Einstein frame quantities.  The necessary calculations, expressed in the variables defined above, are given in Appendix \ref{s:curvature}.  The  Einstein frame scale factor $a$ and lapse $n$ are given by
\be\label{e:EinsteinFrameTX}
a(t) = e^{\phi/2} A(t), \quad n(t) = e^{\phi/2} N(t)
\ee
where
\be\label{e:PhiDef}
e^\phi = \ell^{-k}_{4+k} \int e^{2\Omega(t,y^\alpha)} \; \deteM \, \ud^k y
\ee
and $\deteM$ is the determinant of the vielbeins $\eu a$ that describe the metric of $\MM$, and $\ell_{4+k}$ is the $(4+k)$-dimensional Planck length.
In an unwarped compactification, $e^\phi$ would be the volume of 
the compact space measured in $(4+k)$-dimensional Planck lengths, and $\phi$ would be proportional to the universal Kaluza-Klein breathing mode modulus. To obtain a sensible Einstein-frame action in the warped case, this definition must be modified by mixing the breathing mode modulus and warp factor.  Another way to think about $\phi$ is that it measures the volume of $\MM$ not in the naive metric defined by the $\eu a$, but the volume of $\MM$ in an auxiliary conformally related metric on $\MM$.  This auxiliary metric has vielbeins $\teu a$ defined by
\be
\teu a = e^{2\Omega/k} \eu a
\ee
We will sometimes refer to the metric defined by $\eu a$ as the ``naive" metric on $\MM$, and the one defined by the $\teu a$ as the ``auxiliary" or ``warped" metric.  For the purposes of Kaluza-Klein reduction, it is more natural to think of this auxiliary metric as \emph{the} metric on $\MM$.  The specific combination of factors  appearing in (\ref{e:PhiDef}) is characteristic of warped compactifications.
As one example, it is precisely the combination of factors that appears in the Randall-Sundrum models \cite{Randall:1999ee,Randall:1999vf}, and is responsible for the nonstandard relation between the four- and higher-dimensional Planck lengths that these models exhibit.

As mentioned above, averaging over $\MM$  provides one of the essential tools employed in this work, and having defined the metric we can now define the average.  The auxiliary metric on $\MM$ defines  averages $\mavg{Q}$ of functions $Q(t,y^\alpha)$ by
\be\label{e:DefAvg}
\mavg{ Q } = e^{-\phi}\int Q(t,y^\alpha) \, e^{2\Omega} \; \deteM \, \ud^k y
\ee
This is nothing more than the average value of the quantity $Q$, but in the auxiliary ``warped" metric on $\MM$ instead of the naive one.  This average is canonical because the integral that appears in (\ref{e:DefAvg}) is precisely the one that appears when one ``integrates out" the extra dimensions in Kaluza-Klein reduction.   Since
\be
\mavg{\mavg{ Q}} = \mavg{Q}
\ee
the averaging process defines a projection operator, acting on the space of functions on $\MM$, and projecting to the subspace of constant functions on $\MM$.  Any quantity $Q$ is split into a constant mode $Q_0$ plus a ``fluctuating" mode $Q_\perp$ given by
\be
Q_0(t) = \mavg{ Q(t,y^\alpha) }, \quad Q_\perp(t,y^\alpha) = Q(t,y^\alpha) - Q_0(t)
\ee
Starting from the identity $0 = \mavg{Q_\perp}$ and differentiating leads to
\be\label{eq:DotQIdentity}
\mavg{\dot Q_\perp} = -\mavg{ 2\dot \Omega_\perp Q_\perp + \xi_\perp Q_\perp }
\ee
Therefore, unless $2\dot\Omega_\perp + \xi_\perp = 0$, $\mavg{Q_\perp} = 0$ does not imply $\mavg{\dot Q_\perp} = 0 $.  The relation (\ref{eq:DotQIdentity}) is precisely analogous to the connection coefficients of differential geometry. Splitting functions on $\MM$ into constant and fluctuation modes is similar to choosing a basis of vector fields in a manifold.  Under parallel transport, the components of a given vector change due to both ``real" changes in the vector and changes in the basis vectors at different points.  Likewise, the splitting of $Q$ into $Q_0$ and $Q_\perp$ will change not only because of changes to $Q$ itself, but also because of changes in the projection as the underlying metric evolves with time, as reflected by $\dot\Omega_\perp$ and $\xi_\perp$.

While the average weighted by volume in the auxiliary metric on $\MM$ is in a sense the most ``natural" one, it is only one member of a larger family of averages.  While other members of the family lack as clear a physical interpretation as the canonical one (\ref{e:DefAvg}), they are essential in the arguments below.  This family is parameterised by a constant $A$ and the averages denoted by $\mavgn A {\cdot}$.  They act on quantities $Q$ by
\be\label{e:DefAAvg}
\mavgn A {Q(t,y^\alpha)} = 
\left(\int e^{A \Omega}\, Q \;\deteM \ud^k y\right)
\left(\int e^{A \Omega}\, \deteM \ud^k y\right)^{-1}
\ee
The case $A=2$ is the canonical average, and so $\mavg \cdot = \mavgn 2 \cdot$.  These averages do not depend on the unphysical zero mode (constant piece) of the warp factor $\Omega$, nor do they depend on the overall volume of $\MM$ in any metric.  For different values of $A$, these more general averages are sensitive to quantities in different parts of the manifold $\MM$.
As in the case of $\mavg\cdot$, each member of this family defines a constant mode and a fluctuation.  We denote these with the subscript $A$, so
\be
Q(t)_{0|A} = \mavgn A {Q(t,y^\alpha)}, \quad
Q(t,y^\alpha)_{\perp|A} = Q(t,y^\alpha) - Q(t)_{0|A}
\ee
When comparing the components defined by two different members of this family of averages we have
\be\label{eq:AverageComparison}
Q(t)_{0|A} = Q(t)_{0|B} + f(t) \quad
Q(t,y^\alpha)_{\perp|A} = Q(t,y^\alpha)_{\perp|B} - f(t)
\ee
where $f(t)$ depends on $Q$, $A$ and $B$ but is constant over $\MM$.  This means that the definitions of the constant and perpendicular modes differ by constants when switching between different averages.  The analogue of (\ref{eq:DotQIdentity}) is 
\be\label{eq:GeneralDotQIdentity}
\mavgn A {\dot Q_{\perp|A}} = - \mavgn A { A \dot\Omega_{\perp|A} Q_{\perp|A} + \xi_{\perp|A} Q_{\perp|A} }
\ee
As in the case of (\ref{eq:DotQIdentity}), we can interpret this (\ref{eq:GeneralDotQIdentity}) as giving the connection coefficients as the decomposition basis changes with time.  Different values of $A$ correspond to different choices of the decomposition basis.  Just as connection coefficients change under a change of basis, so too does the relation (\ref{eq:GeneralDotQIdentity}).

To obtain a sensible four-dimensional theory it is necessary to place a single restriction on the evolution of the metric on $\MM$.  The restriction is
\be\label{e:TheRestriction}
2\dot\Omega_\perp + \xi_\perp = 0
\ee
Detailed arguments supporting this restriction are given in detail in Appendix \ref{ss:RestrictionDiscussion}, but here we merely describe the equations that result when (\ref{e:TheRestriction}) is assumed.  The restriction eliminates scalar deformations of $\MM$ that preserve its total volume in the auxiliary ``warped" metric.  It is essential to note that this does \emph{not} mean that the $\MM$ can only evolve by uniform ``breathing-mode" transformations, since shear-type transformations, which are volume-preserving, are allowed.  The only allowed scalar transformation which changes the volume density of $\MM$ as measured by $\teu a$ is the one associated with $\phi$, the breathing mode (dilation) of the warped metric on $\MM$. The restriction also does not mean that the metric cannot undergo any scalar transformations: the naive unwarped metric on $\MM$ can undergo volume-preserving scalar deformations, but these must be associated with changes to the warp factor as dictated by (\ref{e:TheRestriction}).  

There are two useful formulas which enable us to simplify averages of derivatives.  For general $A$ the restriction (\ref{e:TheRestriction}) is expressed differently.  Using (\ref{eq:AverageComparison}) we see that, for any other choice of average $A$, there is a function $f(t)$ which is constant on $\MM$ so that 
\be
2 \dot\Omega_{\perp|A} + \xi_{\perp|A} = f(t)
\ee
Using this in (\ref{eq:GeneralDotQIdentity}) gives
\be
\mavgn A {\dot Q_{\perp|A}} = (A/2-1) \mavgn A {\xi_{\perp|A} Q_{\perp|A} }
\ee
This formula allows us to trade time derivatives of various quantities for factors of $\xi_\perp$ within averages.  Another formula follows from integration by parts and is employed frequently below.  Denoting by $\DDo$ the Laplacian in the naive metric on $\MM$, by integration by parts we have
\be
\mavgn A { e^{B\Omega}\DDo \Omega } = - (A+B) \mavgn A { e^{B\Omega}(\partial \Omega)^2 }
\ee
provided that $\Omega$ satisfies suitable integrability requirements.\footnote{Essentially this requirement is that integration by parts is allowed.  This is where our assumption that $\MM$ has no boundaries is important.  When orbifold boundaries exist, we work on a boundary-free covering space.}  This formula plays a crucial role in dealing with warp factor contributions to the Einstein equations.

\section{Curvature-free compactifications}\label{s:RicciFlat}

In this section we describe the no-go results that obtain when the compactification manifold $\MM$ has a curvature-free metric for which  $\Ro$ vanishes everywhere.  Important examples of such manifolds are one-dimensional manifolds, flat tori realised at $\mathbb{R}^k/\Lambda$ with $\Lambda$ a lattice, tori with everywhere nonnegative Ricci scalar, and $SU(n)$, $Sp(n)$, $G_2$ and $Spin(7)$ special holonomy manifolds \cite{Schoen:1979a,Schoen:1979b,Gromov:1980,Besse:1987,Joyce:2000}.  In Section \ref{ss:RFQuintessence} we derive the e-folding bounds for the case in which the four-dimensional $w$ is constant.  These limits are summarised in Figure \ref{f:SummaryNoTune}.  This method also generalizes in a simple way to time-dependent equations-of-state.  In Section \ref{sss:TimeVaryW} we show how the results for constant $w$ can be used to put bounds on $w$ with arbitrary time-dependence.

The arguments here rest on a useful lemma, which is stated and proven in Appendix \ref{ss:Lemma}.  There is a related proof regarding NEC violation with precisely de Sitter expansion which we give in Appendix \ref{ss:RFdeSitter}.  Some technical details concerning the $k=4$ theorems are given in Appendix \ref{sss:NEC_k4}.







\subsection{Constant $w$}\label{ss:RFQuintessence}

According to the lemma proven in Appendix \ref{ss:Lemma}, if the NEC is satisfied we must have $\rho^D + P_3^D \ge 0$ and $\rho^D + P_k^D \ge 0$, with $\rho^D$, $P_3^D$ and $P_k^D$ defined in Appendix \ref{ss:EinsteinEquations}.  Using the Einstein equations (\ref{eq:4dEOM}) and averaging over $\MM$ using $\mavgn A \cdot$ yields
\be\label{eq:QuinRhoPlusP3}
n^2 e^{-\phi}
\mavgn A { e^{2\Omega} (\rho^D + P_3^D) } = n^2 (\rho_T + P_T)
- \frac{k+2}{2k} \xi_{0|A}^2 - \frac{k+2}{2k} \mavgn A {\xi_{\perp|A}^2}  - \mavgn A {\sigma^2}
\ee
where $\rho_T$ and $P_T$ are the total four-dimensional energy density and pressure, as defined in (\ref{eq:4dEOM}). This expression is
independent of the curvature and warp on $\MM$.  When the four-dimensional energy density $\rho_T$ and pressure $P_T$ satisfy $\rho_T + P_T > 0$, the NEC requirement $\rho^D + P_3^D \ge 0$ is consistent with nonvanishing kinetic energy.  This has important consequences for the $\rho^D+P_k^D \ge 0$ condition, which after averaging over $\MM$ is
\begin{align}\label{eq:QuinRhoPlusPk}
n^2 e^{-\phi}
\mavgn A { e^{2\Omega} (\rho^D + P_k^D) } = & 
\frac{n^2}{2}\left( \rho_T +3P_T \right) + 2\left(\frac{A}{4} - 1\right)\frac{k+2}{2k}\mavgn A {\xi_{\perp|A}^2} \notag \\
& - \frac{k+2}{2k} \xi_{0|A}^2 
  - \mavgn A {\sigma^2} \notag \\
& + \left[\left(\frac{4}{k} - 1\right) (A+2) + \left( 4-\frac 4 k \right) \right]
\mavgn A {e^{2\Omega} (\partial\Omega)^2} \notag \\
& + \frac{k+2}{2k} \frac{n}{a^3} \totd {} t \left( \frac{a^3}{n} \xi_{0|A} \right)
\end{align}
Some of these terms are negative definite, but some have no definite sign -- especially the last term.   Since $\ud \xi / \ud t$ can be arbitrarily large, it can compensate for any negative contributions in the previous lines, no matter how large.  Therefore there cannot be a no-go theorem for instantaneous non-de Sitter acceleration if $\MM$ is dynamical.  However, this observation  suggests a strategy for proving no-go theorems for \emph{eternal} acceleration.  It is true one can ``beat" the energy conditions if $\ud \xi / \ud t$ is large:  but if maintained indefinitely, $\xi$ itself becomes large, eventually violating the NEC condition (\ref{eq:QuinRhoPlusP3}).  In the following sections, we use this observation to bound the amount of accelerated expansion that is possible without violating the NEC.




We use these observations to prove our no-go theorems by introducing the concept of an ``optimal" solution for $\xi_{0|A}$ at a specific value of $A$.  In the context of a specific model the time evolution of $\xi_0$, $\xi_\perp$, $\sigma$ and $\Omega$ is fixed, but we imagine that we are free to choose these functions as desired.  The optimal solution is the one which allows the NEC inequalities to be satisfied for the largest number of e-foldings.  Our strategy is to seek a value of $A$ for which the coefficients of the free terms in (\ref{eq:QuinRhoPlusPk}) involving $\xi_{\perp}$, $\sigma$ and $\Omega$ are all negative, or a value of $A$ for which their sum is guaranteed to be negative.   We call such a value of $A$ and ``optimising" value.  When $A$ is fixed an an optimising value, the optimal solution for $\xi_{0|A}$ is the one in which the sum of terms involving $\xi_{\perp}$, $\sigma$ and $\Omega$ is fixed at its maximum, and for which (\ref{eq:QuinRhoPlusPk}) vanishes.  Any other choice for $\xi_{\perp}$, $\sigma$ and $\Omega$ forces $\xi_{0|A}$ to evolve more rapidly with time to satisfy the NEC condition involving (\ref{eq:QuinRhoPlusPk}), and leads to a violation of the NEC condition involving (\ref{eq:QuinRhoPlusP3}) in less time.


\subsubsection{$0 < k \le 4$ and $k \ge 10$}\label{sss:NEC04and10plus}

To begin proving the no-go theorems, we must find an optimising value of $A$.
The simplest possibility is that all terms except the final one in (\ref{eq:QuinRhoPlusPk}) are nonpositive.  The $\mavgn A {\xi_{\perp|A}^2}$ term in the top line of (\ref{eq:QuinRhoPlusPk}) requires that 
\be\label{e:Aless4cond}
A\le 4
\ee
The third line of (\ref{eq:QuinRhoPlusPk}) requires 
\be\label{eq:NonPosWarpCond}
A \le \frac{2k+4}{k-4} \; \text{for} \; k < 4
\quad \text{and} \quad
A \ge \frac{2k+4}{k-4} \; \text{for} \; k > 4
\ee
These two nonpositivity conditions are only compatible for $k<4$ and $k \ge 10$, which we study in this section, addressing the $4 < k < 10$ case in Section \ref{sss:NEC4to10}.  The $k=4$ case is a limit of the $k<4$ cases, and establishing this fact requires some technical arguments given in Appendix \ref{sss:NEC_k4}.

For $k<4$ and $k \ge 10$, where (\ref{e:Aless4cond}) and (\ref{eq:NonPosWarpCond}) can be satisfied simultaneously, an optimal solution must have $\xi_{\perp|A} = 0$, since the coefficient of $\mavgn A {\xi_{\perp|A}^2}$ is nonpostive.  The $\mavgn A \sigma^2$ coefficient is always negative so $\sigma = 0$.  Likewise, the optimal solution corresponds to taking $\Omega$ to be constant over $\MM$.  In other words the optimal solution is the one associated with a cosmology in which only the breathing mode of an unwarped $\MM$ is dynamical.  By focusing on the optimal solution we can say something about a very complicated general case by treating a much simpler situation.  The breathing mode dynamics of curvature-free unwarped compactifications have been described in Section \ref{ss:BreathingMode}.  The analysis in this section is more general, because unlike Section \ref{ss:BreathingMode} we do not assume that the breathing mode field $\xi_{0|A}$ drives the accelerated expansion.

We can now formulate a differential equation whose solution is the slowest evolution of $\xi_{0|A}$ compatible with the NEC in the higher-dimensional theory.  To put these equations in a simple form we take the lapse and scale factor to be 
\be
n = 1, \qquad a(t) = (t/t_{\rm ref})^{2/3(1+w)}
\ee
corresponding to evolution in physical time $t$ and a constant equation of state $w$.  (Time dependent $w$ is discussed in Section \ref{sss:TimeVaryW}).  We set the reference time $t_{\rm ref}=1$ as it drops out of our final results. The Friedmann equations imply
\be
\rho_T = \frac{4}{3(1+w)^2 t^2}
\ee
where $\rho_T$ is the total energy density in the four-dimensional universe, and not just that associated with the modular dynamics of $\MM$.
With these definitions the optimal solution for $\xi_{0|A}$ obeys
\be\label{e:NECDiffEqk1_4_10}
\totd {} t \xi_{0|A} 
+ \frac{2}{(1+w) t} \xi_{0|A} = 
  \xi_{0|A}^2
- \frac{2k}{k+2}\frac{2(1+3w)}{3(1+w)^2  t^2}
\ee 
which comes from the averaged NEC condition (\ref{eq:QuinRhoPlusPk}) after setting $\xi_{\perp|A}=\sigma=\Omega=0$.  This differential equation is independent of $A$ so long as (\ref{e:Aless4cond}) and (\ref{eq:NonPosWarpCond}) are satisfied.  The initial conditions for this equation are given by (\ref{eq:QuinRhoPlusP3}).  If the accelerated period of expansion starts at $t=t_0$, then the second NEC inequality involving (\ref{eq:QuinRhoPlusP3}) is saturated if
\be\label{e:XiBC1}
\xi(t_0)_{0|A} = - \frac{1}{t_0}\sqrt{ \frac{2k}{k+2} \frac{4}{3(1+w)} }
\ee
The initial value of $\xi_{0|A}$ is negative since the differential equation 
(\ref{e:NECDiffEqk1_4_10}) indicates $\xi_{0|A}$ is increasing.
The NEC condition (\ref{eq:QuinRhoPlusP3}) indicates that 
if there exists a time $t_1 > t_0$ with
\be\label{e:XiBC2}
\xi(t_1)_{0|A} = + \frac{1}{t_1}\sqrt{ \frac{2k}{k+2} \frac{4}{3(1+w)} }
\ee
then the NEC is violated.  If such a $t_1$ exists, then satisfying higher-dimensional NEC is only compatible with a transient period of acceleration.
To solve (\ref{e:NECDiffEqk1_4_10}) we introduce a new variable $v$ defined by
\be
\frac{v(t)}{t} =  \xi(t)_{0|A}
\ee
the new variable $v(t)$ is proportional to the fractional energy density contributed by $\xi_{0|A}$.  In terms of this new variable, (\ref{e:NECDiffEqk1_4_10}) becomes
\be \label{eq:vEq}
t \totd {v} t =  v^2 + \alpha v  + \beta
\ee
where
\be\label{e:alphaDef}
\alpha = \frac{w-1}{w+1}
\ee
and
\be
\beta=-\frac{4k(1+3w)}{3(k+2)(1+w)^2}
\ee
The boundary conditions (\ref{e:XiBC1}) and (\ref{e:XiBC2}) define boundary values $\pm v_F$ of $v$ by
\be\label{e:vBC}
v_F = \sqrt{\frac{2k}{k+2}\frac{4}{3(1+w)}}
\ee
These equations are only relevant when $A$ is chosen so that (\ref{e:Aless4cond}) and (\ref{eq:NonPosWarpCond}) are satisfied.  We have established that values of $A$ satisfying these conditions  exist for all $0<k<4$ and $k \ge 10$, but since the differential equation (\ref{eq:vEq}) and initial conditions (\ref{e:vBC}) are independent of $A$, the precise value chosen for $A$ is irrelevant.

Some useful information can be gleaned from the phase structure of the differential equation (\ref{eq:vEq}).  If there is to be a finite number of e-foldings, then the right-hand side of (\ref{eq:vEq}) must be nonzero for all values of 
$-v_F < v < v_F$.  The zeros of the right-hand side are located at
\be
v_0^\pm = -\alpha/2 \pm \sqrt{(\alpha/2)^2-\beta} = -\alpha/2 \pm i\Delta
\ee
So long as $\Delta$ is real, there are no zeros in the right-hand side of (\ref{eq:vEq}) between $-v_F$ and $v_F$.  $\Delta$ is real if $w < w_\Delta$, with $w_\Delta$ defined by
\be
w_\Delta = \frac{6-21k + 8 \sqrt{6}\sqrt{k^2-k}}{3(2+k)}
\ee
The value of $w_\Delta$ does not completely determine the range of $w$ for which acceleration is transient. The range is larger than $-1 \le w < w_\Delta$ if
 $\Delta$ is imaginary and the zeros lie outside the range $[-v_F,v_F]$. Then, as $v$ moves from $-v_F$ to $v_F$, it never encounters a zero, and acceleration is transient.  For fixed $k$, the position of the zeros $v_0^\pm$ and $v_F$ are both  functions of $w$: therefore there is a critical value of $w$, denoted $w_\times$, at which one of the zeros first crosses in (or out) of the range $[-v_F,v_F]$.  The right-hand side of (\ref{eq:vEq}) is positive until the smaller zero crosses into the range, which always occurs at 
\be
w_\times = -\frac{k+6}{3k+6}
\ee
This is precisely the equation of state $w_k$ obtained by compactifying a $(4+k)$-dimensional cosmological constant to four dimensions.  At some $w \in [-1,w_\times]$ there exists a zero between $-v_F$ and $v_F$.   For this reason, the transient range of $w$ cannot be larger than $-1 \le w < w_\times$.  
This is an important consistency check, for since $w_\times = w_k$, and we can always construct a model with eternal acceleration and $w=w_k$, it stands to reason that there should be no transience constraint for $w > w_\times = w_k$.

\begin{figure}
  \begin{center}
  \includegraphics[width=1.0\textwidth]{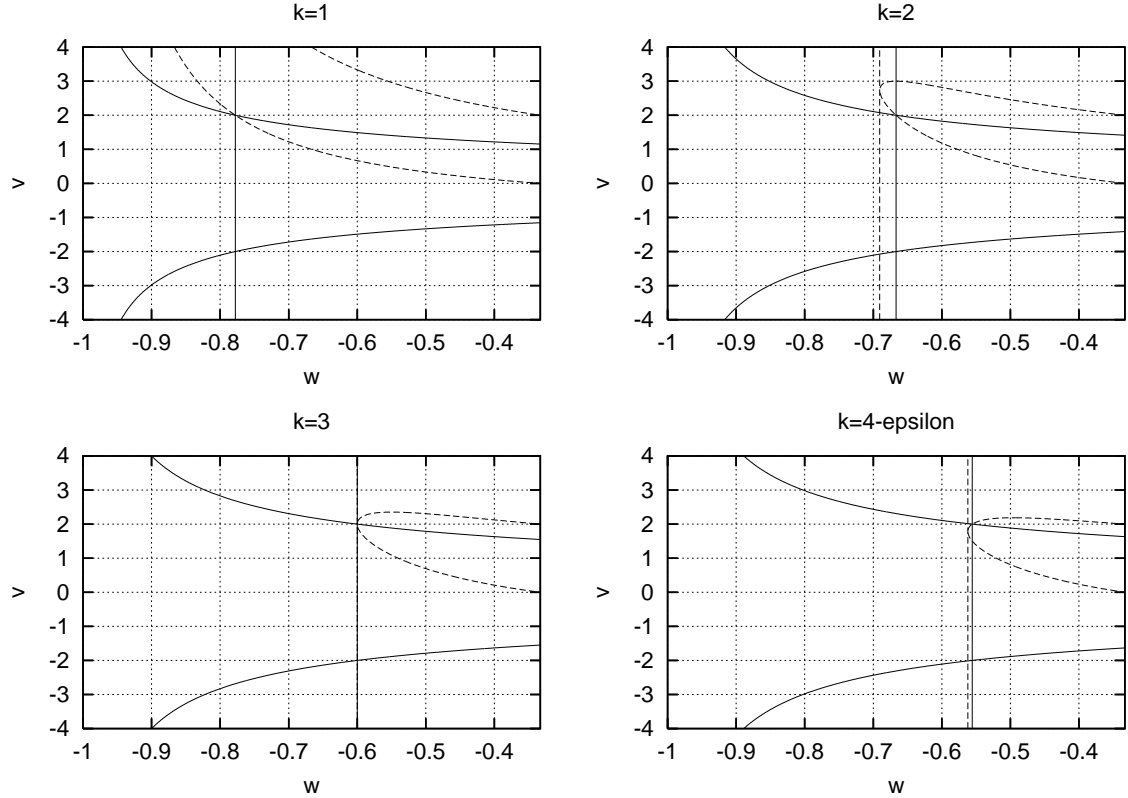}
  \end{center}
   \caption{The solid curves bound the region $[-v_F,v_F]$ outside of which the NEC is violated. For fixed $w$, trajectories with the maximal number of e-foldings begin with $v$ at the bottom solid curve, and move vertically upwards.  The dashed curves show  zeros $v_0^\pm$ of the right-hand side of (\ref{eq:vEq}), representing fixed points of the evolution of $v$.  The solid vertical lines denote the value $w_\times$ at which the fixed point moves into the region $[-v_F,v_F]$.  The dashed vertical lines at $w_\Delta$ separate regions where $\Delta$ is real (to the left) and imaginary (to the right).
   For $k=1,2,3$ any value of $w$ to the left of this line allows $v$ to reach the top curve in finite time, while for values of $w$ to the right $v$ reaches a fixed point and is trapped.  For $k=4$, the relevant limit is $w_\Delta$ which is slightly less than $w_\times$. }
     \label{f:k1-4}
\end{figure}

By studying the position of the zeros we can determine the precise range in $w$ for which there is a transience constraint.
For $k \le 3$ it turns out that $w_\Delta \le w_\times$, but the zeros of the right-hand side lie outside the range $[-v_F,v_F]$.  In these cases the right-hand side of (\ref{eq:vEq}) does not vanish unless $w \ge w_\times$, and so acceleration must be transient for all $w < w_\times$.  Since $w_\times = w_k$, this constraint is optimal, for eternally accelerating universes  with $w=w_k$ exist by construction.  For $k \ge 3$, we still have $w_\Delta < w_\times$, but now  the zero lies within the range $[-v_F,v_F]$.  For these $k$ acceleration must be transient for $w < w_\Delta$ and can be eternal otherwise.
Therefore the limit is set by $w_\times$ for $k \le 3$, and by $w_\Delta$ for $k \ge 3$.  This is illustrated in Figure \ref{f:k1-4}.

We obtain an explicit expression for the allowed number of e-folds by solving the differential equation (\ref{eq:vEq}).  
The solution is
\be
v(t) = -\frac{\alpha}{2} + \Delta \tan \left[ \Delta ( \ln t + \gamma ) \right]
\ee
where $\gamma$ is an integration constant.  This form of the solution is valid when $\Delta$ is real: otherwise we must analytically continue this expression.  
The beginning and ending conditions for $v$ are symmetrical
\be
v(t) = \pm v_F
\ee
Using these ingredients we can find the total number $N$ of e-foldings that are possible. We use
\be
N = \frac{2}{3(1+w)} \ln \left(\frac{t_1}{t_0}\right)
\ee
where $t_0$ and $t_1$ are the times at the beginning and end of acceleration, so that 
$v(t_0) = -v_F$ and $v(t_1)=+v_F$. Then by integrating (\ref{eq:vEq}) we obtain
\be\label{e:NECefolds1}
N= \frac{2}{3(1+w)} \frac{1}{\Delta}
\text{Tan}^{-1} \left[ \frac{2v_F \Delta}{ \beta - v_F^2}\right]
\ee
This is only valid for $0<k<4$ and $k \ge 10$ when (\ref{e:Aless4cond}) and  (\ref{eq:NonPosWarpCond}) is satisfied, and in this regime is independent of $A$.

\begin{figure}
  \begin{center}
  \includegraphics[width=1.0\textwidth]{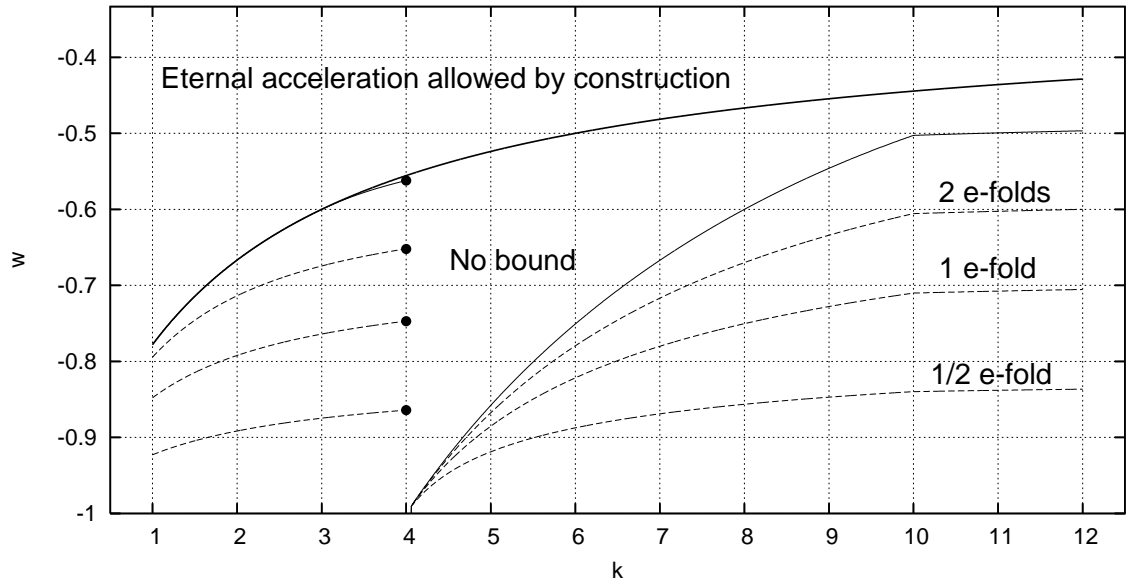}
  \end{center}
   \caption{A summary of the constraints on curvature-free models.  The upper curve denotes the maximal possible exclusion region for no-go theorems, since explicit eternally accelerating models can be constructed with these values of $w$.  Below the lower lines, models are forced to have transient acceleration or violate the NEC in the higher-dimensional theory.  Below the lower solid line, acceleration must be transient, and below the dashed lines the number of e-foldings is bounded as shown.  The  $k=4$ case is a continuation of the $k<4$ cases. }
\label{f:SummaryNoTune}
\end{figure}

\begin{figure}
  \begin{center}
  \includegraphics[width=1.0\textwidth]{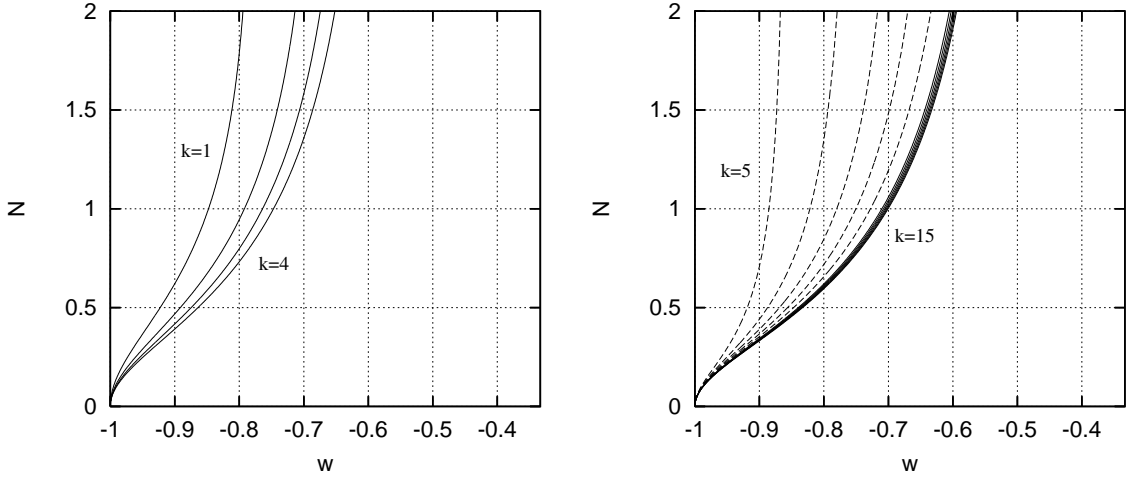}
  \end{center}
   \caption{Summary of e-folding constraints for curvature-free $\MM$,  showing the maximum number of e-folds allowed at a given value of $w$.  Left panel: from top to bottom curve, $k=1$ to 4.  Right panel: $k=5$ to 15 from left to right, dotted curves for $k=5-9$, and solid curves are for $k \ge 10$.}
   \label{f:NECeFolds}
\end{figure}

One must be careful when using (\ref{e:NECefolds1}) because of the branch cut in $\text{Tan}^{-1}$ when its argument goes to infinity at $w=-3/5$.  (This is really only an issue when $w_\Delta > -3/5$ which holds for $k > 3$).  We can circumvent the branch cut problem by using the equivalent expression
\be\label{e:RFefoldBoundNoCuts}
N = \frac{2}{3\Delta(1+w)}\left[
\text{Tan}^{-1} \left( \frac{\alpha/2 + v_F}{\Delta} \right)
- \text{Tan}^{-1} \left( \frac{\alpha/2- v_F}{\Delta} \right)
\right]
\ee
in which the arguments of both $\text{Tan}^{-1}$ are finite.  When $w > w_\Delta$ then $\Delta$ is imaginary and we should use the analytic continuation of (\ref{e:NECefolds1})
\be
N = \frac{2}{3(1+w)} \frac{1}{|\Delta |}
\text{Tanh}^{-1} \left[ \frac{2v_F |\Delta |}{ \beta - v_F^2}\right]
\ee
This is an issue for $k < 3$, where it is the crossing point $w_\times$ which determines the range of $w$ for which there is a transience constraint.
For the latter expression to be valid, the argument of $\text{Tanh}^{-1}$ must be in $[-1,+1]$, which is true provided that 
\be
w \le - \frac{k+6}{3k+6} = w_k
\ee
This is always satisfied, since we for $w > w_k$ the maximum number of e-folds is infinite.  These limits are summarised, along with those in other dimensions, in Figures \ref{f:SummaryNoTune} and \ref{f:NECeFolds}.

As discussed in detail in Appendix \ref{sss:NEC_k4}, for $k=4$ we can use the limit $k \to 4^-$ of the $k<4$ constraints described here.

\subsubsection{$4 < k < 10$}\label{sss:NEC4to10}

For $4 < k < 10$ the coefficients of the warp terms, of $\xi^2_{\perp|A}$ and of $\sigma^2$  are not simultaneously nonpositive for any $A$.  We study these cases by choosing $A$ so that the sum of multiple terms is nonpositive, even though some individual terms are themselves positive.
The essential observation is that, while the coefficient of $\mavgn A {\xi_{\perp|A}^2}$ is positive,  the term cannot be arbitrarily large thanks to (\ref{eq:QuinRhoPlusP3}). Using (\ref{eq:QuinRhoPlusP3}) reveals
\be\label{e:XiPerpBound}
\frac{k+2}{2k} \mavgn A { \xi_{\perp|A}^2 } \le n^2(\rho_T + P_T)
\ee
By saturating this inequality (\ref{eq:QuinRhoPlusPk}) gives
\be
n^2 e^{-\phi}
\mavgn A { e^{2\Omega} (\rho^D + P_k^D) } = 
\frac{n^2 \rho_T}{2}\left[  (A-3) + w (A-1) \right] 
+ \text{other terms}
\ee
Since $\rho_T \ge 0$ then the first line in (\ref{eq:QuinRhoPlusPk}) is nonpositive provided that
\be\label{eq:NonPosKinCond2}
A \le \frac{3+w}{1+w}
\ee
For $4 < k < 10$ the weaker condition (\ref{eq:NonPosKinCond2}) replaces the condition (\ref{e:Aless4cond}) which ensured the $\xi^2_{\perp|A}$ term was nonpositive.  The other condition on $A$, coming from the warp factors, remains the same as (\ref{eq:NonPosWarpCond}).  There is also the condition $A \ge 2$ which arises from the $\xi_0^2$ term.   This condition is never important, for if we could  choose $A<2$ then we would certainly have $A < 4$, but this choice is already excluded by the other constraints on $A$.  Therefore the constraints on the optimising $A$ in the $4 < k < 10$ case are given by (\ref{eq:NonPosWarpCond}) and (\ref{eq:NonPosKinCond2}).

Before establishing the existence of optimising values of $A$, we derive the differential equations satisfied by the optimal solution. If an optimising $A$ is chosen which satisfies  (\ref{eq:NonPosWarpCond}) and (\ref{eq:NonPosKinCond2}), then the optimal solution has $\sigma=\partial\Omega=0$, yielding 
\begin{subequations}
\begin{align}
\mavgn A { \xi_{\perp|A}^2 } &= -\xi_{0|A}^2 + \frac{2k}{k+2}\frac{4}{3(1+w)} t^{-2} \\
\totd {} t \xi_{0|A} 
+ \frac{2}{1+w} t^{-1} \xi_{0|A} &= \left(\frac{A}{2}-1\right) \xi_{0|A}^2
- \frac{4k\left[ (A-3) + w(A-1)\right]}{3(k+2)(1+w)^2}
\end{align}
\end{subequations}
The second equation is similar to (\ref{e:NECDiffEqk1_4_10}), but now depends on $A$.  We define a parameter $u(t)$ which is closely related to $v(t)$ by 
\be
\xi_{0|A} = \frac{1}{A/2 - 1}\frac{u(t)}{t}
\ee
The function $u(t)$ obeys 
\be
t \frac{\ud u}{\ud t} = u^2 + \alpha u + \beta_A
\ee
which is similar to the differential equation (\ref{eq:vEq}) for $v(t)$, but with
\be
\beta_A = -\frac{2 k (A-2) \left[ (A-3) + w(A-1) \right]}{3(k+2)(1+w)^2}
\ee
and boundary conditions $u(t) = \pm u_F$ with
\be
u_F = (A/2-1) \sqrt{\frac{2k}{k+2}\frac{4}{3(1+w)}} = (A/2-1) v_F
\ee
Unlike the situation in Section \ref{sss:NEC04and10plus}, after satisfying the constraints  (\ref{eq:NonPosWarpCond}) and (\ref{eq:NonPosKinCond2}), our specific choice of optimising $A$ influences the bounds of $w$ in the no-go theorem. The zeros of the right-hand side of the $u$-equation are located at
\be
u_0^\pm = -\alpha/2 \pm i \Delta_A
\ee
where $\alpha$ is given by (\ref{e:alphaDef}), and
\be
\Delta_A = \sqrt{\beta_A - (\alpha/2)^2}
\ee
where $\Delta_A$ depends on $A$.  The $w$ at which $\Delta_A=0$, denoted $w_\Delta (A)$, gives an estimate of the $w$ for which there is a transience constraint: the analysis of the last section suggests that acceleration must be transient for $w<w_\Delta(A)$, and may possibly have to be transient for even larger $w$ if the zeros $u_0^\pm$ lie outside the range $[-u_F,u_F]$.

\begin{figure}
  \begin{center}
  \includegraphics[width=1.0\textwidth]{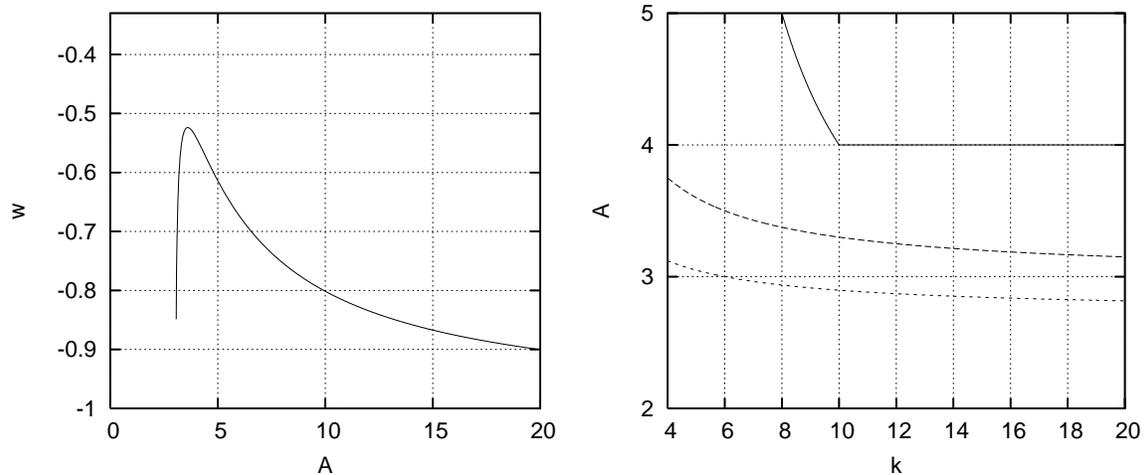}
  \end{center}
   \caption{Left panel: Value of $w_\Delta$ at which $\Delta_A$ becomes imaginary, as a function of $A$, plotted for $k=5$.  The maximum value of $w_\Delta$ is always equal to $w_k$.  Right panel: The solid curve shows the lower bound on $A$ coming from positivity requirements in (\ref{eq:NonPosWarpCond}): $A$ must lie above this line.  After $k=10$, we can choose $A=4$ which gives an optimal average. The long-dashed line shows $A_*$, the value of $A$ for which $w_\Delta$ is largest.  The short-dashed line shows $A_{\rm min}$, and for $A < A_{\rm min}$  $\Delta$ is always imaginary.}
   \label{f:ChoosingA}
\end{figure}

We employ the strategy of choosing $A$ so that $w_\Delta(A)$, and therefore the interval in $w$ for which there is a transience constraint, is as large as possible.  Other strategies can be envisioned -- for example, choosing $A$ so that the transience bound for $w$ is lower, but the number of allowed e-folds at fixed $w$ is smaller.  A plot of $w_\Delta(A)$ for various $A$ is given in 
 the left panel of Figure \ref{f:ChoosingA}. The function  has a maximum value at $A_*$ given by
\be
A_* = 3\left( 1 + \frac{1}{k}\right)
\ee
The quantity $w_\Delta(A_*)$ is the largest value of $w$ below which acceleration must be transient, and is
\be
w_\Delta(A_*) = -\frac{k+6}{3k+6} = w_k
\ee
This is perfectly consistent, since we should not be able to adjust $A$ and obtain a constraint on transient acceleration for models which have eternal acceleration and satisfy the NEC by construction.  The quantity $w_\Delta(A)$ is real provided that $A > A_{\rm min}$ with
\be
A_{\rm min} = 1 + \sqrt{3}\left[ 1 + \frac{2}{k} \right]^{1/2}
\ee
Since $2.73 \lesssim A_{\rm min} \lesssim 4$ this constraint is redundant with those we have already studied.  The two functions $A_{\rm min}$ and $A_*$ are plotted in the right panel of Figure \ref{f:ChoosingA}.

\begin{figure}
  \begin{center}
  \includegraphics[width=1.0\textwidth]{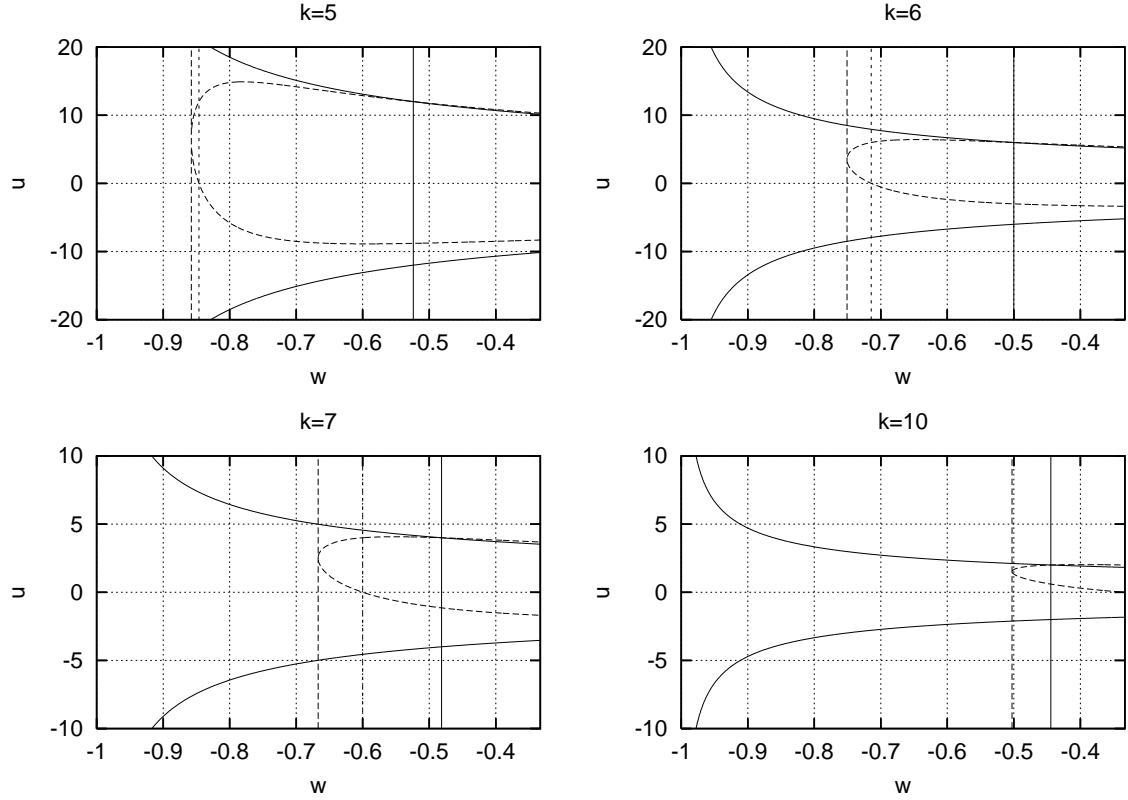}
  \end{center}
   \caption{Curves for the ``best" choice $A=A_{\rm best}$.  (Note different vertical scale on bottom panels) The solid curves show $\pm u_{F{\rm best}}$, so $u$ starts on the lower curve and travels directly upward to the upper curve.  The dashed curves show the zeros $u^\pm_{0 {\rm best}}$ of the right-hand side of the $u$-equation, and are fixed points.  The solid vertical lines shows the intersection of these two curves at $w_\times$.  The long-dashed vertical line shows $w_{\Delta {\rm best}}$ at which $\Delta$ is switching from real to imaginary.  The short-dashed vertical line shows the largest $w$, denoted $w_+$, allowed from positivity constraints, which is always larger than $w_{\Delta {\rm best}}$.}
\label{f:vs_k5-10}
\end{figure}

To make $w_\Delta(A)$ as large as possible, we should choose $A$ to be as close to $A_*$ as possible.  As can be seen in the right panel of Figure \ref{f:ChoosingA}, the warping constraint (\ref{eq:NonPosWarpCond}) prevents us from choosing $A=A_*$.  In fact the values of $A$ allowed by this constraint are always larger than $A_*$, and lie above the solid line shown in the right panel of Figure \ref{f:ChoosingA}.  The ``best" choice of $A$, which gives the largest value of $w_\Delta$, is the one which saturates the warp constraint.
Therefore we choose $A=A_{\rm best}$, with
\be\label{e:ChoiceOfA}
A_{\rm best} = \frac{2k+4}{k-4}
\ee
which means $A$ lies on the solid line in the right panel of Figure \ref{f:ChoosingA}.  We use the subscript ``best" in $A$-dependent quantities to indicate that we are evaluating them with $A=A_{\rm best}$.
Then for $4 < k \le 10$ we have $w_{\Delta{\rm best}} = w_\Delta(A_{\rm best})$ given by 
\be
w_{\Delta{\rm best}} = \frac{32 - 128 k - 22 k^2 + k^3 + 16 \sqrt{2}\sqrt{-k^4 + 17 k^3 + 32 k^2 - 48k}}{(k-4)^2(k+2)}
\ee
As a check, there is a constraint on $A$ which requires that the combination of $\rho_T$, $P_T$ and $\xi^2_{\perp|A}$ terms are nonpositive.  For our choice of $A$ the maximum value of $w$ allowed by (\ref{eq:NonPosKinCond2}) is 
\be
w_+ = \frac{k-16}{k+8}
\ee
For our choice of optimum $A$, we have $w_+ > w_{\Delta{\rm best}}$ and so this constraint is automatically satisfied. This is illustrated in Figure \ref{f:vs_k5-10}.

Our choice of $A=A_{\rm best}$ in (\ref{e:ChoiceOfA}) therefore satisfies all available constraints.  As a final consistency check, 
the phase space analysis indicates that $w_{\Delta{\rm best}}$ completely controls when there is a transience constraint.  That is, the crossing points $w_\times$ are all at higher $w$ than $w_{\Delta{\rm best}}$.  There are no subleties with zeros of the right-hand side of the $u$-equation crossing in and out of the region $[-u_F,u_F]$ as in the cases studied in Section \ref{sss:NEC04and10plus}.  So by maximising $w_\Delta(A)$ as a function of $A$ we have obtained transience constraints for the widest possible range of $w$ values.    Whenever $w <w_{\Delta{\rm best}}$, the number of e-folds in the four-dimensional cosmology must be finite to satisfy the higher-dimensional NEC. To illustrate these points, phase space plots for several of the $4 < k <  10$ cases are shown in Figure \ref{f:vs_k5-10}.

Since the differential equation satisfied by $u$ is essentially the one we found for $v$ in Section \ref{sss:NEC04and10plus}, the solutions are also similar.  The e-fold bound is  
\be\label{e:NECefolds4to10}
N= \frac{2}{3(1+w)} \frac{1}{\Delta_{\rm best}}
\text{Tan}^{-1} \left[ \frac{2u_{F{\rm best}} \Delta_{\rm best}}{ \beta_{\rm best} - u_{F{\rm best}}^2}\right]
\ee
where
\be
\Delta_{\rm best} = \sqrt{\beta_{\rm best} - (\alpha/2)^2 }
\ee
with
\be
\beta_{\rm best} = -\frac{8 k \left[ k(w-1) + 8(2+w) \right]}{(k-4)^2 (k+2) (1+w)^2}
\ee
and
\be
u_{F {\rm best}} = \frac{4 \sqrt{6}}{k-4} \sqrt{\frac{k}{(2+k)(1+w)}}
\ee
The argument to the $\text{Tan}^{-1}$ is real when $w < w_{\Delta {\rm best}}$, and since the region in $w$ for which there is a transience constraint is bounded by $w_{\Delta {\rm best}}$, we need not worry about analytically continuing (\ref{e:NECefolds4to10}).  The argument to the $\text{Tan}^{-1}$ in (\ref{e:NECefolds4to10}) does have a pole at
\be
w_{\rm pole} = \frac{k-28}{k+20} 
\ee
which leads one over the branch in the $\text{Tan}^{-1}$ function.  Since $w_{\rm pole} < w_{\Delta {\rm best}}$ for $4 < k < 10$ this pole must always be dealt with.  One way to avoid the branch choice is to use
\be\label{e:RFefoldBounds410noCut}
N = \frac{2}{3(1+w)\Delta_{\rm best}}\left[
\text{Tan}^{-1} \left( \frac{\alpha/2 + u_{F {\rm best}}}{\Delta_{\rm best}} 
\right) -
\text{Tan}^{-1} \left( \frac{\alpha/2 - u_{F {\rm best}}}{\Delta_{\rm best}} 
\right)
\right]
\ee
The arguments to the $\text{Tan}^{-1}$ are real and finite in the full range $-1 \le w < w_{\Delta {\rm best}}$.  These e-folding bounds are illustrated in Figures \ref{f:SummaryNoTune} and \ref{f:NECeFolds}.
 
\subsection{Time-varying $w$}\label{sss:TimeVaryW}

To obtain simple expressions for the e-folding bounds in Section \ref{ss:RFQuintessence}, we have assumed that $w$ is constant. Nonetheless, this allows us to place e-folding bounds on time-varying $w$, as we show below.  By this we mean the total four-dimensional effective $w(t) = P_T/\rho_T$ and not only the $w$ of energy components of the four-dimensional universe which come from moduli dynamics.  We relax the subscript ``$T$" in the following for clarity of notation, but we are always referring to the total energy density and pressure.

We claim that, if the four-dimensional effective $w(t)$ varies with time over the interval $t \in [t_0,t_1]$, but always satisfies $w(t) \le w_\star$, with $w_\star$ a constant, then
\be\label{e:ConstantBoundsBest}
N[ w(t) ] \le N(w_\star)
\ee
where $N[w(t)]$ is the maximum allowed number of e-foldings given the time-varying $w(t)$, and $N(w_\star)$ is the e-folding bound derived for constant $w=w_\star$ in the previous sections.

Heuristically, the claim (\ref{e:ConstantBoundsBest}) is a consequence of the monotonic nature of the e-folding bounds.  As is evident in Figure \ref{f:NECeFolds}, when $w$ is smaller the allowed number of e-folds $N(w)$ is as well.  Therefore one cannot extract more e-foldings by decreasing $w$.

Less heuristically, the transience constraints work because, to satisfy the NEC, the time derivative of $\xi_{0|A}$ has to balance the negative-definite terms in (\ref{eq:QuinRhoPlusPk}).  When $w$ is time-dependent, these terms are never smaller. To show this, along with the universe with energy density $\rho$, we construct an auxiliary universe with $w=w_\star$, normalised so that $\rho(t_0) = \rho_\star$.  Since
\be\label{e:EvolAsW}
\frac{\ud \ln \rho}{\ud \ln a} = -3(1+w)
\ee
and $w(t) \le w_\star$, then $\rho(t) \ge \rho_\star$ for $t_0 < t < t_1$.  Taking (\ref{e:EvolAsW}) for $\rho$ and $\rho_\star$ and dividing yields
\be\label{e:howRhosEvolveRelatively}
\frac{\ud \ln \rho}{\ud \ln \rho_\star} =  \frac{1+ w(t)}{1+w_\star}
\ee
By the assumption that $w(t) \le w_\star$, the right hand side is always $\le 1$, and so we can rewrite (\ref{e:howRhosEvolveRelatively}) as
\be\label{e:VaryingNECbound}
(1 + w(t)) \rho  \le (1 + w_\star) \rho_\star 
\ee
This also implies 
\be\label{e:VaryingSECbound}
(1 + 3 w(t)) \rho  \le (1 + 3 w_\star) \rho_\star 
\ee
which follows from (\ref{e:VaryingNECbound}) since $\rho$ and $\rho_\star$ are positive and $w(t) \le w_\star$.  The results (\ref{e:VaryingNECbound}) and (\ref{e:VaryingSECbound}) show that, while $\rho$ is larger than $\rho_\star$, its $w$ is sufficiently more negative that the combinations $\rho+P$ and $\rho+3P$ are in fact smaller.  This means that the time derivative of $\xi_{0|A}$ must be larger.

We prove the claim (\ref{e:ConstantBoundsBest}) by contradiction.  We assume that $\xi_{0|A}$ is the optimal solution for the profile $w(t)$, and that it allows more e-foldings than the optimal solution for $w=w_\star$. Then $\xi_{0|A}$ obeys the differential equation obtained by saturating the inequality (\ref{eq:QuinRhoPlusPk}) after setting $\xi_\perp=\sigma=0$.  This differential equation, written for non-constant $w$,  is
\be\label{e:VaryWDE1}
\frac{\ud  \xi_{0|A} }{\ud t} + 3 H \xi_{0|A} - \xi_{0|A}^2 + \left(\frac{1+3 w}{2} \right)\left(\frac{2k}{k+2}\right)\rho = 0
\ee 
We define a variable $\zeta$ by
\be
\xi_{0|A} = H \zeta
\ee
and since the Friedmann equations imply
$\dot H = -3H^2(1+w)/2$, we transform (\ref{e:VaryWDE1}) to
\be\label{e:VaryWDE2}
H^{-1} \frac{\ud \zeta}{\ud t} + \frac{3(1-w)}{2} \zeta - \zeta^2 + \frac{3k(1+3w)}{k+2} = 0
\ee
Next we trade the proper time variable $t$ for the number of e-foldings $N$ by
\be
\frac{\ud}{\ud t} \to \frac{\ud N}{\ud t}\frac{\ud}{\ud N} = H \frac{\ud}{\ud N}
\ee
Which transforms (\ref{e:VaryWDE2}) into
\be\label{e:VaryWDE3}
\frac{\ud \zeta}{\ud N} = \zeta^2 + \frac{3(w-1)}{2}\zeta
- \frac{3k(1+3w)}{k+2}
\ee
The boundary conditions for this differential equation are, as before, defined by saturating (\ref{eq:QuinRhoPlusP3}).  In terms of the variables defined here,  the boundary conditions are $\zeta = \pm \zeta_F$ with
\be\label{e:zetaF}
\zeta_F = \left[ \frac{6k(1+w)}{k+2} \right]^{1/2}
\ee
where $\zeta_F$ is a function of time when $w$ is. To satisfy the NEC, the inequality (\ref{eq:QuinRhoPlusPk}) implies that $-\zeta_F \le \zeta \le +\zeta_F$.  The bounding value $\zeta_F$ is never larger for time-dependent $w$ under our assumption that $w(t) \le w_\star$.

We establish the claim (\ref{e:ConstantBoundsBest}) by showing that (\ref{e:VaryWDE3}) implies that $\ud \zeta / \ud N$ is never smaller for $w(t)$ than for $w_\star$.  Of the three terms on the right-hand side of (\ref{e:VaryWDE3}), the first is independent of $w$, the second decreases if $w$ does, and the last increases as $w$ decreases.  For the same value of $\zeta$, we subtract the version of (\ref{e:VaryWDE3}) for time-varying $w$ from the version of (\ref{e:VaryWDE3}) for constant $w=w_\star$.  This gives
\be\label{e:VaryWDiff}
\frac{3(w_\star - w)}{2}\left(\zeta - \frac{6k}{k+2} \right)
\ee
If this difference is positive, then the right-hand side of (\ref{e:VaryWDE3}) is larger for the constant $w=w_\star$, and if negative, the right-hand side is larger for time-varying $w$. Since $w_\star \ge w$, the difference (\ref{e:VaryWDiff}) is only positive if $\zeta > 6k/(k+2)$.  However the boundary conditions -- which is really the NEC condition (\ref{eq:QuinRhoPlusP3}) -- imply that $\zeta \le \zeta_F$.  Furthermore, using our definition (\ref{e:zetaF}) of $\zeta_F$ we have
\be\label{e:zetaFbound}
\zeta_F 
\le 2 \left(\frac{k}{k+2}\right)^{1/2} < 2
\ee
where the first inequality follows from (\ref{e:zetaF}) because $\zeta_F$ assumes its maximum possible value when $w=-1/3$, and the second obtains because the parenthetical expression is always less than unity for finite positive $k$.  The bound (\ref{e:zetaFbound}) means that $\zeta$ can never satisfy $\zeta > 6k/(k+2)$, and so (\ref{e:VaryWDiff}) is never positive.  Since (\ref{e:VaryWDiff}) is never positive $\ud \zeta / \ud N$ is never smaller for time-dependent $w$ than it is for constant $w=w_\star$.  But since $\ud \zeta / \ud N$ is never smaller and $\zeta_F$ never larger in the time-dependent case, the total number of e-foldings cannot be larger than in the constant-$w$ case.  This proves the assertion (\ref{e:ConstantBoundsBest}).

The argument above has been given for the curvature-free case when $k <4$ or $k \ge 10$, though similar proofs hold for other values of $k$ and the curved cases.
In cases where the bound (\ref{e:ConstantBoundsBest}) does not give useful constraints, or more precision is required, a bound can be obtained by directly integrating the differential equations arising from (\ref{eq:QuinRhoPlusP3}) and (\ref{eq:QuinRhoPlusPk}) given a specific function $w(t)$.

\section{Curved compactifications}\label{s:Curved}

In Section \ref{s:RicciFlat} we assumed that the Ricci scalar of $\MM$ vanishes everywhere.  This is sufficient for special manifolds that are guaranteed to be curvature-free.  As described in Section \ref{ss:BreathingMode}, introducing curvature weakens the no-go theorem, but implies a certain degree of fine-tuning.  It may be that this tuning is well-motivated in specific models.  Or  we could also regard the whole problem of ``tuning" as essentially a matter of opinion and aesthetics. With this in mind we describe the best that one can hope to do when $\MM$ is curved.  There are two approaches to this issue.  

The first approach involves constructing quantities that are independent of the curvature of $\MM$ and leads to a no-go theorem involving the SEC.  In Section \ref{ss:CurvedQuintessence} we construct a one-parameter family of curvature-independent averages, and by using a specific member of the family, we show that the SEC puts limits on the number of e-folds of expansion with $w > -1$, irrespective of the curvature of $\MM$.  This is similar to the non-de Sitter cases studied in Section \ref{ss:RFQuintessence}, but the NEC is replaced by the SEC, and we obtain different threshold $w$ and different e-folding constraints.  This no-go theorem is a natural extension of the SEC de Sitter no-go described in \cite{Gibbons:1985,Maldacena:2000mw} to cases where extra dimensions are dynamical.

The second approach involves a limit process and leads to a no-go theorem involving the NEC.  In Section \ref{ss:WarpeddeSitter} we show that, although the SEC must be violated by four-dimensional de Sitter expansion, it is possible to construct de Sitter models that satisfy the NEC if the curvature of $\MM$ is carefully tuned and the warp factor vanishes.  For this case the constraints on curved $\MM$ are much weaker than those we derived in Appendix \ref{ss:RFdeSitter} for curvature-free $\MM$.  In Section \ref{ss:WarpeddeSitter}, we show that NEC violation can be proven if the Ricci scalar and warp factor satisfy a ``bounded average condition."

\subsection{Non-de Sitter and the SEC}\label{ss:CurvedQuintessence}

To prove no-go theorems for curved compactification manifolds $\MM$, we choose to study quantities that are independent of the curvature of $\MM$.  There is family of linear combinations of $\rho^D$, $P_3^D$ and $P^D_k$ which fit this requirement, which we construct in Appendix \ref{sss:CurveIndependent}.  The family is parameterized by a single parameter $\gamma$. In Appendix \ref{sss:CurveIndependent} we also study the positivity requirements and find optimal values of $A$.  The technique of proof is very similar to that employed in Section \ref{ss:RFQuintessence}, and  we refer the reader there for details of the methodology.


%


Using the results in Appendix \ref{sss:CurveIndependent}, the SEC inequality (\ref{e:SECdefinition}), when $t^a = (1,0,\dots)$, is equivalent to (\ref{e:abEnergyCondition}) and (\ref{e:bOfa}) for the specific choice $\gamma=\gamma_\star$ with
\be
\gamma_\star = \frac{3}{1+k}
\ee
So by taking $\gamma=\gamma_\star$, we can use the machinery of Appendix \ref{sss:CurveIndependent} to probe the SEC.\footnote{In the NEC case we proved a lemma which allowed us to study only the trace parts of the stress-energy.  Here such a lemma is unnecessary, for we are considering only $t^a = (1,0,\dots)$ and only the trace parts of $T^D_{ab}$ and a single component $T_{00} = \rho^D$ appear in the SEC condition with this choice.}  
For $\gamma=\gamma_\star$ the differential equation (\ref{e:abDiffEq}) describing the optimal solution becomes
\be\label{e:SECDE}
t\frac{\ud v}{\ud t} = \frac{k+2}{k} v^2 + \frac{w-1}{w+1} v - \frac{4(1+3w)}{3(1+w)^2}
\ee
For this choice of $\gamma$, the arguments in the previous section show that there exists an $A$ such that the pair $(\gamma_\star,A)$ satisfy the nonpositivity constraints.  Neither the differential equation (\ref{e:SECDE}) nor the boundary conditions (\ref{e:abBoundaryConds}) depend on $A$, so the precise value chosen is immaterial.

\begin{figure}
  \begin{center}
  \includegraphics[width=1.0\textwidth]{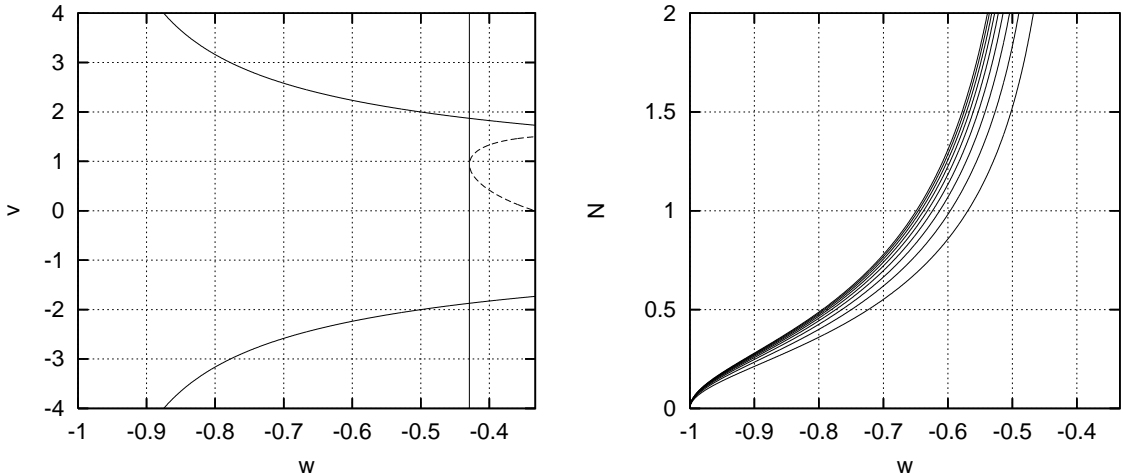}
  \end{center}
   \caption{Left panel: Phase plane for the curved case when $k=6$.  The beginning and ending values of $v=\pm v_F$ are shown by the solid curves, and the locus $v=v_0^\pm$ where the right-hand side is zero by the dashed lines.  The lowest value of $w$ for which the right-hand side vanishes is shown by the vertical line.  Right panel: The maximum number of e-foldings as a function of $w$, for various values of $k$.  From bottom to top curve these are $k=2,3,\dots,10$.}
\label{f:TuneCurves}
\end{figure}

We can now carry out the phase plane analysis as in Section \ref{ss:RFQuintessence}. A typical plot is shown in Figure \ref{f:TuneCurves}.  First we seek the zeros of the right-hand side of (\ref{e:SECDE}) in order to determine when there is a transience constraint.  The zeros are located at
\be\label{e:v0pmStar}
v_{0\star}^\pm = -\alpha_\star /2 \pm i\Delta_\star 
\ee
with
\be
\alpha_\star = \frac{k(w-1)}{(k+2)(w+1)}
\ee
and
\be
\Delta_\star = -\frac{32k+19k^2+96kw+42k^2w+3k^2w^2}{2\sqrt{3}(k+2)(w+1)}
\ee
So long as $\Delta_\star$ is real, there are no zeros on the right-hand side of (\ref{e:SECDE}).  This is the case for
\be
w < w_{\Delta\star} = \frac{-48 -21k + 8\sqrt{6}\sqrt{6+5k+k^2}}{3k}
\ee
In the absence of zeros, $\ud v /\ud t> 0$ and so it seems likely that $v$ goes from $-v_F$ to $+v_F$ in finite time. 
In earlier sections, we showed that 
 the range of $w$ for which there is a transience constraint can be no smaller than $-1 \le w \le w_{\Delta\star}$ but could be larger if the zeros exist but are located outside of $[-v_F,v_F]$.  For the NEC cases it was important to take account of this possibility, but here the only $w_{\times\star}$ at which $v_{0\star}^\pm$ crosses $\pm v_F$ occur at
\be\label{e:SECcrossingWs}
w_{\times\star} = 1, -\frac{1}{3} + \frac{4}{3k}
\ee
These values of $w$ are outside the accelerating range of $w$, so when the zeros first appear at $w=w_{\Delta\star}$, if they are within the range $[-v_F,v_F]$, they do not cross outside this range for any accelerating $w$.  Likewise if the zeros appear outside the range $[-v_F,v_F]$ they cannot cross in for any accelerating $w$. The location $v_{\rm appear}$ of the zeros when they first appear is found by evaluating (\ref{e:v0pmStar}) at $w=w_{\Delta\star}$, which gives
\be
v_{\rm appear} = \frac{2k}{5k+16}\left[ 1 + \sqrt{6}\sqrt{\frac{k+3}{k+2}} \right]
\ee
We should compare this to the value of $v_F$ when the zeros first appear, denoted by $v_{F{\rm appear}}$ and obtained by evaluating (\ref{e:abBoundaryConds}) at $w=w_{\Delta\star}$, yielding
\be
v_{F{\rm appear}} = \frac{2k}{\sqrt{(2+k)\left[ -24 -9k + 4\sqrt{6}\sqrt{(k+2)(k+3)} \right]}}
\ee
For all positive $k$, $v_{\rm appear} \in [-v_{F{\rm appear}},v_{F{\rm appear}}]$, so when the zeros appear they do so inside the range $[-v_{F},v_{F}]$.  This means that $w_{\Delta\star}$ controls the range of $w$ for which there is a transience constraint: for $-1 \le w < w_{\Delta\star}$ satisfying the SEC implies that accelerated expansion must be transient.

To get a precise constraint on the number of e-foldings, we solve the differential equation (\ref{e:SECDE}) using the boundary conditions (\ref{e:abBoundaryConds}), and derive the e-folding constraint as in Section \ref{ss:RFQuintessence}.  This gives
\be\label{e:SECEfoldFormula}
N = \frac{2k}{3(1+w)(k+2) \Delta_\star} \text{Tan}^{-1} \left[ \frac{2 \Delta_\star v_F}{\Delta_\star^2+(\alpha_\star/2)^2 - v_F^2} \right]
\ee
When $w$ passes through $-3/5$ the argument to $\text{Tan}^{-1}$ goes through a pole, taking us over a branch cut.  The expression (\ref{e:SECEfoldFormula}) assumes that we stay on the Riemann sheet which keeps $N$ continuous, which can also be accomplished by splitting the $\text{Tan}^{-1}$ into two terms, giving
\be\label{e:CurvedEfoldBoundsNoCut}
N = \frac{2k}{3(1+w)(k+2) \Delta_\star} \left[
\text{Tan}^{-1} \left( \frac{\alpha_\star/2+v_F}{\Delta_\star} \right) -
\text{Tan}^{-1} \left( \frac{\alpha_\star/2-v_F}{\Delta_\star} \right)
\right]
\ee
These functions are plotted for various values of $k$ in Figure \ref{f:TuneCurves}.  As can be seen from the picture, the constraint curves are relatively insensitive to $k$.

Thanks to the monotonic nature of $N$ as a function of $w$, when $w$ is time-dependent we can conclude that if $w(t) \le w_\star$ then 
\be
N[w(t)] \le N(w_\star)
\ee
through arguments substantially identical to those given in Section \ref{sss:TimeVaryW}.  The differential equations defining the optimal solution can be readily integrated for a preferred function $w(t)$ if more precision is required.

\subsection{Warped de Siter and the NEC}\label{ss:WarpeddeSitter}

Here we consider whether the SEC-violation can be strengthened to a NEC-violation condition, and find that for this is sometimes possible for pure de Sitter expansion. We begin by constructing a model which gives de Sitter expansion in four dimensions and satisfies the NEC.  This shows that the energy condition for the SEC no-go theorem cannot be naively weakened from SEC to NEC.  Then, we show that NEC violation is required in some warped de Sitter compactifications: if $\MM$ satisfies a ``bounded average condition" then de Sitter expansion violates the NEC.

For de Sitter expansion consistent with the  $\rho^D + P^D_3$ NEC condition, $\MM$ must be static. Temporarily we take $\Omega=0$, which gives pointwise Einstein equations
\begin{subequations}
\begin{align}
G_{00} = +\lambda_3 &= + 3 e^{\phi} H_0^2 +\frac{1}{2} \Ro \label{e:NECtuned00} \\
\frac{1}{3} \delta^{mn} G_{mn} = -\lambda_3 &= - 3 e^{\phi} H_0^2-\frac{1}{2} \Ro  \label{e:NECtunedmn} \\
\frac{1}{k} \delta^{ab} G_{ab} =  -\lambda_k &= - 6 e^{\phi} H_0^2 + \left( \frac{1}{k} -\frac{1}{2} \right) \Ro \label{e:NECtunedab}
\end{align}
\end{subequations}
where we have taken $n=1$ so that $t$ is the proper time coordinate, and $H_0$ is the four-dimensional Einstein-frame Hubble constant.  The quantities $\lambda_3$ and $\lambda_k$ are the traces of parts of the $(4+k)$-dimensional stress-energy tensor.  They are free to vary with position in an arbitrary way, but they cannot vary with time, for all ``metric moduli" are frozen so $\Ro$ is constant in time. Equations (\ref{e:NECtuned00},b) are satisfied if
\be\label{e:RoInTermsOfOther}
\Ro = 2\lambda_3 - 6 e^\phi H_0^2
\ee
which determines the curvature in terms of the other two variables.  The $\rho^D + P_3^D$ condition is trivially satisfied, while the $\rho^D + P_k^D$ NEC condition is equivalent to $\lambda_3 - \lambda_k \ge 0$ which implies
\be\label{e:RoNEC}
\Ro \ge 3 k e^\phi H_0^2
\ee
Taking (\ref{e:RoInTermsOfOther}) and (\ref{e:RoNEC}) together we have 
\be
\lambda_3 \ge \frac{3(k+2)}{2}e^\phi H_0^2
\ee
with a similar requirement on $\lambda_k$. This demonstrates that it is possible to construct a de Sitter universe which satisfies the higher-dimensional NEC by tuning the curvature.  One apparent limitation of this approach is that if $\Ro$ is negative anywhere on $\MM$ then (\ref{e:RoNEC}) will be violated: therefore this tuning argument can only work on manifolds $\MM$ of everywhere nonnegative curvature.\footnote{In \cite{Carroll:2001ih} a similar argument showed that the Ricci curvature must be nonnegative in all directions to satisfy the NEC.}  This excludes toy models built from compactification on tori, for  a torus of everywhere nonnegative Ricci curvature must be flat.

Before reinstating the warp factor, we round out the argument in the previous paragraph by explicitly constructing a model which gives de Sitter expansion in four dimensions and satisfies the NEC.  We take the higher-dimensional stress-energy to be a cosmological constant, so that $\lambda_k = \lambda_3 = \lambda$. By adding (\ref{e:NECtuned00},c) this requires
\be
\Ro  = 3k e^\phi H_0^2
\ee
Therefore  the compactification manifold $\MM$ must have constant positive scalar curvature, which can be realised in any dimension as a sphere $S^k$.
Using (\ref{e:RoInTermsOfOther}) implies
\be
\lambda = \frac{3(k+2)}{2} e^\phi H_0^2
\ee
which gives the required value of the higher-dimensional cosmological constant.
This model satisfies the NEC because the higher-dimensional cosmological constant does.  We saw in Section \ref{ss:BreathingMode} that compactifying a cosmological constant on a Ricci-flat $\MM$ cannot give de Sitter expansion and satisfy the NEC.  We have just shown that this is possible if $\MM$ is curved in a specific way.  This mechanism allows some models to satisfy the NEC in unwarped compactifications on curved spaces (\emph{e.g.} \cite{Carroll:2003db}).

Now we reinstate the warp factor and show that the NEC is violated.  In light of the construction in the previous paragraph, it may seem that warped compactifications should make it easier to satisfy the NEC, since warping essentially introduces another free function.  We show that this is not necessarily so.  To prove this we need to assume that the warp factor and curvature each satisfy the bounded average condition.  The bounded average condition requires that there exist values $A_\Omega$, $A_R$ and $B_R$, and a positive number $B_\Omega$  such that
\begin{subequations}
\begin{align}
\mavgn A {e^{2\Omega} (\partial\Omega)^2} & > B_\Omega \quad 
\text{when} \quad
\begin{cases}
\quad k > 4 & \quad \text{and} \quad {A > A_\Omega} \\
\quad k \le 4 & \quad \text{and} \quad {A < A_\Omega}
\end{cases} 
\label{e:warpUnboundCondition} \\
\mavgn A {e^{2\Omega} \Ro } & < B_R \quad 
\text{when} \quad
\begin{cases}
\quad k > 4 & \quad \text{and} \quad {A > A_R} \\
\quad k \le 4 & \quad \text{and} \quad {A < A_R}
\end{cases} 
 \label{e:RUnboundCondition}
\end{align}
\end{subequations}
This is to say that for $k > 4$ in the $A\to +\infty$ limit, or for $k \le 4$ in the $A \to -\infty$ limit, the warp average is bounded away from zero and the curvature average is bounded above.  The $k=4$ case requires some additional discussion for the same reasons as in Section \ref{sss:NEC_k4}, and we return to this case momentarily.\footnote{We are grateful to Juan Maldacena for pointing out that the bounded average condition is not satisfied if $\Omega$ and $R$ are smooth functions of $\MM$, as we erroneously claimed in an earlier version of this work.}

We now show that if the bounded average condition (\ref{e:warpUnboundCondition},b) is satisfied then de Sitter expansion must violate the NEC.  
In the de Sitter limit we must have $\xi_{0|A} = \xi_{\perp|A} = \sigma = 0$.  
The averaged $\rho^D + P_k^D$ condition is then
\be\label{eq:QuinRhoPlusPkB}
\mavgn A { e^{2\Omega} (\rho^D + P_k^D) } = 
- 3 e^\phi H_0^2
+ \frac{1}{k} \mavgn A { e^{2\Omega} \Ro } 
 + \left[ \left(\frac{4}{k}-1\right) A + \left( 2 + \frac{4}{k}\right)\right]
\mavgn A {e^{2\Omega} (\partial\Omega)^2} 
\ee
The key observation is that the warp term coefficient is a function of $A$ while the coefficients of the other terms are constant in $A$.  If the averages themselves are in some sense bounded in $A$, then we can make the warp term as large as we like by making $A$ very positive (or negative).  To make this work we need three things to be true.  First, the coefficient of the warp term must be negative, which is exactly the nonpositivity condition  (\ref{eq:NonPosWarpCond}).  Second, we need the bounded average conditions  (\ref{e:warpUnboundCondition},b) to be satisfed.  Last, we need to choose $A$ so that, depending on dimension, it is either greater than or less than both $A_\Omega$ and $A_R$.  Once these three conditions are satisfied, we can choose $A$ sufficiently positive (or negative) so that the last two terms in (\ref{eq:QuinRhoPlusPkB}) are negative. 
This amounts to choosing
\begin{subequations}
\begin{align}
A \le \text{Inf}\left[ A_\Omega, A_R, \frac{B_R/B_\Omega + (2k+4)}{k-4}  \right] & \quad \text{when} \quad k \le 4 \\
A \ge \text{Sup}\left[ A_\Omega, A_R, \frac{B_R/B_\Omega + (2k+4)}{k-4}  \right]  & \quad \text{when} \quad k > 4
\end{align}
\end{subequations}
which guarantees that the last two terms in (\ref{eq:QuinRhoPlusPkB}) are nonpositive.  Because the first term is negative, the NEC must be violated.

The method here provides an alternative to the proof in Section \ref{ss:RFdeSitter} that de Sitter expansion violates the NEC in the curvature-free case.  When $\MM$ is de Sitter then we can relax the condition that $B_\Omega$ is positive, and can allow it to be zero.  (In the curved case, we needed $B_\Omega$ to be positive to ensure that we could always cancel off the curvature term, but in the curvature-free case there is no curvature to cancel).  We only require that there exists an $A$ such that the final term in (\ref{eq:QuinRhoPlusPkB}) is nonpositive to prove NEC violation, which amounts to 
\begin{subequations}
\begin{align}
A \le \text{Inf}\left[ A_\Omega, \frac{2k+4}{k-4}  \right] & \quad \text{when} \quad k \le 4 \\
A \ge \text{Sup}\left[ A_\Omega, \frac{2k+4}{k-4}  \right]  & \quad \text{when} \quad k > 4
\end{align}
\end{subequations}
since such a value of $A$ always exists then de Sitter expansion with curvature-free $\MM$ must violate the NEC.

We have postponed dealing with the $k=4$ case, but this can be managed in the same way as in Section \ref{sss:NEC_k4}.  As in (\ref{e:EpsilonEC}) we consider the averaged inequality
\be\label{e:4kdSNECineq}
\mavgn A { e^{2\Omega}\left( \rho^D + [1+\epsilon] P_k^D \right) } \ge 0
\ee
which has the same warp term coefficient as does (\ref{eq:QuinRhoPlusPkB}) but the coefficient contains an additional $\epsilon$-dependent term
\be
\epsilon\left[ \left( \frac{4}{k}-4 \right) A + \left( 2 + \frac{4}{k} \right) \right]
\ee
precisely as in (\ref{e:EpsilonECWarpCoeff}). When $k=4$ the coefficient of the warp term is
\be
-3\epsilon A + 3(\epsilon + 1)
\ee
As in Section \ref{sss:NEC_k4} we can allow $\epsilon$ to approach zero from positive or negative values as desired.  As previously, we take $\epsilon$ to approach from negative values.  Then for any $\epsilon < 0$ we take
\be\label{e:k4dSNECineq}
A \le \text{Inf}\left[ A_\Omega, A_R, 1 + \frac{1}{\epsilon}\left(1 + \frac{B_R}{12 B_\Omega} \right)   \right]
\ee
We have already seen that when $\Ro$ and $\Omega$ are smooth and bounded on $\MM$, then the $A \to -\infty$ limit exists.  This implies that an $A$ satisifying (\ref{e:k4dSNECineq}) exists for all $\epsilon < 0$, and the assertion is proven by taking the double $\epsilon \to 0^-$, $A \to -\infty$ limit while satisfying (\ref{e:k4dSNECineq}). When $\Ro$ or $\Omega$ are singular it may be that there is a limit to how close to zero $\epsilon$ may be adjusted.  If the $\epsilon \to 0^-$ limit of (\ref{e:4kdSNECineq}) exists, then  (\ref{e:4kdSNECineq}) excludes everything the NEC excludes except for a higher-dimensional AdS cosmological term.  As in Section \ref{sss:NEC_k4} this is not a serious issue in practice, since the AdS energy density is the wrong sign to give accelerated expansion in four dimensions.  So while violating  (\ref{e:4kdSNECineq}) is not precisely equivalent to violating the NEC, it is close enough.

The ability to prove NEC violation appears to be unique to the de Sitter case, for only then can we set all of the kinetic terms to zero using the $\rho^D + P_3^D$ NEC condition.  We cannot extend away from de Sitter and prove a NEC transience bound for non-de Sitter expansion without allowing for these terms to be nonzero.  Then, the conditions for nonpositivity do not always allow us to make $A$ unboundedly large (or small).  Nonetheless, it may be possible to use the techniques here to prove bounds in special cases.

\section{Conclusions}\label{s:Conclusions}

We have proven no-go theorems that show that one must violate certain energy conditions if accelerating cosmologies are to be accommodated in theories with extra dimensions.  In some cases the relevant energy condition is the strong energy condition (SEC), while in others it is the weaker null energy condition (NEC).  The no-go theorems apply to both exact de Sitter expansion in the four-dimensional Einstein frame, as well as acceleration with an effective $w > -1$.  These results improve existing no-go theorems by a weakening of the energy condition from the SEC to the NEC, by treating cases where acceleration is not exactly de Sitter, and by including situations where the extra dimensions are dynamical.

The no-go theorems  lead us to three interesting conclusions.  The first conclusion is that one can escape the no-go theorems of \cite{Gibbons:1985,Maldacena:2000mw} by four-dimensional acceleration which is not de Sitter, but that one can only do so transiently,  and the new theorems put quantitative bounds on the amount of accelerated expansion that is allowed.  The second conclusion is that NEC violation is necessary for cosmic acceleration in many interesting models that fall into the curvature-free category, such as braneworld models and simple Calabi-Yau compactifications.  The third conclusion is that simple experimental measurements can tell us a great deal about possible-extra dimensional physics: if observations can show that the universe violates the bounds derived here, then large families of extra-dimensional models can be ruled out.

To what extent the theorems constrain accelerating cosmologies from compactification of higher-dimensional theories is an interesting question.  The SEC is a rather strict energy condition, but it is satisfied by the fields present in the classical action for M-theory and other supergravities.  This means that one must appeal to other elements of these theories which violate the SEC, such as D or M branes, to obtain accelerating universes.  Interestingly, there are supergravity no-go theorems of an entirely different nature which hold in the absence of these extended objects.  These theorems forbid the presence of warp terms in pure supergravity compactifications, and show that warping is only possible when extended objects with sufficently negative pressure are introduced \cite{Maldacena:2000mw,de Wit:1986xg}.  These extended objects play an essential role in the rich variety of warped compactifications currently under study \cite{Giddings:2001yu,DeWolfe:2002nn,Kachru:2003aw,Kachru:2003sx,Douglas:2006es,Burgess:2006mn}.
As we have seen in Section \ref{ss:WarpeddeSitter}, it is precisely when we have nonzero warping the SEC no-go theorem extends to a NEC no-go theorem for de Sitter cosmologies obtained from curved $\MM$.  So while introducing extended objects evades the de Sitter SEC no-go theorem, it can lead one afoul of the NEC no-go theorem.

Since it is a weaker energy condition, violating the NEC is more serious than violating the SEC, and the no-go theorems correspondingly more useful.  In the case of two-derivative field theories, there are good reasons to believe that NEC violation goes hand-in-hand with pathologies such as superluminal signal propagation, unitarity violations, and instabilities \cite{Cline:2003gs,Hsu:2004vr,Dubovsky:2005xd,Buniy:2006xf}.  There are mechanisms by which the NEC can be violated in theories of physical interest.  Some objects in string theory (such as orientifold planes) and quantum effects (such as Casimir energies) can violate the NEC.  Higher derivative terms may permit NEC violation without associated pathologies: it may be possible to construct pathology-free quantum field theories with more than two derivatives \cite{ArkaniHamed:2003uy,Creminelli:2006xe}.  String and M theory have a characteristic pattern of higher-derivative terms in the low-energy effective action.  These terms are essential in anomaly cancellation, in finding $\mathcal{N}=1$ string vacua, and in the overall consistency of the theory.  If Einstein's equations are modified, it may be easier to violate the NEC without causing pathologies \cite{Boisseau:2000pr,Bronnikov:2006pt}.  The no-go theorems indicate that, in some circumstances, these NEC-violating mechanisms must play an essential role in accommodating accelerating universes.  But, even if such terms prevent pathologies, by repackaging all of the higher-derivative terms on one side of the field equations and the Einstein tensor on the other, we obtain an ``effective" stress-energy which violates the NEC.  This means that a variety of exotic solutions to Einstein's equations which require NEC-violating matter could potentially be permitted \cite{Morris:1988cz,Visser:2003yf,Alcubierre:1994tu,Krasnikov:1995ad,Morris:1988tu,Hawking:1991nk,Caldwell:1999ew,Caldwell:2003vq}, but this must be checked on a case-by-case basis.    It would be interesting if we were forced to accept the possibility of exotic solutions of Einstein's equations, or modifications to gravity, from observations that the universe is currently accelerating.

The literature provides many examples of the no-go theorems ``in action."   We briefly discuss a sampling of models with some interesting features:
\begin{itemize}  

\item Supersymmetric large extra dimensions (SLED) models in six dimensions provide a very interesting class of examples which evade the conditions of the theorems  (see \emph{e.g.} \cite{Aghababaie:2002be,de Rham:2005ci,Tolley:2005nu,Tolley:2006ht} and \cite{Burgess:2007ui} for a review).  These models provide especially vivid illustrations since they have fully explicit descriptions in both six and four dimensions.  Therefore one can check for NEC violation in the six-dimensional theory.  These models produce de Sitter universes without NEC violation, but are not counterexamples to the theorem of Section \ref{ss:WarpeddeSitter} since the bounded average condition is not satisfied. The models feature codimension-two branes and there are always curvature singularities at the brane locations.  Our no-go theorems imply that warped de Sitter compactifications of six-dimensional supergravity, satisfying the bounded average condition, should not exist.\footnote{Assuming such compactifications contain only matter which satisfies the NEC.}  

\item Braneworld cosmologies, with a single warped extra dimension, satisfy the conditions of the theorems for curvature-free $\MM$ \cite{Randall:1999ee,Randall:1999vf,Shiromizu:1999wj,DeWolfe:1999cp,Brax:2004xh}.  When the extra dimension is compact,  a negative-tension brane, which violates the NEC, must be present even in reductions to Minkowski space.   There are models in which de Sitter reductions are possible without apparent NEC violations in addition the negative-tension brane (\emph{e.g.} \cite{DeWolfe:1999cp}).  On the other hand there are also a variety of solutions which cannot give a four-dimensional de Sitter universe and satisfy the NEC modulo the negative-tension brane without introducing naked singularities and other pathologies \cite{ArkaniHamed:2000ds,ArkaniHamed:2000eg,Kachru:2000hf,Forste:2000ft,Binetruy:2000wn,Csaki:2000dm,Csaki:2001mn,Cline:2001yt,Apostolopoulos:2004ic,Apostolopoulos:2005at,Koroteev:2007yp}.  All of this is consistent with the no-go theorems, and the latter examples suggest that it may be possible to refine the theorems to probe additional NEC violations in accelerating braneworld cosmologies.

\item  The ``flux compactification" de Sitter constructions in string theory provide further examples of the no-go theorems (\emph{e.g.} \cite{Giddings:2001yu,DeWolfe:2002nn,Kachru:2003aw,Kachru:2003sx,Douglas:2006es,Burgess:2006mn}).  In these constructions a fully explicit higher-dimensional description is not yet available.   Nonetheless, in some examples it is known that such a description must contain NEC-violating elements, such as orientifold planes.  These models satisfy the conditions of the no-go theorems, and contain NEC-violating stress-energy as the theorems predict.\footnote{In many of these constructions, the presence of $p$-form flux distorts the Calabi-Yau compactification manifold away from Ricci-flatness.  Nonetheless, in these cases results concerning NEC violation can be obtained which are similar to those which hold for the curvature-free case discussed here.  These results will be described in a forthcoming publication \cite{PJSWesley}.}

\end{itemize}
In each of these examples, the no-go theorems give a sense of the price that must be paid -- in terms of curvature singularities or NEC violation -- in order to obtain accelerating cosmologies in four dimensions.

It seems likely that the no-go theorems presented here can be significantly improved.  Presently the weakest results are for $w > -1$ universes when $\MM$ is curved.  In this case we can only show SEC violation.  We have argued that models of this type which give accelerating universes must balance the curvature of $\MM$ against the warp terms and matter stress-energy.  A number of models in a particular subclass, which balance the curvature of $\Ro$ to obtain accelerating universes, have been constructed.  Many of these are based on compactifications on (possibly non-compact) hyperbolic manifolds \cite{Hull:1988jw,Starkman:2000dy,Kaloper:2000jb,Starkman:2001xu,Gibbons:2001wy,Townsend:2001ea} which are related to S brane solutions \cite{Gutperle:2002ai,Chen:2002yq,Kruczenski:2002ap,Deger:2002ie,Ivashchuk:2002ge,Ohta:2003uw,Ohta:2003pu,Ohta:2003ie,Ohta:2004wk}, and often give only transient acceleration \cite{Chen:2003ij,Chen:2003dca,Wohlfarth:2003ni,Roy:2003nd,Townsend:2003fx}. Some related non-hyperbolic models also give only transient acceleration \cite{Neupane:2003cs,Wohlfarth:2003kw,Neupane:2005ms,Neupane:2005nb}.  In many of these models acceleration is transient because they have a scalar field potential which is too steep to support eternal acceleration: the scalar field climbs its potential, and as it turns around accelerating solutions are possible \cite{Emparan:2003gg}.  This has led to conjectures that this is the only possible mechanism for accelerated expansion in realistic models \cite{Townsend:2003qv}.
Since many of the models above give only transient acceleration, there may be some problem with using $\Ro$ to drive accelerated expansion which is not visible using the techniques employed in this work.  It would be interesting if further work could expose these problems, or show that nearly de Sitter eternal acceleration is possible without violating the anything weaker than the SEC. 

The no-go theorems show that satisfying higher-dimensional energy conditions introduces a tension between cosmic acceleration and moduli stabilisation with interesting experimental consequences.  While some of the models we have studied can accommodate several $w > -1$ e-foldings, solutions below the threshold $w$ manage to satisfy the relevant energy conditions by balancing a rapidly increasing $\xi_0$ against other terms in the Einstein equation.  Since $\xi_0$ is the rate of change of the breathing mode of $\MM$, the volume of $\MM$ must vary significantly over a Hubble time.  We cannot significantly slow this evolution, for any solution in which $\MM$ evolves more slowly will not be optimal and will permit fewer e-foldings of accelerated expansion.  In typical Kaluza-Klein reductions, the volume of $\MM$ controls the values of coupling constants in a model-dependent way. Part of the moduli stabilisation problem is keeping these moduli fixed so that their associated couplding constants do not vary too much with time. It therefore seems likely that constraints on varying couplings could be very effective at ruling out cosmic acceleration in certain classes of models.

Observationally, the new no-go theorems are useful for they show that measurements of $w$ in the present-day universe can give useful information about the higher-dimensional theory.  The case is sometimes made that experiments which attempt to measure the current $w$ to great precision are useless for discriminating between different sources of dark energy.  As one example, the dark energy could be a scalar field with potential $V$.  Since in the slow-roll limit $1 + w_{\rm eff} \simeq V'/V  $, by making $V$ sufficiently flat we can make $w_{\rm eff}$ arbitrarily close to the de Sitter value of $-1$.  The no-gos indicate that there are thresholds in $w$, relatively far from $w=-1$, beyond which the nature of higher-dimensional physics must change significantly.   Experiments which could constrain $w$ to lie below these thresholds therefore provide a promising avenue to learn about fundamental physics from observations of the present-day universe.


\section*{Acknowledgements}
Over the course of this project we have benefited greatly from discussions with Latham Boyle, Gary Gibbons, Juan Maldacena, Kate Marvel, Claudia de Rham, David Seery, Paul Steinhardt, Andrew Tolley, and Amanda Weltman.  We are grateful to Scott Dodelson for corrections, and to Juan Maldacena for pointing out an error in Section \ref{ss:WarpeddeSitter} in an earlier version of this manuscript.  We thank the Perimeter Institute, the University of Cape Town, and the Center for Theoretical Sciences at Princeton University for their hospitality while completing parts of this work.

\appendix

\section{Curvature computations}\label{s:curvature}

Taking the metric parameterisation of
\begin{subequations}
\begin{align}
\eu 0 & = e^{\Omega(t,y)} N(t) \, \ed t \\
\eu m & = e^{\Omega(t,y)} a(t) \, \ed x^\mu \\
\eu a & = \eud a \alpha (t,y) \, \ed y^\alpha
\end{align}
\end{subequations}
The first Mauer-Cartan structure equation defines the spin connection in terms of the derivatives of vielbeins
\be\label{eq:MC1}
\ed \eu A + \wud A B \wedge \eu B = 0
\ee
if one parameterises this as
\be\label{eq:MC1_p}
\wud A B = \wudd A B C \, \eu C, \qquad \ed \eu A = {c^A}_{BC} \, \eu B \wedge \eu C
\ee
then using the useful identity
\be
\wddd A B C = \left(
c_{ABC} + c_{BCA} - c_{CAB} - c_{ACB} - c_{BAC} + c_{CBA}
\right) / 2
\ee
one finds
\begin{subequations}
\begin{align}
\wddd m 0 n &= \delta_{mn} \frac{e^{-\Omega}}{N} \left( \frac{\dot A}{A} + \dot\Omega\right) \\
\wddd 0 a 0 &= - \partial_a \Omega \\
\wddd a 0 b &= \frac{e^{-\Omega}}{N} \xi_{ab} \\
\wddd a b c &= \woddd a b c
\end{align}
\end{subequations}
The second Mauer-Cartan structure equation gives the curvature $\thud A B$ by
\be
\thud A B = \ed \wud A B + \wud A C \wedge \wud C B
\ee
The curvature $\thud A B$ is related to the Riemann curvature tensor through
\be
\thud A B = \frac{1}{2} {R^A}_{BCD} \, \eu C \wedge \eu D
\ee
Computing the curvature requires the identity
\be
\totd {}{t} \woddd a b c = \Dod b \, \xidd a c - \Dod a \,\xidd c b - \woddd a b d \,\xiud d c
\ee
which can be proven as follows.  Differentiating (\ref{eq:MC1}) leads to the expression
\be
\left( - \totd {}{t} \woddd a b c + \Dod b \, \xidd a c + \woddd a c d \, \xiud d b \right) \, \eu b \wedge \eu c = 0
\ee
which is exactly the expression (\ref{eq:MC1_p}) with the derivative of $\woddd a b c$ playing the role of $\wddd A B C$ and 
\be
c_{abc} = \Dod b \, \xidd a c + \woddd a c d \,\xiud d b
\ee
The identity can be written in more geometrical language by defining
$\xiu a = \xiud a b \eu b$, in which case
\be
\totd {}{t} \wodd a b = \Dod b \,\xid a - \Dod a \,\xid b
\ee
In any event using this identity straightforward computation reveals
\begin{subequations}
\begin{align}
R_{0m0n} &=  \left[ \frac{e^{-2\Omega}}{N^2}\left( -\frac{\ddot A}{A} + \frac{\dot A \dot N}{A N} - \ddot \Omega - \frac{\dot A}{A}\dot \Omega + \frac{\dot N}{N} \dot \Omega \right) + (\partial \Omega)^2
\right] \delta_{mn}\\
R_{mnrs} &=  \left[
\frac{e^{-2\Omega}}{N^2} \left( \frac{\dot A}{A} + \dot \Omega \right)^2 - (\partial\Omega)^2 \right]\left(\delta_{mr}\delta_{ns} - \delta_{ms}\delta_{nr}\right) \\
R_{0a0b} &= \frac{e^{-2\Omega}}{N^2} \left[ - \dotxidd a b - \xidd a c \xiud c b + \left( \frac{\dot N}{N} + \dot\Omega \right) \xidd a b \right] + \Dod b \partial_a \Omega + \partial_a \Omega \partial_b \Omega \\
R_{0abc} &= \frac{e^{-\Omega}}{N}\left( \Dod c \xidd a b - \Dod b \xidd a c + \partial_b \Omega \xidd a c - \partial_c \Omega \xidd a b \right) \\
R_{manb} &= \left[ \frac{e^{-2\Omega}}{N^2}\left( \frac{\dot A}{A} + \dot \Omega \right)\xidd a b -\Dod b \partial_a \Omega - \partial_a \Omega \partial_b \Omega \right]\delta_{mn} \\
R_{abcd} &= \Rodddd a b c d + \frac{e^{-2\Omega}}{N^2} \left( \xidd a c \xidd b d - \xidd a d \xidd b c \right) \\
R_{0man} &= - \frac{e^{-\Omega}}{N} \totd{\partial_a \Omega}{t} \delta_{mn}
\end{align}
\end{subequations}
For the Ricci tensor and scalar
\begin{subequations}
\begin{align}
R_{00} &= \frac{e^{-2\Omega}}{N^2} \left[ -3 \frac{\ddot A}{A} + 3\frac{\dot A \dot N}{AN} - 3\ddot \Omega - 3\frac{\dot A}{A}\dot \Omega + 3 \frac{\dot N}{N}\dot \Omega - \dot \xi - \xidd a b \xiuu a b + \left( \frac{\dot N}{N} + \dot \Omega \right)\xi \right] \notag \\
& \qquad + 4 (\partial \Omega)^2 + \DDo \Omega \\
R_{mn} &= \frac{e^{-2\Omega}}{N^2} \left[ \frac{\ddot A}{A}  - \frac{\dot A\dot N}{AN} + 2 \left(\frac{\dot A}{A}\right)^2 + \left( \frac{\dot A}{A} + \dot \Omega\right) \xi + \ddot \Omega + 5 \frac{\dot A}{A} \dot \Omega - \frac{\dot N}{N}\dot \Omega + 2\dot \Omega^2 \right]\delta_{mn} \notag \\
& \qquad - \left[ \DDo \Omega + 4(\partial\Omega)^2 \right]\delta_{mn} \\
R_{ab} &= \Rodd a b + \frac{e^{-2\Omega}}{N^2} \left[ \dotxidd a b + \left( 3\frac{\dot A}{A} - \frac{\dot N}{N} + 2\dot\Omega\right)\xidd a b + \xi \xidd a b \right] - 4\left( \Dod a \partial_b \Omega + \partial_a \Omega \partial_b \Omega \right) \\
R_{0a} &= \frac{e^{-\Omega}}{N} \left[ -3 \partial_a \dot \Omega + \Dod b \left( e^\Omega \xiud b a \right) - \Dod a \left( e^\Omega \xi \right) \right]
\end{align}
\end{subequations}
Finally
\begin{align}
R & = \Ro + \frac{e^{-2\Omega}}{N^2} \Bigg{[} 6 \left( \frac{\ddot A}{A} - \frac{\dot A \dot N}{AN} + \left(\frac{\dot A}{A}\right)^2 \right) +
6 \left( \ddot \Omega + 3 \frac{\dot A}{A} \dot \Omega - \frac{\dot N}{N} \dot \Omega + \dot \Omega^2 \right) \notag \\
& \qquad  + \frac{n+1}{n} \xi^2 + \sigma^2 + 2 \frac{N e^{-2\Omega}}{A^3}\totd {}{t} \left( \frac{A^3 e^{-2\Omega}}{N} \xi \right) \Bigg{]} - 20 (\partial \Omega)^2 + 8 \DDo \Omega
\end{align}
We require these expressions in terms of the Einstein frame $a$ and $n$, instead of the Jordan frame $A$ and $N$.  Using (\ref{e:EinsteinFrameTX}) gives
\begin{subequations}\label{e:RicciTensorInEF}
\begin{align}
R_{00} & = \frac{e^{-2\Omega + \phi}}{n^2} \Bigg{[}
-3 \frac{\ddot a}{a} + 3 \frac{\dot a \dot n}{an} - \frac{k+2}{2k} \xi_0^2 - \sigma^2 - \frac{1}{k} \xi_\perp^2 + \dot\Omega\xi_\perp + \frac{1}{2} \frac{n}{a^3} \frac{\ud}{\ud t} \left( \frac{a^3}{n} \xi_0 \right) \notag \\ 
& \qquad - 3 \ddot \Omega - 3 \frac{\dot a}{a} \dot \Omega + 3 \frac{\dot n}{n} \dot\Omega + \dot\Omega\xi_0 - \dot\xi_\perp - \frac{k+2}{2k} \xi_0 \xi_\perp + \frac{\dot n}{n} \xi_\perp \Bigg{]} + 4(\delta \Omega)^2 + \DDo \Omega \\
R_{mn} &= \delta_{mn} \frac{e^{-2\Omega + \phi}}{n^2} \Bigg{[}
\frac{\ddot a}{a} -\frac{\dot a \dot n}{an} + 2 \left(\frac{\dot a}{a}\right)^2 + \dot\Omega\xi_\perp - \frac{1}{2} \frac{n}{a^3} \frac{\ud}{\ud t} \left( \frac{a^3}{n} \xi_0 \right) + 2 \dot\Omega^2 + \ddot \Omega + 5 \frac{\dot a}{a} \dot\Omega - \frac{\dot n}{n} \dot\Omega \notag \\
& \qquad - \dot\Omega \xi_0 + \frac{\dot a}{a} \xi_\perp - \frac{1}{2} \xi_0 \xi_\perp  \Bigg{]} - \delta_{mn} \left[ 4(\delta\Omega)^2 - \DDo \Omega \right] \\
R_{ab} &= \delta_{ab} \frac{e^{-2\Omega + \phi}}{n^2} \Bigg{[}
\frac{1}{k} \frac{n}{a^3} \frac{\ud}{\ud t} \left( \frac{a^3}{n} \xi_0 \right) + \frac{1}{k} \xi_\perp^2 + \frac{2}{k} \xi_\perp \dot\Omega + \frac{1}{k}\xi_0 \xi_\perp + \frac{2}{k} \xi_0 \dot\Omega + \frac{1}{k} \frac{n}{a^3} \frac{\ud}{\ud t} \left( \frac{a^3}{n} \xi_\perp \right)
 \Bigg{]} \notag \\
 & \qquad + 
\frac{e^{-2\Omega + \phi}}{n^2} \Bigg{[} \frac{n}{a^3} \frac{\ud}{\ud t} \left( \frac{a^3}{n} \sigma_{ab} \right) + \xi_\perp \sigma_{ab} + 2 \dot\Omega \sigma_{ab}  \Bigg{]}
\end{align}
\end{subequations}
and
\begin{align}\label{e:RicciScalarInEF}
R & = \Ro - 8 \DDo \Omega - 20 (\partial \Omega)^2 + \frac{e^{-2\Omega + \phi}}{n^2} \Bigg{[} 6 \left( \frac{\ddot a}{a} - \frac{\dot a \dot n}{an} + \left(\frac{\dot a}{a}\right)^2 \right) - \frac{n}{a^3}\frac{\ud}{\ud t} \left(\frac{a^3}{n} \xi_0 \right) \notag \\
& \qquad + \frac{k+2}{2k} \xi_0^2 + \sigma^2 + \frac{k+1}{k} \xi_\perp^2 + 4\xi_\perp \dot\Omega + 6\dot\Omega^2 + 2\frac{n}{a^3}\frac{\ud}{\ud t}\left( \frac{a^3}{n} \xi_\perp \right) + 6 \frac{n}{a^3}\frac{\ud}{\ud t}\left( \frac{a^3}{n}\dot\Omega \right) \notag \\
& \qquad - 2\dot\Omega\xi_0 + \frac{2}{k} \xi_0 \xi_\perp \Bigg{]}
\end{align}
Decomposing the terms into constant and $\perp$ components gives the four-dimensional action (\ref{e:NoRestrictEHTerm}).  Applying the  restriction (\ref{e:TheRestriction}) yields the physical four-dimensional action (\ref{e:Phys4Daction}).  Forming the Einstein tensor from (\ref{e:RicciTensorInEF}) and (\ref{e:RicciScalarInEF}) and applying the restriction (\ref{e:TheRestriction}) gives the Einstein equations (\ref{e:Einstein_00}), (\ref{e:Einstein_mn}), and (\ref{e:Einstein_ab}).

\section{Four-dimensional action and higher-dimensional Einstein equations}\label{ss:EinsteinEquations}

To obtain a sensible four-dimensional theory, and to prove the no-go theorems, it is necessary to place a single restriction on the evolution of the metric on $\MM$.  
The restriction is
\be\label{e:TheRestriction2}
2\dot\Omega_\perp + \xi_\perp = 0
\ee
Detailed arguments supporting this restriction are given in detail in Appendix \ref{ss:RestrictionDiscussion}, but here we merely describe the equations that result when (\ref{e:TheRestriction}) is assumed.  


The four-dimensional action is computed in the standard way using the metric (\ref{e:TheVielbeins}a-c) and transforming to the Einstein conformal frame. 
This calculation is given in Appendix \ref{s:curvature}.  The result is
\begin{align}\label{e:Phys4Daction}
S = & S_{other} + \ell_4^{-2} \int \left[
-6 \left( \frac{\dot a}{a}\right)^2 + \frac{k+2}{2k} \xi_0^2 + \mavg{\sigma^2} + \frac{k+2}{2k} \mavg{\xi_\perp^2} \right] \frac{a^3}{n} \; \ud t \notag \\
& +  \ell_4^{-2} \int e^{-2\phi} \left[ \int \left( \Ro + 12(\partial \Omega)^2 \right) e^{4\Omega} \; \deteM \; \ud^k y \right] a^3 n \; \ud t
\end{align}
where $S_{other}$ represents the part of the action that does not come solely from the $(4+k)$-dimensional Einstein-Hilbert term, and describes the other degrees of freedom that are present in the higher-dimensional theory.  This is precisely analogous to Einstein gravity with scalar fields $\phi^I$, if we take the ``velocities" $\xi_0,\xi_\perp$ and $\sigma$ to be analogues of  $\dot\phi^I$. There is an effective potential $U$ for the metric modes and a kinetic energy $T$.  The kinetic and potential terms are uniquely identified in the general case through their scaling with the lapse $n$ and  scale factor $a$ in the four-dimensional Einstein frame, and are
\begin{subequations}
\begin{align}
n^2 T &  = \frac{k+2}{4k}\left( \xi_0^2 + \mavg{\xi_\perp^2}\right) + \frac{1}{2}\mavg{\sigma^2} \\
U & =  - \ell_4^{-2} e^{-2\phi} \int \left[ \frac{\Ro}{2} + 6(\partial \Omega)^2 \right] \; e^{4\Omega} \deteM \; \ud^k y
\end{align}
\end{subequations}
If there were no other physics involved, these would be the only kinetic energies and potentials in the four-dimensional effective theory, but in the general case they are only one of many contributions.  When $\MM$ has a sensible moduli space, then $T$ would be the kinetic term for the scalars present in the lower-dimensional theory.  These scalars would have a potential that comes from the dependence of $\Ro$ on their expectation values, as well as from other higher-dimensional physics. 

Varying with respect to the lapse $n$ and scale factor $a$ gives the standard Friedmann and acceleration equations of a FRW universe
\begin{subequations}\label{eq:4dEOM}
\begin{align}
3 \left(\frac{\dot a}{an}\right)^2 &=  \rho_{T} = T + U + \rho_X \\
-2 \left( \frac{\ddot a}{a} - \frac{\dot a \dot n}{an} \right) - \left(\frac{\dot a}{an}\right)^2 &=P_{T} = T - U + P_X
\end{align} 
\end{subequations}
which defines the total effective energy density $\rho_{T}$ and pressure $P_{T}$ in four dimensions. This includes the energy density $\rho_X$ and pressure $P_X$ coming from the unspecified physics encoded in $S_{other}$.  Throughout this paper, when we describe the four-dimensional energy density and pressure, we are always referring to $\rho_T$ and $P_T$.  It is crucial to note that this automatically includes all sources of stress-energy and not just those associated with metric moduli.

Many of our arguments rest on the components of the Einstein tensor in $(4+k)$ dimensions.  It is convenient to express these components in terms of four-dimensional Einstein frame quantities.  The tensor components given below assume that the restriction (\ref{e:TheRestriction}) has been applied: without this restriction (or with a different one of a similar nature) the tensors  have different components.   The components are given with tangent space indices: to convert to coordinate indices it is necessary to multiply by the vielbein components (\ref{e:TheVielbeins}a-c).  The energy density in this basis is
\begin{align}\label{e:Einstein_00}
G_{00} =& \frac{1}{2} \Ro - 3 \DDo \Omega - 6 (\partial \Omega)^2 \notag \\
& + \frac{e^{-2\Omega + \phi}}{n^2} \left[ 3 \left(\frac{\dot a}{a}\right)^2 - \frac{k+2}{4k} \left( \xi_0 + \xi_\perp \right)^2  - \frac{1}{2} \sigma^2 \right] \end{align}
where $\Ro$ is the intrinsic Ricci scalar for the compactification manifold $\MM$ in the unwarped metric, and $\DDo$ is the Laplacian defined by this metric ($g_{\alpha\beta}^{(k)}$ in (\ref{e:KKmetric})).  The pressure along the three noncompact directions is isotropic with components
\begin{align}\label{e:Einstein_mn}
G_{mn} =& - \frac{1}{2} \delta_{mn} \Ro + 3 \delta_{mn} \DDo \Omega + 6\delta_{mn} (\partial \Omega)^2 \notag \\
& + \delta_{mn} \frac{e^{-2\Omega + \phi}}{n^2} \left[ -2 \frac{\ddot a}{a} 
 + 2\frac{\dot a \dot n}{an} - \left(\frac{\dot a}{a}\right)^2 -\frac{k+2}{4k} \left( \xi_0 + \xi_\perp \right)^2  - \frac{1}{2} \sigma^2 \right]
\end{align}
If we ignore the curvature and warp terms, then by combining (\ref{e:Einstein_00}) and (\ref{e:Einstein_mn}) we obtain the standard Friedmann and acceleration equations for the scale factor $a$, with scalar field kinetic energy in the form of $\xi_0,\xi_\perp$ and $\sigma$.  The resulting equations agree with (\ref{eq:4dEOM}a-b) since the cross terms $\xi_0\xi_\perp$ average to zero upon integration over $\MM$.  This is another sign that the reduction is self-consistent.  The term with the most complex structure is the Einstein tensor along the compact directions, which is 
\begin{align}\label{e:Einstein_ab}
G_{ab} = & \Rodd a b - \frac{1}{2}\delta_{ab} \Ro - 4 \Dod a \Dod b \Omega + 4 \delta_{ab} \DDo \Omega - 4 \partial_a \Omega \partial_b \Omega  + 10 \delta_{ab} (\partial \Omega)^2 \notag \\
& + \delta_{ab} \frac{e^{-2\Omega + \phi}}{n^2} \left[ - 3\frac{\ddot a}{a} + 3 \frac{\dot a \dot n}{an}  - 3 \left(\frac{\dot a}{a}\right)^2 -\frac{k+2}{4k} \left( \xi_0 + \xi_\perp \right)^2 - \frac{1}{2} \sigma^2 \right] \notag \\
& + \frac{e^{-2\Omega + \phi}}{n^2} \left[ \delta_{ab} \frac{k+2}{2k} \frac{n}{a^3} \totd {} t \left( \frac{a^3}{n} [\xi_0 + \xi_\perp] \right)  + \frac{n}{a^3}\totd {} t \left( \frac{a^3}{n} \sigma_{ab} \right) \right]
\end{align}
There are also nonzero components $G_{0a}$ which are proportional to gradients of scalar fields, but these can be ignored: a detailed discussion can be found in the proof of the lemma in Appendix \ref{ss:Lemma}.
Again ignoring the warp and curvature terms, there are some parts of (\ref{e:Einstein_ab}) which appear to be combinations of the four-dimensional Einstein equations, including scalar field kinetic energy.  There are also additional terms that do not seem to fit with the scalar field interpretation.  They are present because the extra dimensions ``see" the scalar fields as distortions of $\MM$.  Taking $\xi_0$, $\xi_\perp$ and $\sigma$ as analogues of $\dot\phi^I$, then these terms would be $\ddot \phi^I$.  This fact is crucial for the constraints on transient acceleration described below.

For the most part the index structure in the Einstein equations can be ignored, and reduced to three scalar quantities $\rho^D$, $P_3^D$ and $P_k^D$.  
The effective pressure $P^D_k$ along the extra dimensions is defined through a trace average of $G_{ab}$ by
\be
P_k^D = \frac{1}{k} \delta^{ab} G_{ab}
\ee
where $\delta^{ab}$ is the Kronecker delta along the $k$ compact dimensions.
In terms of the four-dimensional Einstein frame variables this gives
\begin{align}
P_k^D &= \left( \frac{1}{k} - \frac{1}{2} \right)\Ro + 4 \left( 1 - \frac{1}{k}\right) \DDo \Omega + \left(10 - \frac{4}{k}\right) (\partial \Omega)^2 \notag \\\ 
& + \frac{e^{-2\Omega + \phi}}{n^2} \left[ - 3\frac{\ddot a}{a} + 3 \frac{\dot a \dot n}{an}  - 3 \left(\frac{\dot a}{a}\right)^2 -\frac{k+2}{4k} \left( \xi_0 + \xi_\perp \right)^2 - \frac{1}{2} \sigma^2 \right] \notag \\
& + \frac{e^{-2\Omega + \phi}}{n^2} \left[ \frac{k+2}{2k} \frac{n}{a^3} \totd {} t \left( \frac{a^3}{n} [\xi_0 + \xi_\perp] \right) \right] \label{e:PdDdef}
\end{align}
The effective energy density $\rho^D$ and three-dimensional pressure $P^D_3$ are defined in a similar way
\be\label{e:rDP3Ddef}
\rho^D = G_{00} \qquad
P_3^D = \frac{1}{3}\delta^{mn} G_{mn} 
\ee
where $\delta_{mn}$ is the Kronecker delta along the three noncompact spatial dimensions. Since $G_{mn}$ is isotropic the trace average just picks out one of the diagonal components.  The expressions given in (\ref{e:PdDdef}) and (\ref{e:rDP3Ddef}) use the decomposition of $\xi$ into $\xi_\perp$ and $\xi_0$ defined by the $A=2$ average.  We have occasion below to use different values of $A$ in the averaging process, and these different values of $A$ divide $\xi$ into different components.  As described by (\ref{eq:AverageComparison}), for generic expressions involving $\xi$, $\xi_0$, $\xi_\perp$ and their averages, switching between different $A$s introduces unknown functions of $t$.  In (\ref{e:Einstein_00}), (\ref{e:Einstein_mn}) and (\ref{e:Einstein_ab}), $\xi_0$ and $\xi_\perp$ only appear in the combination $\xi_0+\xi_\perp$.  The unknown functions cancel in the sum, so expressions for different values of $A$ can be obtained after the simple substitution $\xi_{0} \to \xi_{0|A}$ and $\xi_{\perp} \to \xi_{\perp|A}$, and no additional functions appear.

\section{The scalar mode restriction}\label{ss:RestrictionDiscussion}


This section is largely devoted to a discussion of why (\ref{e:TheRestriction}) is reasonable in the context of Kaluza-Klein dimensional reduction.  In this section we argue that a restriction such as (\ref{e:TheRestriction}) is always necessary to obtain a sensible Kaluza-Klein compactification.

Our first argument in favor of (\ref{e:TheRestriction}) is that scalar transformations which preserve the total volume of an isolated manifold $\MM$ can always be eliminated by a coordinate transformation. By ``isolated" we mean that $\MM$ is not treated as a factor in a space of higher dimension.  In the usual Kaluza-Klein picture, the lower-dimensional spectrum is obtained by expanding fluctuations of fields on $\MM$, modulo gauge transformations, into eigenfunctions of harmonic operators on $\MM$.  The arguments here indicate that since the deformation in question can be gauged away, it is unphysical and can be set to any desired value -- in this case, zero. 

We suppose that the metric on $\MM$ is determined by vielbeins $\heu A$, so that the volume density is $\dethe$.  (For this section only, we use $M,N,\dots$ and $A,B,\dots$ for coordinate and tangent-space indices on $\MM$, respectively).  In addition to the coordinates $x^M$ on the manifold, we suppose that the vielbeins depend on a parameter $\lambda$, which is a model for time in the physical case.  If
\be
\frac{\ud \heud A M }{\ud \lambda} = {\xi^A}_B \heud B M
\ee
then  the fractional change in volume density is
\be
\totd {\,\ln \dethe}{\lambda} = \xi  
\ee
with $\xi = {\xi^A}_B {\delta^B}_A$.  Only the scalar deformations can influence the volume of $\MM$.  Next, consider an infinitesimal coordinate transformation 
$x^M \to x^M + \delta x^M$, under which the vielbeins transform as
\be\label{e:CoordTX}
\delta \heud A M =  \heud A {M,N} \delta x^N + \heud A N \partial_M{\delta x^N}
\ee
this means that 
\be
\delta \,\dethe =  {\hat{E}_A}^M \delta \heud A M  \,\dethe  = \widehat\nabla \cdot \delta x \, \dethe
\ee
Now we focus on scalar coordinate transformations of the form
\be
\delta x^M = g^{MN}\widehat\nabla_N \delta\chi
\ee
where $g_{MN}$ is the metric and $\widehat\nabla$ is the gradient associated with the vielbeins $\heu A$, and $\delta\chi$ an infinitesimal scalar function.  Under the combined metric change and coordinate transformation the volume density transforms as
\be\label{e:InfiniVolTransform}
\delta \ln \,\dethe = \xi\delta\lambda+ \widehat\Lap \delta\chi
\ee
with $\widehat\Lap = \widehat\nabla \cdot \widehat\nabla$ the Laplacian associated with $\heu A$.  Now we prove the assertion that any transformation $\xi$ which preserves the total volume can be gauged away.  Given any function $\delta s$ on $\MM$, we can always solve the equation
\be
\widehat\Lap \delta\chi = \delta s
\ee
provided we satisfy the consistency condition
\be\label{e:PoissonConsistency}
0 = \int \delta s \, \dethe \,\ud^k x = \int \widehat\Lap \delta\chi\, \dethe \,\ud^k x
\ee
This is the statement that we can solve the Poisson equation on a compact manifold $\MM$ provided that the total charge vanishes.  By defining
\be
\langle \xi \rangle = \left( \int \xi \, \dethe \,\ud^k x \right)
\left( \int \, \dethe \,\ud^k x \right)^{-1}
\ee 
then (\ref{e:PoissonConsistency}) indicates that the equation
\be\label{e:SolvablePoissonEquation}
-\widehat\Lap \delta\chi = \left(\xi - \langle \xi \rangle \right)\delta\lambda
\ee
can be solved for $\delta\chi$ given arbitrary $\xi$.  By solving  (\ref{e:SolvablePoissonEquation}) equation (\ref{e:InfiniVolTransform}) becomes
\be
\delta \ln \,\dethe = \langle\xi\rangle \delta\lambda
\ee
which means that the change in the volume density is a constant function over $\MM$.   If the transformation preserves the total volume, which implies $\langle\xi\rangle = 0$, then the change to the volume density $\dethe$ can be completely gauged away by a coordinate transformation. Up to now we have taken the vielbeins $\heu a$ to be completely arbitrary, but by taking them to be the auxiliary warped metric on $\MM$, or $\heu a = \teu a = e^{2\Omega/k} \eu a$ we have proven the original assertion:  any scalar transformation which preserves the total volume in the auxiliary ``warped" metric on $\MM$ can be gauged away by a coordinate transformation.

There is a separate, geometrical argument in favor of the restriction (\ref{e:TheRestriction}), framed in terms of locally-defined quantities on $\MM$, as opposed to nonlocal ones such as the total volume of $\MM$.  The evolution of the manifold $\MM$ with time can be visualised as the motion of a point $p$, representing the specific metric on $\MM$, through the space $\text{Met}(\MM)$ of all metrics on $\MM$.  We have a well-defined moduli space if $\text{Met}(\MM)$, modulo coordinate transformations $\text{Diff}(\MM)$, has a manifold structure.  Usually this is only possible if $\MM$ has some special property.  But in the general case where $\text{Met}(\MM)/\text{Diff}(\MM)$ is not well-defined, we can nonetheless say where we are going even if we can't say exactly where we are. That is, we can describe a velocity of $p$ through the space of metrics, even if we cannot give a sensible coordinatisation in an open neighborhood of $p$.  This velocity is just the change in the metric per time, which is precisely the $\xidd ab $ defined above.  The possible velocities at $p$ span a vector space $\text{Vel}_p(\MM)$. This is an infinite-dimensional function space, but as a linear space it splits into two subspaces
\be
\text{Vel}_p(\MM) = \text{Diff}_p(\MM) \oplus \text{Phys}_p(\MM)
\ee
where the first summand contains those $\xidd ab$ that are pure coordinate transformations, and the second summand is its complement, the physical metric transformations.  The subscript $p$ is a reminder that we are dealing with quantities defined at $p$: when a moduli space description exists $\text{Phys}_p(\MM)$ is the tangent space of the moduli space at $p$.  One way to ensure that a given velocity $\xidd a b \in \text{Vel}_p(\MM)$ is physical is by demanding that it is orthogonal to all infinitesimal coordinate transformations in a suitable metric on the velocity space.  Viewed as a function space the natural requirement is
\be\label{e:OrthoToGauge}
\int \xidd a b \, \delta \eud a \alpha E^{b \alpha} \; \dethe \, \ud^n y = 0
\ee
for any $\delta \eud a \alpha$ that is obtained by a coordinate transformation. Using the  formula (\ref{e:CoordTX}) for $\delta \eud a \alpha$ and integrating by parts yields the condition
\be\label{e:OrthoToGaugeCondition}
\widehat \nabla_A \xiud A B = 0
\ee
For scalar coordinate transformations $\xidd A B = \xi \delta_{A B}$, so the condition (\ref{e:OrthoToGaugeCondition}) becomes $\partial_M \xi = 0$.  This implies that, to be a physical metric transformation, $\xi$ must be a constant, which is precisely the conclusion reached previously via a different route.  

These arguments cannot be naively extended to the time-dependent Kaluza-Klein case.  If the parameter $\lambda$ is promoted to the time coordinate of the full Kaluza-Klein spacetime, then the coordinate transformations employed here would introduce $\ud t\, \ud y^\alpha$ components of the metric, taking us out of the canonical form (\ref{e:KKmetric}).  On the other hand, our arguments suggest that a restriction such as (\ref{e:TheRestriction}) is necessary.  If the extra-dimensional manifold $\MM$ evolves along a sequence of metrics that are related by nothing more than a coordinate transformation, it seems that within the context of the Kaluza-Klein philosophy we should not see any difference in the four-dimensional effective theory.  Indeed, within the context of the ``moduli space approximation," where one considers only adiabatic evolution of $\MM$ through a sequence of approximately static configurations, our arguments show that there is no problem imposing (\ref{e:TheRestriction}) as a gauge choice.  Problems only appear when one attempts to go beyond the moduli space approximation consistently, as we do here.

Our third argument in favor of the restriction (\ref{e:TheRestriction}) is based on the apparent pathologies that appear when it is relaxed.
In this case, before integrating over $\MM$, the part of the higher-dimensional action originating from the Einstein-Hilbert term is
\be\label{e:NoRestrictEHTerm}
\int \left[ -6 \left(\frac{\dot a}{a}\right)^2 + \frac{k+2}{2k} \xi_0^2
- \frac{k-1}{k}\xi_\perp^2 -6 \xi_\perp\dot\Omega_\perp - 6\dot\Omega_\perp^2 \right] \frac{a^3}{n} e^{2\Omega} \dete \; \ud^{k} y\, \ud t
\ee
where we have ignored terms that average to zero, as well as some others that are irrelevant here.  The action is that of a flat FRW universe coupled to three scalars $\xi_0$, $\xi_\perp$ and $\dot\Omega_\perp$, where the latter two have a non-diagonal kinetic term.
Writing the kinetic term for $\xi_\perp$ and $\dot\Omega_\perp$ as $\Phi^T M \Phi$, with $\Phi^T = (\xi_\perp , \dot\Omega_\perp)$ and
\be
M = \left(
\begin{matrix}
-\frac{k-1}{k}   & -3 \\
-3               & -6
\end{matrix}
\right)
\ee
then diagonalising $M$ gives two eigenvalues $\lambda_\pm$ and defines two fields $\theta_\pm$ that are linear combinations of $\xi_\perp$ and $\dot\Omega_\perp$, with $M\theta_\pm = \lambda_\pm \theta_\pm$.  Since $\text{tr}(M) < 0 $ and
$\text{det}(M) < 0$ for all $k > 0$, we have $\lambda_+ > 0 $ and $\lambda_- <0$. Integrating over $\MM$ yields the four-dimensional action
\be
\int \left[ -6 \left(\frac{\dot a}{a}\right)^2 + \frac{k+2}{2k} \xi_0^2
+\lambda_+ \mavg {\theta^2_+} +\lambda_- \mavg {\theta^2_-}\right] \frac{a^3}{n} \,\ud t
\ee
Because $\lambda_- < 0$ the kinetic terms have a Lorentzian signature $(-++)$.  If we were to interpret these kinetic terms as moduli kinetic terms, then fluctuations in $\theta_-$ give rise to apparent ghosts.  Of course, these are not real ghosts, since the higher-dimensional theory is ghost-free and we should not be able to introduce any by dimensional reduction.  But it does mean that our interpretation of the Kaluza-Klein fields as scalars in four dimensions is breaking down.

A problem like this always arises in Kaluza-Klein reductions unless a restriction such as (\ref{e:TheRestriction}) is made.  We can illustrate this by  working within a framework which contains only the minimal elements for dimensional reduction.  To eliminate features that come from specific choices of metric or gauge, it is convenient to use an Arnowitt-Deser-Misner (ADM) decomposition of the $(4+k)$-dimensional metric, which simplifies managing gauge freedom \cite{MTW,ADM:62,DeWitt:1967yk}.  The ADM decomposition of the metric is
\be
\ud s^2 = - (N \ud t)^2 + ( \ud X^I + N^I \ud t )( \ud X^J + N^J \ud t ) \gamma_{IJ}
\ee
where $N$ is the lapse function, $N^I$ the shift vector, $I=1,\dots 3+k$ denotes purely spatial indices, and $\gamma_{IJ}$ the induced metric on spatial surfaces.  The extrinsic curvature is defined by
\be
K_{IJ} = \frac{1}{2N}\left( N_{I|J} + N_{J|I} - \frac{\ud}{\ud t} \gamma_{IJ} \right)
\ee
where ``$|$" denotes covariant derivatives with respect to the metric $\gamma_{IJ}$.  In these variables the Einstein-Hilbert action is
\be
\int \left( \,^{(3+k)}R + K_{IJ} K^{IJ} - [\text{tr}\,K]^2 \right ) N \sqrt{\gamma} \, \ud^{3+k} x \,\ud t
\ee
where $\text{tr}\,K = \gamma^{IJ} K_{IJ}$, $\,^{(3+k)}R$ is the (intrinsic) Ricci scalar of $\gamma_{IJ}$, and some total derivatives and Lagrange multiplier terms have been dropped.  

The full ADM action shows that $N$ and $N^I$ are nondynamical, and their equations of motion are constraint equations arising from the coordinate freedom in the problem. This freedom manifests itself in the ADM action through the freedom to specify $N$ and $N^I$ as desired.  We do not fix $N$ or $N^I$ at all, except as dictated by some symmetry requirements.  By framing the Kaluza-Klein reduction in terms of the extrinsic curvature, we can argue without reference to $N$ or $N^I$ at all, thus ensuring our argument is independent of any specific choice of gauge.

To have a sensible four-dimensional cosmology, we should minimally require that the metric is invariant under three-dimensional rotations.  Taking $\mu,\nu$ as three-dimensional indices and $\alpha,\beta$ as $k$-dimensional ones, this implies $K_{\mu\alpha} = 0$, which means $K_{IJ}$ is block-diagonal, with a purely ``three-dimensional" block and a purely extra-dimensional one. Rotational invariance also implies that $K_{\mu\nu}$ is of the form
\be
K_{\mu\nu} = \frac{1}{3}\delta_{\mu\nu} \tilde\theta_3
\ee
where $\tilde\theta_3$ depends on $t$ and $y$.  We decompose
\be
K_{\alpha\beta} = \frac{1}{k} \gamma_{\alpha\beta} \tilde\theta_k + \tilde\Sigma_{\alpha\beta} \qquad \text{where}\quad \gamma^{\alpha\beta} \tilde\Sigma_{\alpha\beta} = 0
\ee
Here $\tilde\theta_3$ encodes the four-dimensional Jordan-frame Hubble parameter and rate of change of warp factor, and $\tilde\theta_k$ encodes the rate of change of the extra-dimensional volume.

The other minimal element of a Kaluza-Klein reduction is a conformal transformation to obtain Einstein frame gravity in four dimensions.  To achieve this we set
\be
\gamma_{\mu\nu} = e^{2\Psi} g_{\mu\nu},\qquad N = e^\Psi n
\ee
with $g_{\mu\nu}$ and $n$ the Einstein frame metric and lapse,
and $\Psi$ a function of $t$ and $y^\alpha$.  Then we have also
\be
\tilde\theta_3 = e^{-\Psi} \left( \theta_3 - \frac{3}{n}\frac{\ud \Psi}{\ud t} \right), \qquad \tilde\theta_k = e^{-\Psi}\theta_k, \qquad \tilde\Sigma_{\alpha\beta} = e^{-\Psi}\Sigma_{\alpha\beta}
\ee
where $\theta_3$, $\theta_k$ and $\Sigma_{\alpha\beta}$ are associated with the ``Einstein frame" extrinsic curvature that is naturally defined with $n$ and $g_{\mu\nu}$.  Finally, it is convenient to define a quantity $\Delta$ by
\be
2\dot\Psi + \theta_k + \Delta = 0.
\ee
Rewriting the original ADM action in these variables gives
\be\label{e:ADMinVariables}
\int \left[ -\frac{2}{3} \theta_3^2 + \frac{k+2}{2k} \theta_k^2 +
\Sigma_{\alpha\beta}\Sigma^{\alpha\beta}
+ \Delta \left( 2\theta_3 - \frac{3}{2}\Delta\right) \right]
e^{2\Psi} n \sqrt{g}\sqrt{\gamma^k} \ud^{3+k} x \,\ud t
\ee
where $\gamma^k$ is the determinant of the $\alpha\beta$-block of the metric.  

After integrating over the extra dimensions, the action (\ref{e:ADMinVariables}) should describe Einstein gravity.  If $\gamma_{IJ}$ were independent of the $y^\alpha$ then $\theta_3 = -3H$, with $H$ the four-dimensional Einstein frame Hubble parameter.  So it is reasonable that the first term would give the canonical $-6H^2$ in the four-dimensional action.  The second and third terms give the kinetic energy terms appropriate for a system of scalar fields, with the appropriate signs.  The last term represents a nonstandard coupling between the Hubble parameter and scalar field kinetic energy, of the schematic form $H\dot\Psi$.  It can be eliminated by choosing $\Delta = 0$ or $\Delta = 4\theta_3/3$.  In either case, integrating over the extra dimension defines the four-dimensional Hubble parameter by the requirement that we obtain Einstein gravity.  Denoting the integral over the compact dimensions by $[\cdot]$, we would have $9H^2 = [\theta_3^2]$.

Regardless of the averaging, there is always an apparent ghost mode in the four-dimensional theory.  The problem comes from the first term.  Since $H$ is a function of time only but $\theta_3$ is a function of both time and space, we should decompose
\be
\theta_3 = H + \delta\theta_3
\ee
Inserting this in the action (\ref{e:ADMinVariables}) we obtain a cross term $H\delta\theta_3$ which potentially integrates to zero, but we also obtain a nonpositive term $-(2/3)\delta\theta_3^2$.  This term cannot be interpreted as a sensible scalar field in four dimensions, because it has a ghostlike kinetic term with the wrong sign.  Its presence is a signal that the Kaluza-Klein dimensional reduction has broken down.  In the reduction studied here, the restriction (\ref{e:TheRestriction}) eliminates this ghost mode.

The presence of these ``wrong-sign" kinetic terms is guaranteed in an unrestricted Kaluza-Klein reduction because it is related to the conformal factor problem which has been extensively studied in the context of Euclidean quantum gravity \cite{Gibbons:1978ac,Page:1978zz}.  The gravitational action has a negative mode because the Einstein-Hilbert term can be made arbitrarily negative through a suitable conformal transformation of the metric.  In four-dimensional gravity we are accustomed to this as a fact of life.  It does not cause any serious problems because the ADM lapse constraints prevent the gravitational Hamiltonian from becoming unboundedly negative -- as evidenced by the existence of positive mass theorems \cite{Schon:1979rg,Schon:1981vd,Witten:1981mf,Ludvigsen:1981gf,Horowitz:1981uw,Gibbons:1982jg}.  In unwarped Kaluza-Klein reductions the Weyl transformation after integrating out the extra dimensions manages to put this ``negative mode" entirely in the four-dimensional gravitational degrees of freedom, so the Kaluza-Klein scalars have a kinetic term with the correct sign.  But in the general case this cannot be accomplished for all of  the negative modes.  The wrong-sign scalar in four dimensions is a remnant of the higher-dimensional conformal mode that could not be repackaged as a four-dimensional conformal mode.  When dimensional reductions are carried out on manifolds which are restricted in some way this problem does not arise. But unless we have more information about the warp factors or the compactification manifold a condition such as (\ref{e:TheRestriction}) is always required.

\section{A useful lemma}\label{ss:Lemma}

The lemma we prove here holds that, to prove NEC violation, we only need the traces of the various components in the higher-dimensional stress-energy tensor. The symmetries of Kaluza-Klein \emph{ansatz} (\ref{e:KKmetric}) indicate that\footnote{We remind the reader that these tensors use the vielbein indices defined by (\ref{e:TheVielbeins}).}
\be
T_{00} = \rho^D \quad T_{mn} = \delta_{mn} P^D \quad T_{0a} = J_a
\ee
and while $T_{ab}$ is arbitrary we define $P^D_k$ by
\be
P^D_k = \frac{1}{k} \delta^{ab}T_{ab}.
\ee
where $\delta_{mn}$ and $\delta_{ab}$ are the Kronecker deltas on the three noncompact spatial dimensions and the $k$ compact dimensions, respectively. We claim the NEC is violated if either
\be\label{e:NECviolationConditions}
\rho^D + P_3^D < 0 \qquad \text{or} \qquad \rho^D + P_k^D < 0
\ee
To prove the first part of the claim, we consider any null vector $n^A = (1,\hat u,0)$ with $\hat u$ a unit vector pointing along the ``large" three spatial dimensions.  Then
\be
T_{MN} n^M n^N = \rho^D + P_3^D
\ee
as defined above.  If the right hand side is negative, then $n^A$ is a null vector that shows the NEC is violated.  

Next we focus on the extra-dimensional parts of $T_{MN}$.  At a fixed point in the spacetime, since $T_{ab}$ is symmetric we can diagonalise it by a matrix in $O(k)$ and the diagonalised stress energy tensor will have real eigenvalues $(\lambda_1, \dots \lambda_k)$.  To each eigenvalue $\lambda_j$ there is an associated unit eigenvector $\hat n_{\lambda j}$.  Taking the null vector $n^A = (1,0,0,0, \epsilon \hat n_{\lambda j})$, with $\epsilon = \pm 1$, yields
\be\label{e:LemmaRhoPk}
T_{MN} n^M n^N = \rho^D + \lambda_j + \epsilon J_a \hat n_{\lambda j}^a
\ee
The last term is nonpositive for at least one of the choices for $\epsilon$. Using the null vector corresponding to this choice of $\epsilon$, then (\ref{e:LemmaRhoPk}) shows that there exists a null vector $n^A$ such that
\be
T_{MN} n^M n^N \le \rho^D + \lambda_j
\ee
This further implies that, if there exists any eigenvalue $\lambda_j$ of $T_{ab}$ such that $\rho^D + \lambda_j < 0$, the NEC is violated. Since the trace of $T_{ab}$ is the sum of the eigenvalues $\lambda_1 \dots \lambda_n$, by definition $P_k^D$ is the average of these eigenvalues.  Therefore there exists an eigenvalue $\lambda_*$ with $\lambda_* \le P_k^D$ and
\be
\rho^D + \lambda_* \le \rho^D + P_k^D 
\ee
Therefore if $\rho^D + P_k^D < 0$, the required eigenvalue exists and the NEC is violated as claimed.

\section{The curvature-free de Sitter case}\label{ss:RFdeSitter}


When the four-dimensional spacetime is exactly de Sitter, it is possible to give a simple proof that the NEC is violated without employing the full machinery of Section \ref{ss:RFQuintessence}.
Using the lemma proven in Appendix \ref{ss:Lemma}, we see that to satisfy the NEC we must have $\rho^D + P_3^D \ge 0$, where
\be\label{e:deSitterRhoPlusP3}
\rho^D + P_3^D = 
\frac{e^{-2\Omega+\phi}}{n^2} \left[ 2\left(\frac{\dot a}{a}\right)^2 - 2\frac{\ddot a}{a} + 2\frac{\dot a \dot n}{an} 
- \frac{k+2}{2k} \left(\xi_0 + \xi_\perp\right)^2 -\sigma^2 
\right]
\ee
Exact de Sitter expansion is defined by
\be\label{e:dSdef}
\frac{\dot a}{an} = H_0 = \text{constant}
\ee
which implies the sum of the first three terms on the right hand side of (\ref{e:deSitterRhoPlusP3}) vanishes. Since the remaining terms in (\ref{e:deSitterRhoPlusP3}) are negative semidefinite, they must vanish, and all of the velocity components $\xi_0$, $\xi_\perp$ and $\sigma_{ab}$ are zero.\footnote{The pointwise condition implies that $\xi_0 = - \xi_\perp$, but since $\xi_0$ is constant over $\MM$ and $\xi_\perp$ averages to zero the only solution to this condition is $\xi_0 = \xi_\perp = 0$.}  This is similar to the four-dimensional situation, where for exact de Sitter expansion, we can have no scalar kinetic energy at all.  To prove that this is so in this Kaluza-Klein case we must assume the NEC in the higher-dimensional theory.  The conclusion that the kinetic terms must vanish is valid regardless of the curvature $\Ro$ and warp $\Omega$ on $\MM$, since these terms precisely cancel in the $\rho^D + P_3^D$ NEC condition.

Next we turn to the other NEC condition.  Since all of the kinetic terms are zero, we have
\begin{align}\label{e:deSitterNeck}
\rho^D + P^D_k &= -3 \frac{e^{-2\Omega+\phi}}{n^2}\left[ \frac{\ddot a}{a} - \frac{\dot a \dot n}{an} \right]
+ \left(1-\frac{4}{k}\right) \DDo \Omega + \left(4-\frac{4}{k}\right) (\partial \Omega )^2 \notag \\
& = -3 e^{-2\Omega+\phi} H_0^2 + \left(1-\frac{4}{k}\right) \DDo \Omega + \left(4-\frac{4}{k}\right) (\partial \Omega )^2
\end{align}
To show that de Sitter expansion implies NEC violation, we show that for any dimension $k$ there exists a point $q \in \MM$, at which the terms involving derivatives of $\Omega$ are nonpositive.  At this point $q$, the right hand side is negative definite, and so the NEC is violated by de Sitter expansion.

To account for the warp terms, we first suppose $k \ne 4$, and show that the warp terms make a nonpositive contribution to the NEC condition  (\ref{e:deSitterNeck}).  We rewrite the warp terms as
\be
\left(1-\frac{4}{k}\right) \DDo \Omega + \left(4-\frac{4}{k}\right) (\partial \Omega )^2 = \left(1-\frac{4}{k}\right)
e^{-A\Omega} \Dod {} \cdot \left[ e^{A \Omega} \Dod {} \Omega \right]
\ee
where
\be
A = 4 \left( \frac{k-1}{k-4} \right)
\ee
We next consider the term
\be
W = \Dod {} \cdot \left[ e^{A \Omega} \Dod {} \Omega \right]
\ee 
and show that we can find a point $q\in \MM$ at which $W \ge 0$ when $k<4$, and a point $q\in \MM$ at which $W \le 0$ when $k>4$.  First consider the $k<4$ case, and suppose that the assertion is false. This would mean that $W < 0$ everywhere on $\MM$, but since
\be\label{eq:div_eA_eq_0}
\int W \, \deteM \; \ud^k  = 
\int \Dod {} \cdot \left[ e^{A \Omega} \Dod {} \Omega \right] \, \deteM \; \ud^k y = 0
\ee
then we have a contradiction.  Thus $W \ge 0$ somewhere on $\MM$ and the point $q$ with the desired properties exists.  In the $k > 4$ case a precisely analogous argument establishes the existence of a point $q$ with $W \le 0$.  When $k=4$ the warp terms can no longer be written as a total derivative, for
\be\label{e:NECcondkis4}
\rho^D + P_k^D = -3e^{-2\Omega+\phi}H_0^2 + 3 (\partial \Omega)^2
\ee
If  $\Omega$ is smooth, then
since $\MM$ is compact  $\Omega$ has an extremum at some point $q$.  At this point $q$ we have $\partial \Omega = 0$, and so the derivative terms vanish.  If $\Omega$ is not smooth then there may be no points with $\partial\Omega = 0$ and this argument does not apply.  There are two alternative arguments which yield similar conclusions: the de Sitter case appears as the $w \to -1$ limit of the arguments presented for transient acceleration in  Section \ref{sss:NEC_k4}, and a NEC no-go is proven using entirely different techniques in Section \ref{ss:WarpeddeSitter}.

\section{The $k=4$ curvature-free case}\label{sss:NEC_k4}

When $k=4$ the constraint on $A$ which ensures that the warp terms are nonpositive is undefined, so this case must be treated separately.  We show here that, by carefully taking a series of limits, the $k=4$ case is merely a continuation of the $k < 4$ cases studied in Section \ref{sss:NEC04and10plus}.
We consider an energy condition which slightly displaces the pathology at $k=4$.  The condition
\be\label{e:EpsilonEC}
\rho^D + (1+\epsilon)P^D_k \ge 0
\ee
becomes one of the NEC conditions when $\epsilon \to 0$.  Throughout this discussion we assume  $|\epsilon | \ll 1 $.  It addition to the usual coefficient of the warp terms in the third line of (\ref{eq:QuinRhoPlusPk}) the new condition (\ref{e:EpsilonEC}) adds
\be\label{e:EpsilonECWarpCoeff}
\epsilon\left( 2 - 4A + \frac{4}{k}+\frac{4A}{k}\right)
\ee
The value of $A$ for which the combined coefficient vanishes forms the boundary of the allowed values of $A$. The coefficient vanishes at $A=A_0$ given by
\be
A_0 = \frac{2k+4}{k-4} - \frac{6k(k+2)}{(k-4)^2}\epsilon + \mathcal{O}(\epsilon^2)
\ee
where the presence of the additional term shifts the vanishing value of $A$ slightly.  The inequalities become undefined when this expression has a pole, which is located at
\be\label{e:k4PoleLoc}
k_\epsilon = 4 - 12\epsilon
\ee
so by adjusting $\epsilon$ we can move the boundary of the inequality to either side of $k=4$.  This means that we can make the $k=4$ case well-defined by an appropriate choice of the sign of $\epsilon$.  To the left of $k_\epsilon$, the warp terms contribute nonpositively if $A \le A_0$, just as for the $0 < k < 4$ cases.  To the right of $k_\epsilon$ the warp terms are nonpositive if $A \ge A_0$, as for the $4 < k < 10$ cases.

The sign of $\epsilon$ is an important choice.  It determines whether the $k=4$ case mimics the $k>4$ or $k<4$ cases.  We show in Section \ref{sss:NEC4to10} that the $k>4$ averaging techniques lead to constraints that are weaker than the $k<4$ constraints: using the $k>4$ techniques, as $k\to 4^+$ only the de Sitter case $ w = -1$ is constrained.  Therefore we should mimic the $k<4$ cases: this amounts to taking $\epsilon < 0$ so that by (\ref{e:k4PoleLoc}) the pole occurs at a $k$ larger than four, so the $k=4$ case is well-defined.  All of the other terms appearing in the $\rho^D + P^D_k$ condition have nonzero coefficients, so as $\epsilon\to 0$ we can neglect the contributions of $\mathcal{O}(\epsilon)$ and use their $\epsilon=0$ values.
Choosing $\epsilon < 0$ also fixes the nature of the energy condition probed with (\ref{e:EpsilonEC}).  The usual NEC condition is that $\rho^D \ge -P^D_k$.  Here we can choose $\epsilon$ arbitrarily close to zero but it cannot vanish.  When $\rho^D \ne -P^D_k$ this new condition is equivalent to the condition $\rho^D > -P^D_k$, where ``$\ge$" in the usual NEC condition has been replaced by ``$>$".  When $\rho^D = P^D_k$ then the case $\rho^D \ge 0$ is allowed by the condition (\ref{e:EpsilonEC}) but $\rho^D < 0$ is forbidden.  In other words, the condition (\ref{e:EpsilonEC}) allows a de Sitter cosmological term but forbids an anti-de Sitter one.

To summarize, in the $k=4$ case, the arguments in the previous paragraph show that either the NEC is violated \emph{or} that the higher-dimensional spacetime has a purely AdS stress energy.  In practice this latter case is not a significant limitation: if the higher-dimensional spacetime is curvature-free and possesses only an AdS cosmological constant, then the resulting energy density in four dimensions has the wrong sign, and cannot account for the positive energy density required by an accelerating Friedmann universe.

\section{Curvature independent averages}\label{sss:CurveIndependent}

In this Appendix we construct a family of combinations of $\rho^D$, $P_3^D$ and $P_k^D$  which do not involve the Ricci curvature $\Ro$.  We also construct the differential equation which gives optimal solutions.  The SEC corresponds to one member of this family, so these results are used in Section \ref{ss:CurvedQuintessence} to prove the SEC no-go theorems for curved $\MM$.

The inequality
\be\label{e:abEnergyCondition}
\rho^D + \gamma P_3^D + \gamma_k P_k^D \ge 0
\ee
is independent of $\Ro$ provided
\be\label{e:bOfa}
\gamma_k = \frac{k(1-\gamma)}{k-2}
\ee
while $\gamma$ is a free parameter. When $k=1$ there can be no curvature, since all one-dimensional manifolds have zero Ricci scalar, so we assume that $k \ge 2$.  We need to be careful since certain expressions (for example, (\ref{e:bOfa})) become undefined at $k = 2$, but the fundamental equations used in  the no-go theorems will turn out to be well-defined in the $k\to 2$ limit.

\begin{figure}
  \begin{center}
  \includegraphics[width=1.0\textwidth]{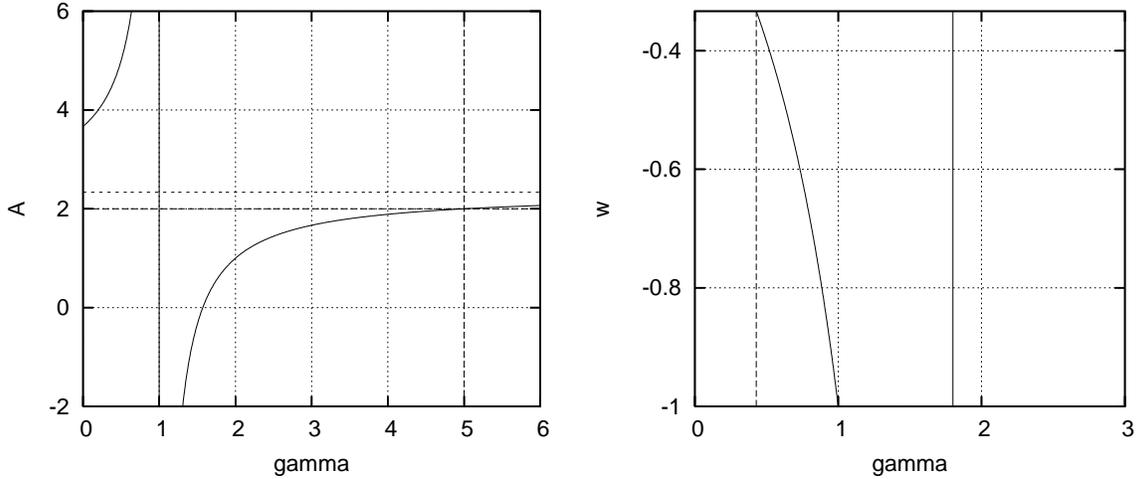}
  \end{center}
   \caption{Summary of positivity constraints for $k=6$ extra dimensions.  The left panel shows the $(\gamma,A)$ plane for $c_\perp$, which is negative if it below the solid black curve.  The lower horizonatal line is the boundary of the $A>2$ region required by $c_\Omega$, and the upper horizontal line the asymptote of the curve.  The right vertical line shows where the constraint curve passes above the $A=2$ line.
Right panel: the $(\gamma,w)$ plane for $c_\rho$.  The left vertical line is $\gamma_{\rho-1/3}$, the curve is $w_{\rho\star}$, an the right vertical line is $\gamma_{\rho{\rm pole}}$.  We have $c_\rho < 0$ for points to the left of the curve, and to the right of $\gamma=1$.}
\label{f:TuneParams}
\end{figure}

We proved no-go theorems in Section \ref{ss:RFQuintessence}  by constructing differential equations from energy condition inequalities, and then using an $A$-average which made as many terms nonpositive as possible.  We follow a similar strategy here, leaving aside for the moment the question of how we should interpret the inequality (\ref{e:abEnergyCondition}) for general $\gamma$. Taking the linear combination parameterised by $\gamma$,  averaging using $\mavgn A \cdot $, setting $n=1$, and dividing through by the coefficient of the $\dot\xi_{0|A}$ term gives 
\be
a^{-3}\totd{}{t}\left[ a^3 \xi_{0|A}\right]
+ c_{0} \xi_{0|A}^2 + c_{\perp}\mavgn A {\xi_{\perp|A}^2}
+ c_\rho \rho_T + c_{\Omega} \mavgn A { e^{2\Omega} (\partial\Omega)^2 }
+ c_\sigma \mavgn A {\sigma^2} = 0
\ee
where we have saturated the inequality in order to find the optimal solution for $\xi_{0|A}$.
The coefficients are
\begin{subequations}
\begin{align}
c_0 &= \frac{1+\gamma-k}{k(1-\gamma)} \label{e:Allowed_c0}\\
c_\perp &= \frac{2+2\gamma-4k+2\gamma k+Ak-\gamma Ak}{2k(1-\gamma)} \label{e:Allowed_cPerp}\\
c_\sigma &= \frac{2(1+\gamma-k)}{(1-\gamma)(2+k)} \label{e:Allowed_cSigma}\\
c_\rho &= \frac{4+k[ (w-1)\gamma-3w-1 ]}{(2+k)(\gamma-1)} \label{e:Allowed_cRho}\\
c_\Omega &= 4-2A \label{e:Allowed_cOmega}
\end{align}
\end{subequations}
The strategy is to choose values of $(\gamma,A)$ so that the terms over which there is no control are nonpositive, and then obtain and solve a differential equation for $\xi_{0|A}$.  The constraints arising from nonpositivity are:

\begin{itemize}

\item $c_\Omega$: Requires $A \ge 2$.

\item $c_\perp$: When this coefficient is negative, the optimum solution has $\xi_{\perp|A} = 0$. The coefficient vanishes along a curve in the $(\gamma,A)$ plane given by
\be
A_{\perp\star} = \frac{2(1+\gamma-2k+\gamma k)}{k(\gamma -1)}
\ee
and is singular when $\gamma=1$.  The pole in $c_\perp$ and in $A_{\perp\star}$ combine so that $c_\perp$ is negative when 
\be\label{e:AlessAperpStar}
A < A_{\perp\star}
\ee
for all values of $\gamma$.  In order to be consistent with the $c_\Omega$ constraint we must have $A \ge 2$.  This is consistent with (\ref{e:AlessAperpStar}) when $\gamma < 1$, or when $ \gamma \ge k-1$.  This constraint is illustrated in Figure \ref{f:TuneParams}.

\item $c_\sigma$: This coefficient is nonpositive if $\gamma \le 1$ or $\gamma \ge k-1$.

\item $c_0$: Nonpositive if $\gamma \le 1$ or $\gamma \ge k-1$, which is the same allowed range as for the $c_\sigma$ coefficient.

\item $c_\rho$: to analyse the constraints it is helpful to define a quantity $w_{\rho\star}$ by
\be
w_{\rho\star}=\frac{k+\gamma k-4}{4\gamma+ \gamma k-3k}
\ee
which has the following properties, viewed as a function of $\gamma$:
\begin{enumerate}
\item At $\gamma=\pm\infty$ it asymptotes to 
\be
w_{\rho\star\infty} = \frac{k}{k+4}
\ee
\item It has a pole at 
\be
\gamma_{\rho{\rm pole}} = \frac{3k}{4+k}
\ee
and is above the asymptote to the right of the pole, below to the left.

\item We have $w_{\rho\star} = -1$ when $\gamma=1$, regardless of $k$.

\item We have $w_{\rho\star} = -1/3$ when $\gamma=\gamma_{\rho-1/3}$ with
\be
\gamma_{\rho-1/3} = \frac{3}{1+k}
\ee
 
\end{enumerate}
Using these properties we can determine when $c_\rho$ is negative as follows, starting from the largest values of $\gamma$ and working downward:
\begin{itemize}

\item When $\gamma > \gamma_{\rho{\rm pole}}$, we need $w < w_{\rho\star}$.  For these values of $\gamma$, $w_{\rho\star} > w_{\rho\star\infty} > 1$ by properties 1 and 2. Therefore, in this range of $\gamma$,  $c_\rho$ is negative for any value $-1 \le w < -1/3$ corresponding to an accelerating universe.

\item When $1 > \gamma > \gamma_{\rho{\rm pole}}$, we must have $w > w_{\rho\star}$.  Now $w_{\rho\star} = -1$ at $\gamma=1$ by property 3, and falls to $w_{\rho\star} = -\infty$ at $\gamma = \gamma_{\rho{\rm pole}}$ by property 2. So in this range of $\gamma$,  $c_\rho$ is negative for any value $-1 \le w < -1/3$ corresponding to an accelerating universe.

\item For $\gamma_{\rho-1/3} > \gamma > 1$, we need $w < w_{\rho\star}$.  By properties 3 and 4, $w_{\rho\star}$ falls from $w_{\rho\star} = -1/3$ at $\gamma=\gamma_{\rho-1/3}$ to $w_{\rho\star} = -1$ at $\gamma=1$.  Therefore $c_\rho$ is negative for some $w$ in the range $-1 \le w < -1/3$, depending on the value of $\gamma$.

\item When $\gamma < \gamma_{\rho-1/3}$ we need $w < w_{\rho\star}$.  But by properties 2 and 4, for these values of $\gamma$, we have $w_{\rho\star} > -1/3$.  So $c_\rho$ is negative for all $w$ in the range $-1 \le w < -1/3$.

\end{itemize}
To summarise: $c_\rho$ is negative for all accelerating $w$ if 
\be
\gamma > 1 \qquad \text{or} \qquad \gamma < \gamma_{\rho-1/3}
\ee
and in the range $\gamma_{\rho-1/3} < \gamma < 1$ then $c_\rho$ is negative if
\be
w < w_{\rho\star}
\ee
This is illustrated in Figure \ref{f:TuneParams}.

\end{itemize}
In summary, the only constraints on $A$ come from $c_\Omega$ and $c_\perp$, and can be satisfied if $\gamma \le 1$ or $\gamma \ge k-1$.  These ranges of $\gamma$ are precisely those that satisfy the $c_0$ and $c_\sigma$ constraints.  The $c_\rho$ constraint is more subtle.  In the  $\gamma > k-1$ range, the $c_\rho$ constraint is satisfied for any accelerating $w$.  For $\gamma_{\rho-1/3} < \gamma < 1$ it is satisfied for some accelerating $w$, and for $\gamma < \gamma_{\rho-1/3}$ by any accelerating $w$.  This is illustrated in Figure \ref{f:TuneParams}.

When suitable choices of $A$ and $\gamma$ exist which satisfy all of the constraints, 
the optimal solution has $\Omega= \sigma^2 = \xi_{\perp|A}=0$.  The differential equation describing the optimal solution is
\be\label{e:abDiffEq}
t \totd v t + \frac{k-\gamma-1}{k(\gamma-1)} v^2 + \frac{1-w}{1+w} v + \frac{4\left[4 + k (aw - \gamma - 1 - 3w) + 4 \gamma w \right]}{3(\gamma-1)(k+2)(1+w)^2} =0
\ee
where
\be
\frac{v(t)}{t} = \xi(t)_{0|A}
\ee
and none of the coefficients depend on $A$.
This is an analogue of the differential equation that played a central role in Section \ref{ss:RFQuintessence}.  In that case, we had two conditions, coming from the $\rho^D+P_3^D$ and $\rho^D+P_k^D$ NEC conditions.  We used the latter to define the differential equation analogous to (\ref{e:abDiffEq}), and the former to define its boundary conditions.  In the present case, we have only one inequality for each value of $\gamma$, which gives the differential equation (\ref{e:abDiffEq}).  To define the initial conditions for this equation, we re-use the $\rho^D+P_3^D$ NEC condition, which yields
\be\label{e:abBoundaryConds}
v_F = \pm \sqrt{\frac{2k}{k+2}\frac{4}{3(1+w)}}
\ee
This condition is also independent of $\Ro$.  What we are really testing is a kind of combination of the generalised energy condition defined by (\ref{e:abEnergyCondition}) and the NEC.  Since the NEC is the weakest of the classic energy conditions this is not a significant problem.




\begin{thebibliography}{22}

%
%

\bibitem{Riess:1998cb}
  A.~G.~Riess {\it et al.}  [Supernova Search Team Collaboration],
  Astron.\ J.\  {\bf 116}, 1009 (1998)
  [arXiv:astro-ph/9805201].

\bibitem{Perlmutter:1998np}
  S.~Perlmutter {\it et al.}  [Supernova Cosmology Project Collaboration],
  Astrophys.\ J.\  {\bf 517}, 565 (1999)
  [arXiv:astro-ph/9812133].
  
\bibitem{Riess:2004nr}
  A.~G.~Riess {\it et al.}  [Supernova Search Team Collaboration],
  Astrophys.\ J.\  {\bf 607}, 665 (2004)
  [arXiv:astro-ph/0402512].

\bibitem{Spergel:2006hy}
  D.~N.~Spergel {\it et al.}  [WMAP Collaboration],
  Astrophys.\ J.\ Suppl.\  {\bf 170}, 377 (2007)
  [arXiv:astro-ph/0603449].

%
%

\bibitem{Weinberg:1988cp}
  S.~Weinberg,
  Rev.\ Mod.\ Phys.\  {\bf 61} (1989) 1.

\bibitem{Sahni:1999gb}
  V.~Sahni and A.~A.~Starobinsky,
  Int.\ J.\ Mod.\ Phys.\  D {\bf 9}, 373 (2000)
  [arXiv:astro-ph/9904398].

\bibitem{Padmanabhan:2002ji}
  T.~Padmanabhan,
  Phys.\ Rept.\  {\bf 380}, 235 (2003)
  [arXiv:hep-th/0212290].
  
\bibitem{Peebles:2002gy}
  P.~J.~E.~Peebles and B.~Ratra,
  Rev.\ Mod.\ Phys.\  {\bf 75}, 559 (2003)
  [arXiv:astro-ph/0207347].

\bibitem{Copeland:2006wr}
  E.~J.~Copeland, M.~Sami and S.~Tsujikawa,
  Int.\ J.\ Mod.\ Phys.\  D {\bf 15}, 1753 (2006)
  [arXiv:hep-th/0603057].


%
%

\bibitem{Gibbons:1985}
  G. W. Gibbons, in F. del Aguila, J.A. de Azcaí?rraga, L.E. Ibaí?níÄez (eds>), {\it Supersymmetry, supergravity, and related topics}.  World Scientific, Singapore, 1985

\bibitem{Maldacena:2000mw}
  J.~M.~Maldacena and C.~Nunez,
  Int.\ J.\ Mod.\ Phys.\  A {\bf 16}, 822 (2001)
  [arXiv:hep-th/0007018].

%
%

\bibitem{Wesley:2008de}
  D.~H.~Wesley,
  arXiv:0802.2106 [hep-th].

%
%

\bibitem{Randall:1999ee}
  L.~Randall and R.~Sundrum,
  Phys.\ Rev.\ Lett.\  {\bf 83}, 3370 (1999)
  [arXiv:hep-ph/9905221].
  
\bibitem{Randall:1999vf}
  L.~Randall and R.~Sundrum,
  Phys.\ Rev.\ Lett.\  {\bf 83}, 4690 (1999)
  [arXiv:hep-th/9906064].

%
%

\bibitem{Schoen:1979a}
  R. Schoen and S. T. Yau,
  Ann. of Math. {\bf 110} 127--142 (1979)

\bibitem{Schoen:1979b}
  R. Schoen and S. T. Yau,
  Manuscripta Math {\bf 28} 159--183 (1979)

\bibitem{Gromov:1980}
  M. Gromov and H. B. Lawson, 
  Ann. of Math. {\bf 111} 209--230 (1980)


\bibitem{Besse:1987}
  Arthur L. Besse, {\it Einstein Manifolds}, Springer-Verlag, 1987.
  
\bibitem{Joyce:2000}
  Dominic D. Joyce, {\it Compact Manifolds of Special Holonomy}, Oxford University Press, 2000.


%
%

\bibitem{Candelas:1985en}
  P.~Candelas, G.~T.~Horowitz, A.~Strominger and E.~Witten,
  Nucl.\ Phys.\  B {\bf 258}, 46 (1985).

\bibitem{GSW2}
  M. B. Green, J. H. Schwarz, and E. Witten, {\it Superstring theory, volume II: Loop amplitudes, anomalies, and phenomenology}, Cambridge University Press, 1987.

\bibitem{Pol2}
  Joseph Polchinski, {\it String theory, volume II: superstring theory and beyond}, Cambridge University Press, 1999.

%
%

\bibitem{Cvetic:2000dm}
  M.~Cvetic, H.~Lu and C.~N.~Pope,
  Phys.\ Rev.\  D {\bf 62}, 064028 (2000)
  [arXiv:hep-th/0003286].

\bibitem{Cvetic:2003jy}
  M.~Cvetic, G.~W.~Gibbons, H.~Lu and C.~N.~Pope,
  Class.\ Quant.\ Grav.\  {\bf 20}, 5161 (2003)
  [arXiv:hep-th/0306043].

\bibitem{Gibbons:2003gp}
  G.~W.~Gibbons and C.~N.~Pope,
  Nucl.\ Phys.\  B {\bf 697}, 225 (2004)
  [arXiv:hep-th/0307052].

\bibitem{Cvetic:2004km}
  M.~Cvetic, G.~W.~Gibbons and C.~N.~Pope,
  Nucl.\ Phys.\  B {\bf 708}, 381 (2005)
  [arXiv:hep-th/0401151].


%
%

\bibitem{Breitenlohner:1987dg}
  P.~Breitenlohner, D.~Maison and G.~W.~Gibbons,
  Commun.\ Math.\ Phys.\  {\bf 120}, 295 (1988).

\bibitem{Cremmer:1999du}
  E.~Cremmer, B.~Julia, H.~Lu and C.~N.~Pope,
  arXiv:hep-th/9909099.

%
%

\bibitem{HawkingEllis}
S. W.  Hawking and G. F. R. Ellis, {\it The large-scale structure of space-time}, Cambridge University Press, 1973.

%
%

\bibitem{Tipler:1976bi}
  F.~J.~Tipler,
  Phys.\ Rev.\ Lett.\  {\bf 37}, 879 (1976).


\bibitem{Tipler:1977eb}
  F.~J.~Tipler,
  Annals Phys.\  {\bf 108}, 1 (1977).

\bibitem{Friedman:1993ty}
  J.~L.~Friedman, K.~Schleich and D.~M.~Witt,
  Phys.\ Rev.\ Lett.\  {\bf 71}, 1486 (1993)
  [Erratum-ibid.\  {\bf 75}, 1872 (1995)]
  [arXiv:gr-qc/9305017].

\bibitem{Cline:2003gs}
  J.~M.~Cline, S.~Jeon and G.~D.~Moore,
  Phys.\ Rev.\  D {\bf 70}, 043543 (2004)
  [arXiv:hep-ph/0311312].

\bibitem{Hsu:2004vr}
  S.~D.~H.~Hsu, A.~Jenkins and M.~B.~Wise,
  Phys.\ Lett.\  B {\bf 597}, 270 (2004)
  [arXiv:astro-ph/0406043].

\bibitem{Dubovsky:2005xd}
  S.~Dubovsky, T.~Gregoire, A.~Nicolis and R.~Rattazzi,
  JHEP {\bf 0603}, 025 (2006)
  [arXiv:hep-th/0512260].

\bibitem{Buniy:2006xf}
  R.~V.~Buniy, S.~D.~H.~Hsu and B.~M.~Murray,
  Phys.\ Rev.\  D {\bf 74}, 063518 (2006)
  [arXiv:hep-th/0606091].
  

%
%

\bibitem{Morris:1988cz}
  M.~S.~Morris and K.~S.~Thorne,
  Am.\ J.\ Phys.\  {\bf 56}, 395 (1988).

\bibitem{Visser:2003yf}
  M.~Visser, S.~Kar and N.~Dadhich,
  Phys.\ Rev.\ Lett.\  {\bf 90}, 201102 (2003)
  [arXiv:gr-qc/0301003].

%
%

\bibitem{Alcubierre:1994tu}
  M.~Alcubierre,
  Class.\ Quant.\ Grav.\  {\bf 11}, L73 (1994)
  [arXiv:gr-qc/0009013].

\bibitem{Krasnikov:1995ad}
  S.~V.~Krasnikov,
  Phys.\ Rev.\  D {\bf 57}, 4760 (1998)
  [arXiv:gr-qc/9511068].

\bibitem{Everett:1997hb}
  A.~E.~Everett and T.~A.~Roman,
  Phys.\ Rev.\  D {\bf 56}, 2100 (1997)
  [arXiv:gr-qc/9702049].

\bibitem{Pfenning:1997wh}
  M.~J.~Pfenning and L.~H.~Ford,
  Class.\ Quant.\ Grav.\  {\bf 14}, 1743 (1997)
  [arXiv:gr-qc/9702026].

\bibitem{Olum:1998mu}
  K.~D.~Olum,
  Phys.\ Rev.\ Lett.\  {\bf 81}, 3567 (1998)
  [arXiv:gr-qc/9805003].

\bibitem{Low:1998uy}
  R.~J.~Low,
  Class.\ Quant.\ Grav.\  {\bf 16}, 543 (1999)
  [arXiv:gr-qc/9812067].






%
%

\bibitem{Morris:1988tu}
  M.~S.~Morris, K.~S.~Thorne and U.~Yurtsever,
  Phys.\ Rev.\ Lett.\  {\bf 61}, 1446 (1988).
  
\bibitem{Hawking:1991nk}
  S.~W.~Hawking,
  Phys.\ Rev.\  D {\bf 46}, 603 (1992).

%
%

\bibitem{Caldwell:1999ew}
  R.~R.~Caldwell,
  Phys.\ Lett.\  B {\bf 545}, 23 (2002)
  [arXiv:astro-ph/9908168].

\bibitem{Caldwell:2003vq}
  R.~R.~Caldwell, M.~Kamionkowski and N.~N.~Weinberg,
  Phys.\ Rev.\ Lett.\  {\bf 91}, 071301 (2003)
  [arXiv:astro-ph/0302506].

%
%

\bibitem{Rubakov:2004eb}
  V.~A.~Rubakov,
  arXiv:hep-th/0407104.

\bibitem{Dubovsky:2004sg}
  S.~L.~Dubovsky,
  JHEP {\bf 0410}, 076 (2004)
  [arXiv:hep-th/0409124].

\bibitem{Dubovsky:2006vk}
  S.~L.~Dubovsky and S.~M.~Sibiryakov,
  Phys.\ Lett.\  B {\bf 638}, 509 (2006)
  [arXiv:hep-th/0603158].

\bibitem{ArkaniHamed:2007ky}
  N.~Arkani-Hamed, S.~Dubovsky, A.~Nicolis, E.~Trincherini and G.~Villadoro,
  JHEP {\bf 0705}, 055 (2007)
  [arXiv:0704.1814 [hep-th]].

\bibitem{Eling:2007qd}
  C.~Eling, B.~Z.~Foster, T.~Jacobson and A.~C.~Wall,
  Phys.\ Rev.\  D {\bf 75}, 101502 (2007)
  [arXiv:hep-th/0702124].

%
%

\bibitem{Boisseau:2000pr}
  B.~Boisseau, G.~Esposito-Farese, D.~Polarski and A.~A.~Starobinsky,
  Phys.\ Rev.\ Lett.\  {\bf 85}, 2236 (2000)
  [arXiv:gr-qc/0001066].
  
\bibitem{Bronnikov:2006pt}
  K.~A.~Bronnikov and A.~A.~Starobinsky,
  JETP Lett.\  {\bf 85}, 1 (2007)
  [Pisma Zh.\ Eksp.\ Teor.\ Fiz.\  {\bf 85},  (19??)]
  [arXiv:gr-qc/0612032].

%
%

\bibitem{Huterer:2001yu}
  D.~Huterer,
  Phys.\ Rev.\  D {\bf 65}, 063001 (2002)
  [arXiv:astro-ph/0106399].

\bibitem{Huterer:2000mj}
  D.~Huterer and M.~S.~Turner,
  Phys.\ Rev.\  D {\bf 64}, 123527 (2001)
  [arXiv:astro-ph/0012510].

\bibitem{Maor:2001ku}
  I.~Maor, R.~Brustein, J.~McMahon and P.~J.~Steinhardt,
  Phys.\ Rev.\  D {\bf 65}, 123003 (2002)
  [arXiv:astro-ph/0112526].

\bibitem{Tegmark:2003ud}
  M.~Tegmark {\it et al.}  [SDSS Collaboration],
  Phys.\ Rev.\  D {\bf 69}, 103501 (2004)
  [arXiv:astro-ph/0310723].

\bibitem{Upadhye:2004hh}
  A.~Upadhye, M.~Ishak and P.~J.~Steinhardt,
  Phys.\ Rev.\  D {\bf 72}, 063501 (2005)
  [arXiv:astro-ph/0411803].

\bibitem{Ishak:2005we}
  M.~Ishak,
  Mon.\ Not.\ Roy.\ Astron.\ Soc.\  {\bf 363}, 469 (2005)
  [arXiv:astro-ph/0501594].

%
%

\bibitem{Shiromizu:1999wj}
  T.~Shiromizu, K.~i.~Maeda and M.~Sasaki,
  Phys.\ Rev.\  D {\bf 62}, 024012 (2000)
  [arXiv:gr-qc/9910076].

\bibitem{DeWolfe:1999cp}
  O.~DeWolfe, D.~Z.~Freedman, S.~S.~Gubser and A.~Karch,
  Phys.\ Rev.\  D {\bf 62}, 046008 (2000)
  [arXiv:hep-th/9909134].

\bibitem{Brax:2004xh}
  P.~Brax, C.~van de Bruck and A.~C.~Davis,
  Rept.\ Prog.\ Phys.\  {\bf 67}, 2183 (2004)
  [arXiv:hep-th/0404011].

%
%

\bibitem{Carroll:2001ih}
  S.~M.~Carroll, J.~Geddes, M.~B.~Hoffman and R.~M.~Wald,
  Phys.\ Rev.\  D {\bf 66}, 024036 (2002)
  [arXiv:hep-th/0110149].

\bibitem{Teo:2004hq}
  E.~Teo,
  Phys.\ Lett.\  B {\bf 609}, 181 (2005)
  [arXiv:hep-th/0412164].

\bibitem{Wesley:2006jn}
  D.~H.~Wesley,
  ``Classical and quantum features of string cosmology,'' PhD thesis, Princeton University, 2006.
  
%
%

\bibitem{de Wit:1986xg}
  B.~de Wit, D.~J.~Smit and N.~D.~Hari Dass,
  Nucl.\ Phys.\  B {\bf 283}, 165 (1987).

%
%

\bibitem{MTW}
  C. W. Misner, K. S. Thorne, J. A. Wheeler, {\it Gravitation}, W. H. Freeman,   
  1973.
  
%
%

\bibitem{Eguchi:1980jx}
  T.~Eguchi, P.~B.~Gilkey and A.~J.~Hanson,
  Phys.\ Rept.\  {\bf 66}, 213 (1980).

%
%

\bibitem{ADM:62}
  R. Arnowitt, S. Deser and C. W. Misner, in {\it Gravitation: an introduction to current research}. L. Witten (ed).  Wiley, New York, 1962. [arXiv:gr-qc/0405109].

\bibitem{DeWitt:1967yk}
  B.~S.~DeWitt,
  Phys.\ Rev.\  {\bf 160}, 1113 (1967).

%
%

\bibitem{Gibbons:1978ac}
  G.~W.~Gibbons, S.~W.~Hawking and M.~J.~Perry,
  Nucl.\ Phys.\  B {\bf 138}, 141 (1978).

\bibitem{Page:1978zz}
  D.~N.~Page,
  Phys.\ Rev.\  D {\bf 18}, 2733 (1978).

%
%

\bibitem{Schon:1979rg}
  R.~Schon and S.~T.~Yau,
  Commun.\ Math.\ Phys.\  {\bf 65}, 45 (1979).

\bibitem{Schon:1981vd}
  R.~Schon and S.~T.~Yau,
  Commun.\ Math.\ Phys.\  {\bf 79}, 231 (1981).

\bibitem{Witten:1981mf}
  E.~Witten,
  Commun.\ Math.\ Phys.\  {\bf 80}, 381 (1981).

\bibitem{Ludvigsen:1981gf}
  M.~Ludvigsen and J.~A.~G.~Vickers,
  J.\ Phys.\ A  {\bf 14}, L389 (1981).

\bibitem{Horowitz:1981uw}
  G.~T.~Horowitz and M.~J.~Perry,
  Phys.\ Rev.\ Lett.\  {\bf 48}, 371 (1982).

\bibitem{Gibbons:1982jg}
  G.~W.~Gibbons, S.~W.~Hawking, G.~T.~Horowitz and M.~J.~Perry,
  Commun.\ Math.\ Phys.\  {\bf 88}, 295 (1983).

%
%

\bibitem{ArkaniHamed:2003uy}
  N.~Arkani-Hamed, H.~C.~Cheng, M.~A.~Luty and S.~Mukohyama,
  JHEP {\bf 0405}, 074 (2004)
  [arXiv:hep-th/0312099].

\bibitem{Creminelli:2006xe}
  P.~Creminelli, M.~A.~Luty, A.~Nicolis and L.~Senatore,
  JHEP {\bf 0612}, 080 (2006)
  [arXiv:hep-th/0606090].



%
%

\bibitem{Aghababaie:2002be}
  Y.~Aghababaie, C.~P.~Burgess, S.~L.~Parameswaran and F.~Quevedo,
  JHEP {\bf 0303}, 032 (2003)
  [arXiv:hep-th/0212091].

\bibitem{Carroll:2003db}
  S.~M.~Carroll and M.~M.~Guica,
  arXiv:hep-th/0302067.

\bibitem{de Rham:2005ci}
  C.~de Rham and A.~J.~Tolley,
  JCAP {\bf 0602}, 003 (2006)
  [arXiv:hep-th/0511138].
  
\bibitem{Tolley:2005nu}
  A.~J.~Tolley, C.~P.~Burgess, D.~Hoover and Y.~Aghababaie,
  JHEP {\bf 0603}, 091 (2006)
  [arXiv:hep-th/0512218].

\bibitem{Tolley:2006ht}
  A.~J.~Tolley, C.~P.~Burgess, C.~de Rham and D.~Hoover,
  New J.\ Phys.\  {\bf 8}, 324 (2006)
  [arXiv:hep-th/0608083].

\bibitem{Burgess:2007ui}
  C.~P.~Burgess,
  arXiv:0708.0911 [hep-ph].

%
%

\bibitem{ArkaniHamed:2000ds}
  N.~Arkani-Hamed, M.~Porrati and L.~Randall,
  JHEP {\bf 0108}, 017 (2001)
  [arXiv:hep-th/0012148].

\bibitem{ArkaniHamed:2000eg}
  N.~Arkani-Hamed, S.~Dimopoulos, N.~Kaloper and R.~Sundrum,
  Phys.\ Lett.\  B {\bf 480}, 193 (2000)
  [arXiv:hep-th/0001197].

\bibitem{Kachru:2000hf}
  S.~Kachru, M.~B.~Schulz and E.~Silverstein,
  Phys.\ Rev.\  D {\bf 62}, 045021 (2000)
  [arXiv:hep-th/0001206].

\bibitem{Forste:2000ft}
  S.~Forste, Z.~Lalak, S.~Lavignac and H.~P.~Nilles,
  JHEP {\bf 0009}, 034 (2000)
  [arXiv:hep-th/0006139].


\bibitem{Binetruy:2000wn}
  P.~Binetruy, J.~M.~Cline and C.~Grojean,
  Phys.\ Lett.\  B {\bf 489}, 403 (2000)
  [arXiv:hep-th/0007029].
  
\bibitem{Csaki:2000dm}
  C.~Csaki, J.~Erlich and C.~Grojean,
  Nucl.\ Phys.\  B {\bf 604}, 312 (2001)
  [arXiv:hep-th/0012143].

\bibitem{Csaki:2001mn}
  C.~Csaki, J.~Erlich and C.~Grojean,
  Gen.\ Rel.\ Grav.\  {\bf 33}, 1921 (2001)
  [arXiv:gr-qc/0105114].

%
%

\bibitem{Cline:2001yt}
  J.~M.~Cline and H.~Firouzjahi,
  Phys.\ Rev.\  D {\bf 65}, 043501 (2002)
  [arXiv:hep-th/0107198].
 
\bibitem{Apostolopoulos:2004ic}
  P.~S.~Apostolopoulos and N.~Tetradis,
  Phys.\ Rev.\  D {\bf 71}, 043506 (2005)
  [arXiv:hep-th/0412246].

\bibitem{Apostolopoulos:2005at}
  P.~S.~Apostolopoulos and N.~Tetradis,
  Phys.\ Lett.\  B {\bf 633}, 409 (2006)
  [arXiv:hep-th/0509182].
 
\bibitem{Koroteev:2007yp}
  P.~Koroteev and M.~Libanov,
  arXiv:0712.1136 [hep-th].



%
%

\bibitem{Giddings:2001yu}
  S.~B.~Giddings, S.~Kachru and J.~Polchinski,
  Phys.\ Rev.\  D {\bf 66}, 106006 (2002)
  [arXiv:hep-th/0105097].

\bibitem{DeWolfe:2002nn}
  O.~DeWolfe and S.~B.~Giddings,
  Phys.\ Rev.\  D {\bf 67}, 066008 (2003)
  [arXiv:hep-th/0208123].

\bibitem{Kachru:2003aw}
  S.~Kachru, R.~Kallosh, A.~Linde and S.~P.~Trivedi,
  Phys.\ Rev.\  D {\bf 68}, 046005 (2003)
  [arXiv:hep-th/0301240].

\bibitem{Kachru:2003sx}
  S.~Kachru, R.~Kallosh, A.~Linde, J.~M.~Maldacena, L.~P.~McAllister and S.~P.~Trivedi,
  JCAP {\bf 0310}, 013 (2003)
  [arXiv:hep-th/0308055].

\bibitem{Douglas:2006es}
  M.~R.~Douglas and S.~Kachru,
  Rev.\ Mod.\ Phys.\  {\bf 79}, 733 (2007)
  [arXiv:hep-th/0610102].

\bibitem{Burgess:2006mn}
  C.~P.~Burgess, P.~G.~Camara, S.~P.~de Alwis, S.~B.~Giddings, A.~Maharana, F.~Quevedo and K.~Suruliz,
  arXiv:hep-th/0610255.

%
%

\bibitem{PJSWesley}
  Paul J. Steinhardt, Daniel H. Wesley, in preparation.


%
%

\bibitem{Hull:1988jw}
  C.~M.~Hull and N.~P.~Warner,
  Class.\ Quant.\ Grav.\  {\bf 5}, 1517 (1988).

\bibitem{Starkman:2000dy}
  G.~D.~Starkman, D.~Stojkovic and M.~Trodden,
  Phys.\ Rev.\  D {\bf 63}, 103511 (2001)
  [arXiv:hep-th/0012226].

\bibitem{Kaloper:2000jb}
  N.~Kaloper, J.~March-Russell, G.~D.~Starkman and M.~Trodden,
  Phys.\ Rev.\ Lett.\  {\bf 85}, 928 (2000)
  [arXiv:hep-ph/0002001].

\bibitem{Starkman:2001xu}
  G.~D.~Starkman, D.~Stojkovic and M.~Trodden,
  Phys.\ Rev.\ Lett.\  {\bf 87}, 231303 (2001)
  [arXiv:hep-th/0106143].

\bibitem{Gibbons:2001wy}
  G.~W.~Gibbons and C.~M.~Hull,
  arXiv:hep-th/0111072.

\bibitem{Townsend:2001ea}
  P.~K.~Townsend,
  JHEP {\bf 0111}, 042 (2001)
  [arXiv:hep-th/0110072].

%
%

\bibitem{Gutperle:2002ai}
  M.~Gutperle and A.~Strominger,
  JHEP {\bf 0204}, 018 (2002)
  [arXiv:hep-th/0202210].

\bibitem{Chen:2002yq}
  C.~M.~Chen, D.~V.~Gal'tsov and M.~Gutperle,
  Phys.\ Rev.\  D {\bf 66}, 024043 (2002)
  [arXiv:hep-th/0204071].

\bibitem{Kruczenski:2002ap}
  M.~Kruczenski, R.~C.~Myers and A.~W.~Peet,
  JHEP {\bf 0205}, 039 (2002)
  [arXiv:hep-th/0204144].

\bibitem{Deger:2002ie}
  N.~S.~Deger and A.~Kaya,
  JHEP {\bf 0207}, 038 (2002)
  [arXiv:hep-th/0206057].

\bibitem{Ivashchuk:2002ge}
  V.~D.~Ivashchuk,
  Class.\ Quant.\ Grav.\  {\bf 20}, 261 (2003)
  [arXiv:hep-th/0208101].

\bibitem{Ohta:2003uw}
  N.~Ohta,
  Phys.\ Lett.\  B {\bf 558}, 213 (2003)
  [arXiv:hep-th/0301095].

\bibitem{Ohta:2003pu}
  N.~Ohta,
  Phys.\ Rev.\ Lett.\  {\bf 91}, 061303 (2003)
  [arXiv:hep-th/0303238].

\bibitem{Ohta:2003ie}
  N.~Ohta,
  Prog.\ Theor.\ Phys.\  {\bf 110}, 269 (2003)
  [arXiv:hep-th/0304172].

\bibitem{Ohta:2004wk}
  N.~Ohta,
  Int.\ J.\ Mod.\ Phys.\  A {\bf 20}, 1 (2005)
  [arXiv:hep-th/0411230].

%
%

\bibitem{Chen:2003ij}
  C.~M.~Chen, P.~M.~Ho, I.~P.~Neupane and J.~E.~Wang,
  JHEP {\bf 0307}, 017 (2003)
  [arXiv:hep-th/0304177].
  
\bibitem{Chen:2003dca}
  C.~M.~Chen, P.~M.~Ho, I.~P.~Neupane, N.~Ohta and J.~E.~Wang,
  JHEP {\bf 0310}, 058 (2003)
  [arXiv:hep-th/0306291].

\bibitem{Wohlfarth:2003ni}
  M.~N.~R.~Wohlfarth,
  Phys.\ Lett.\  B {\bf 563}, 1 (2003)
  [arXiv:hep-th/0304089].
  
\bibitem{Roy:2003nd}
  S.~Roy,
  Phys.\ Lett.\  B {\bf 567}, 322 (2003)
  [arXiv:hep-th/0304084].
  
\bibitem{Townsend:2003fx}
  P.~K.~Townsend and M.~N.~R.~Wohlfarth,
  Phys.\ Rev.\ Lett.\  {\bf 91}, 061302 (2003)
  [arXiv:hep-th/0303097].

%
%

\bibitem{Neupane:2003cs}
  I.~P.~Neupane,
  Class.\ Quant.\ Grav.\  {\bf 21}, 4383 (2004)
  [arXiv:hep-th/0311071].

\bibitem{Wohlfarth:2003kw}
  M.~N.~R.~Wohlfarth,
  Phys.\ Rev.\  D {\bf 69}, 066002 (2004)
  [arXiv:hep-th/0307179].

\bibitem{Neupane:2005ms}
  I.~P.~Neupane and D.~L.~Wiltshire,
  Phys.\ Lett.\  B {\bf 619}, 201 (2005)
  [arXiv:hep-th/0502003].

\bibitem{Neupane:2005nb}
  I.~P.~Neupane and D.~L.~Wiltshire,
  Phys.\ Rev.\  D {\bf 72}, 083509 (2005)
  [arXiv:hep-th/0504135].

%
%


\bibitem{Emparan:2003gg}
  R.~Emparan and J.~Garriga,
  JHEP {\bf 0305}, 028 (2003)
  [arXiv:hep-th/0304124].

\bibitem{Townsend:2003qv}
  P.~K.~Townsend,
  arXiv:hep-th/0308149.


%
%

\bibitem{Fre:2002pd}
  P.~Fre, M.~Trigiante and A.~Van Proeyen,
  Class.\ Quant.\ Grav.\  {\bf 19}, 4167 (2002)
  [arXiv:hep-th/0205119].

\bibitem{deRoo:2002jf}
  M.~de Roo, D.~B.~Westra and S.~Panda,
  JHEP {\bf 0302}, 003 (2003)
  [arXiv:hep-th/0212216].





%
%

\bibitem{Lucchin:1984yf}
  F.~Lucchin and S.~Matarrese,
  Phys.\ Rev.\  D {\bf 32}, 1316 (1985).

\bibitem{Halliwell:1986ja}
  J.~J.~Halliwell,
  Phys.\ Lett.\  B {\bf 185}, 341 (1987).

%
%

\bibitem{Steinhardt:1999eh}
  P.~J.~Steinhardt,
  Phys.\ Lett.\  B {\bf 462}, 41 (1999)
  [arXiv:hep-th/9907080].

\bibitem{Caldwell:1997ii}
  R.~R.~Caldwell, R.~Dave and P.~J.~Steinhardt,
  Phys.\ Rev.\ Lett.\  {\bf 80}, 1582 (1998)
  [arXiv:astro-ph/9708069].



%
%

\bibitem{Grana:2001xn}
  M.~Grana and J.~Polchinski,
  Phys.\ Rev.\  D {\bf 65}, 126005 (2002)
  [arXiv:hep-th/0106014].
  
\bibitem{Grana:2000jj}
  M.~Grana and J.~Polchinski,
  Phys.\ Rev.\  D {\bf 63}, 026001 (2001)
  [arXiv:hep-th/0009211].

%
%
\bibitem{Brax:2004ne}
  P.~Brax, C.~van de Bruck and A.~C.~Davis,
  Phys.\ Lett.\  B {\bf 609}, 13 (2005)
  [arXiv:hep-th/0411208].







\end{thebibliography}
\end{document}